%% file: stopro.tex
\documentclass[12pt]{article}
\usepackage{amstext,amsmath,amssymb}
\usepackage{epsfig,subeqnar}

\numberwithin{equation}{section}

\input{commands}

\input{newcommands_tesis}

\newcommand{\epspot}{\varepsilon}

\renewcommand{\Dcoeff}{D}
\newcommand{\mynewpage}{\newpage}
\renewcommand{\mc}{{\bf C}}

\begin{document}


\bibliographystyle{amsplain}

\title{ Introduction to the theory of stochastic processes
and Brownian motion problems }

\author{ \small
Lecture notes for a graduate course,
\\ \small
by J. L. Garc\'{\i}a-Palacios
(Universidad de Zaragoza)
}

\date{\small
May 2004
}

\maketitle

\begin{center}
\parbox{0.8\textwidth}{\small
\vspace*{-3.ex}
These notes are an introduction to the {\em theory of stochastic
processes} based on several sources.
The presentation mainly follows the books of van Kampen
\cite{vankampen} and Wio \cite{wio}, except for the introduction,
which is taken from the book of Gardiner \cite{gardiner} and the parts
devoted to the Langevin equation and the methods for solving Langevin
and Fokker--Planck equations, which are based on the book of Risken
\cite{risken}.
}
\end{center}

\tableofcontents
\enlargethispage*{5.ex}


\mynewpage


\section{Historical introduction}
\markboth{Introduction to the theory of stochastic
processes}{Historical introduction}

Theoretical science up to the end of the nineteenth century can be
roughly viewed as the study of solutions of differential equations and
the modelling of natural phenomena by deterministic solutions of these
differential equations.
It was at that time commonly thought that if all initial (and contour)
data could only be collected, one would be able to predict the future
with certainty.

We now know that this is not so, in at least two ways.
First, the advent of {\em quantum mechanics\/} gave rise to a new
physics, which had as an essential ingredient a purely statistical
element (the measurement process).
Secondly, the concept of {\em chaos\/} has arisen, in which even quite
simple differential equations have the rather alarming property of
giving rise to essentially unpredictable behaviours.

Chaos and quantum mechanics are not the subject of these notes, but we
shall deal with systems were limited predictability arises in the form
of fluctuations due to the finite number of their discrete
constituents, or interaction with its environment (the ``thermal
bath''), etc.
Following Gardiner \cite{gardiner} we shall give a semi-historical
outline of how a phenomenological {\em theory of fluctuating
phenomena\/} arose and what its essential points are.
\begin{figure}[!b]
\centerline{\epsfig{figure=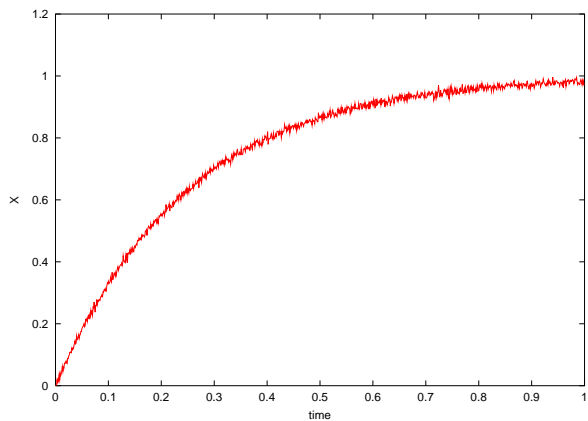,width=8.cm}}
\caption{
Schematic time evolution of a variable $X$ with a well defined
deterministic motion plus fluctuations around it.
}
\label{fluctuations}
\end{figure}

The experience of careful measurements in science normally gives us
data like that of Fig.\ \ref{fluctuations}, representing the time
evolution of a certain variable $X$.
Here a quite well defined deterministic trend is evident, which is
reproducible, unlike the fluctuations around this motion, which are
not.
This evolution could represent, for instance, the growth of the
(normalised) number of molecules of a substance $X$ formed by a
chemical reaction of the form $A\rightleftharpoons X$, or the process
of charge of a capacitor in a electrical circuit, etc.


\subsection{Brownian motion}

\begin{figure}[!b]
\centerline{\epsfig{figure=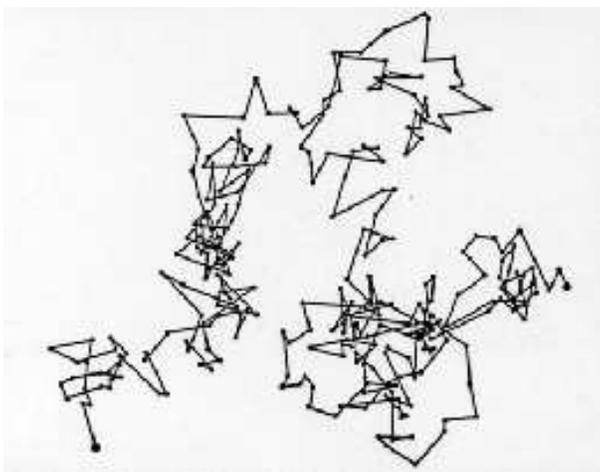,width=8.cm}}
\caption{
Motion of a particle undergoing Brownian motion.
}
\label{brownian-motion}
\end{figure}

The observation that, when suspended in water, small pollen grains are
found to be in a very animated and irregular state of motion, was
first systematically investigated by Robert Brown in 1827, and the
observed phenomenon took the name of {\em Brownian motion}.
This motion is illustrated in Fig.\ \ref{brownian-motion}.
Being a botanist, he of course tested whether this motion was in some
way a manifestation of life.
By showing that the motion was present in any suspension of fine
particles ---glass, mineral, etc.--- he ruled out any specifically
organic origin of this motion.


\subsubsection{Einstein's explanation (1905)}

A satisfactory explanation of Brownian motion did not come until 1905,
when Einstein published an article entitled {\em Concerning the
motion, as required by the molecular-kinetic theory of heat, of
particles suspended in liquids at rest}.
The same explanation was independently developed by Smoluchowski in 1906,
who was responsible for much of the later systematic development of
the theory.
To simplify the presentation, we restrict the derivation to a
one-dimensional system.

There were two major points in Einstein's solution of the problem of
Brownian motion:
\begin{itemize}
\item
The motion is caused by the exceedingly frequent impacts on the pollen
grain of the incessantly moving molecules of liquid in which it is
suspended.
\item
The motion of these molecules is so complicated that its effect on
the pollen grain can only be described probabilistically in term of
exceedingly frequent statistically independent impacts.
\end{itemize}

Einstein development of these ideas contains all the basic concepts
which make up the subject matter of these notes.
His reasoning proceeds as follows:
``It must clearly be assumed that each individual particle executes a
motion which is {\em independent of the motions of all other
particles}: it will also be considered that {\em the movements of one
and the same particle in different time intervals are independent
processes}, as long as these time intervals are not chosen too
small.''

``We introduce a time interval $\tau$ into consideration, which is
very small compared to the observable time intervals, but nevertheless
so large that in two successive time intervals $\tau$, the motions
executed by the particle can be thought of as events which are
independent of each other.''

``Now let there be a total of $n$ particles suspended in a liquid.
In a time interval $\tau$, the $X$-coordinates of the individual
particles will increase by an amount $\Delta$, where for each particle
$\Delta$ has a different (positive or negative) value.
There will be a certain {\em frequency law\/} for $\Delta$; the number
$\D n$ of the particles which experience a shift between $\Delta$ and
$\Delta+\D\Delta$ will be expressible by an equation of the form: $\D
n=n\,\phi(\Delta)\D\Delta$, where
$\int_{-\infty}^{\infty}\phi(\Delta)\D\Delta=1$, and $\phi$ is only
different from zero for very small values of $\Delta$, and satisfies
the condition $\phi(-\Delta)=\phi(\Delta)$.''

``We now investigate how the diffusion coefficient depends on $\phi$.
We shall restrict ourselves to the case where the number of particles
per unit volume depends only on $x$ and $t$.''

``Let $f(x,t)$ be the number of particles per unit volume.
We compute the distribution of particles at the time $t+\tau$ from the
distribution at time $t$.
From the definition of the function $\phi(\Delta)$, it is easy to find
the number of particles which at time $t+\tau$ are found between two
planes perpendicular to the $x$-axis and passing through points $x$
and $x+\D x$.
One obtains:
\begin{equation}
\label{chapmankolmogorov:einstein}
\boxequation{
f(x,t+\tau)\D x
=
\D x
\int_{-\infty}^{\infty}
f(x+\Delta,t)
\phi(\Delta)
\D\Delta
\;.
}
\end{equation}
But since $\tau$ is very small, we can set
\[
f(x,t+\tau)
=
f(x,t)
+
\tau
\frac{\partial f}{\partial t}
\;.
\]
Furthermore, we expand $f(x+\Delta,t)$ in powers of $\Delta$:
\[
f(x+\Delta,t)
=
f(x,t)
+
\Delta
\frac{\partial f(x,t)}{\partial x}
+
\frac{\Delta^{2}}{2!}
\frac{\partial^{2} f(x,t)}{\partial x^{2}}
+
\cdots
\;.
\]
We can use this series under the integral, because only small values
of $\Delta$ contribute to this equation.
We obtain
\begin{equation}
\label{kramersmoyal:einstein}
\boxequation{
f
+
\tau
\frac{\partial f}{\partial t}
=
f
\int_{-\infty}^{\infty}
\phi(\Delta)
\D\Delta
+
\frac{\partial f}{\partial x}
\int_{-\infty}^{\infty}
\Delta\,
\phi(\Delta)
\D\Delta
+
\frac{\partial^{2} f}{\partial x^{2}}
\int_{-\infty}^{\infty}
\frac{\Delta^{2}}{2}
\phi(\Delta)
\D\Delta
\;.
}
\end{equation}
Because $\phi(-\Delta)=\phi(\Delta)$, the second, fourth, etc. terms
on the right-hand side vanish, while out of the 1st, 3rd, 5th, etc.,
terms, each one is very small compared with the previous.
We obtain from this equation, by taking into consideration
\[
\int_{-\infty}^{\infty}
\phi(\Delta)\D\Delta
=
1
\;.
\]
and setting
\begin{equation}
\frac{1}{\tau}
\int_{-\infty}^{\infty}
\frac{\Delta^{2}}{2}
\phi(\Delta)\D\Delta
=
D
\;,
\end{equation}
and keeping only the 1st and 3rd terms of the right hand side,
\begin{equation}
\label{fokkerplanck:einstein}
\boxequation{
\frac{\partial f}{\partial t}
=
D
\frac{\partial^{2} f}{\partial x^{2}}
\;.
}
\end{equation}
This is already known as the differential equation of diffusion and it
can be seen that $D$ is the diffusion coefficient.''

\begin{figure}
\centerline{\epsfig{figure=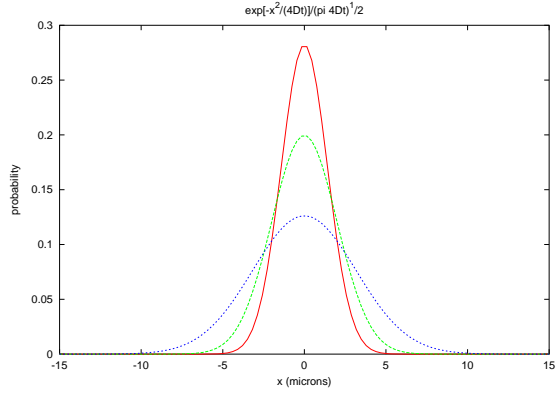,width=7.5cm}}
\vspace*{-1.ex}
\caption{
Time evolution of the non-equilibrium probability distribution
(\ref{pdf-diffusion}).
}
\label{diff}
\end{figure}

``The problem, which correspond to the problem of diffusion from a
single point (neglecting the interaction between the diffusing
particles), is now completely determined mathematically: its solution
is
\begin{equation}
\label{pdf-diffusion}
f(x,t)
=
\frac{1}{\sqrt{\pi\,4 D t}}
\,
e^{-x^{2}/4Dt}
\;.
\end{equation}
This is the solution, with the initial condition of all the Brownian
particles initially at $x=0$; this distribution is shown in Fig.\ \ref{diff}
\label{gaussint}
\footnote{
We can get the solution (\ref{pdf-diffusion}) by using the method of
the integral transform to solve partial differential equations.
Introducing the space Fourier transform of $f(x,t)$ and its inverse,
\[
F(k,t)
=
\int\!\D x\,
e^{-\iu kx}
f(x,t)
\;,
\qquad
f(x,t)
=
\frac{1}{2\pi}
\int\!\D k\,
e^{\iu kx}
F(k,t)
\;,
\]
the diffusion equation (\ref{fokkerplanck:einstein}) transforms
into the simple form
\[
\frac{\partial F}{\partial t}
=
-D\,k^{2}
F
\quad
\Longrightarrow
\quad
F(k,t)
=
F(k,0)
e^{-D\,k^{2}t}
\;.
\]
For the initial condition $f(x,t=0)=\delta(x)$, the above Fourier
transform gives $F(k,t=0)=1$.
Then, taking the inverse transform of the solution in $k$-space, we
finally have
\[
f(x,t)
=
\frac{1}{2\pi}
\int\!\D k\,
e^{\iu kx}
e^{-D\,k^{2}t}
=
\frac{e^{-x^{2}/4Dt}}{2\pi}
\underbrace{
\int\!\D k\,
e^{-Dt\,(k-\iu x/2Dt)^{2}}
}_{\sqrt{\pi/Dt}}
=
\frac{e^{-x^{2}/4Dt}}{\sqrt{\pi\,4Dt}}
\;,
\]
where in the second step we have completed the square in the argument
of the exponential $-Dk^{2}t+\iu kx=-Dt\,(k-\iu x/2Dt)^{2}-x^{2}/4Dt$,
and in the final step we have used the Gaussian integral $\int\!\D
k\,e^{-\alpha(k-b)^{2}}=\sqrt{\pi/\alpha}$, which also holds for
complex $b$.
}

Einstein ends with: ``We now calculate, with the help of this
equation, the displacement $\lambda_{x}$ in the direction of the
$X$-axis that a particle experiences on the average or, more exactly,
the square root of the arithmetic mean of the square of the
displacements in the direction of the $X$-axis; it is
\begin{equation}
\label{rms:einstein}
\lambda_{x}
=
\sqrt{\llangle x^{2}\rrangle-\llangle x_{0}^{2}\rrangle}
=
\sqrt{2\,D\,t}
\;.
\end{equation}

Einstein derivation contains very many of the major concepts which
since then have been developed more and more generally and rigorously
over the years, and which will be the subject matter of these notes.
For example:
\begin{enumerate}
\item[(i)]
The {\em Chapman--Kolgomorov equation\/} occurs as Eq.\
(\ref{chapmankolmogorov:einstein}).
It states that the probability of the particle being at point $x$ at
time $t+\tau$ is given by the sum of the probabilities of all possible
``pushes'' $\Delta$ from positions $x+\Delta$, multiplied by the
probability of being at $x+\Delta$ at time $t$.
This assumption is based on the independence of the push $\Delta$ of
any previous history of the motion; it is only necessary to know the
initial position of the particle at time $t$---not at any previous
time.
This is the {\em Markov postulate\/} and the Chapman--Kolmogorov
equation, of which Eq.\ (\ref{chapmankolmogorov:einstein}) is a
special form, is the central dynamical equation to all Markov
processes.
These will be studied in Sec.\ \ref{sec:stopro-markov}.

\item[(ii)]
{\em The Kramers--Moyal expansion.}
This is the expansion used [Eq.\ (\ref{kramersmoyal:einstein})] to go
from Eq.\ (\ref{chapmankolmogorov:einstein}) (the Chapman--Kolmogorov
equation) to the diffusion equation (\ref{fokkerplanck:einstein}).

\item[(iii)]
{\em The Fokker--Planck equation.}
The mentioned diffusion equation (\ref{fokkerplanck:einstein}), is a
special case of a Fokker--Planck equation.
This equation governs an important class of Markov processes, in which
the system has a continuous sample path.
We shall consider points (ii) and (iii) in detail in Sec.\ \ref{sec:ME-KM-FP}.

\end{enumerate}


\subsubsection{Langevin's approach (1908)}
\label{langevin:explanation}

\newcommand{\vis}{m\gamma}

Some time after Einstein's work, Langevin presented a new method which
was quite different from the former and, according to him,
``infiniment plus simple''.
His reasoning was as follows.

From statistical mechanics, it was known that the mean kinetic energy
of the Brownian particles should, in equilibrium, reach the value
\begin{equation}
\label{equipartition}
\llangle
\half
m
v^{2}
\rrangle
=
\half
\kT
\;.
\end{equation}
Acting on the particle, of mass $m$, there should be two forces:
\begin{enumerate}

\item[(i)]
a viscous force: assuming that this is given by the same formula as in
macroscopic hydrodynamics, this is $-\vis\D x/\D t$, with
$\vis=6\pi\mu a$, being $\mu$ the viscosity and $a$ the diameter of
the particle.

\item[(ii)]
a fluctuating force $\Lan(t)$, which represents the incessant impacts of
the molecules of the liquid on the Brownian particle.
All what we know about it is that is indifferently positive and
negative and that its magnitude is such that maintains the agitation
of the particle, which the viscous resistance would stop without it.

\end{enumerate}

Thus, the equation of motion for the position of the particle is given
by Newton's law as
\begin{equation}
\label{langevin:original}
\boxequation{
m
\frac{\D^{2}x}{\D t^{2}}
=
-\vis\,
\frac{\D x}{\D t}
+
\Lan(t)
\;.
}
\end{equation}
Multiplying by $x$, this can be written
\[
\frac{m}{2}
\frac{\D^{2}(x^{2})}{\D t^{2}}
-mv^{2}
=
-\frac{\vis}{2}
\frac{\D(x^{2})}{\D t}
+
\Lan\,
x
\;.
\]
If we consider a large number of identical particles, average this
equation written for each one of them, and use the equipartition
result (\ref{equipartition}) for $\llangle m v^{2} \rrangle$, we get
and equation for $\llangle x^{2}\rrangle$
\[
\frac{m}{2}
\frac{\D^{2}\llangle x^{2}\rrangle}{\D t^{2}}
+
\frac{\vis}{2}
\frac{\D\llangle x^{2}\rrangle}{\D t}
=
\kT
\;.
\]
The term $\llangle \Lan\,x\rrangle$ has been set to zero because (to
quote Langevin) ``of the irregularity of the quantity $\Lan(t)$''.
One then finds the solution ($C$ is an integration constant)
\[
\frac{\D\llangle x^{2}\rrangle}{\D t}
=
2\kT/\vis
+
C
e^{-\gamma t}
\;.
\]
Langevin estimated that the decaying exponential approaches zero with
a time constant of the order of $10^{-8}$\,s, so that $\D\llangle
x^{2}\rrangle/\D t$ enters rapidly a constant regime $\D\llangle
x^{2}\rrangle/\D t=2\kT/\vis$
Therefore, one further integration (in this asymptotic regime) leads to
\[
\llangle x^{2}\rrangle
-
\llangle x_{0}^{2}\rrangle
=
2(\kT/\vis)t
\;,
\]
which corresponds to Einstein result (\ref{rms:einstein}), provided we
identify the diffusion coefficient as
\begin{equation}
\boxequation{
D
=
\kT/\vis
\;.
}
\end{equation}
It can be seen that Einstein's condition of the independence of the
displacements $\Delta$ at different times, is equivalent to Langevin's
assumption about the vanishing of $\llangle \Lan\,x\rrangle$.
Langevin's derivation is more general, since it also yields the short
time dynamics (by a trivial integration of the neglected $C e^{-\gamma
t}$), while it is not clear where in Einstein's approach this term is
lost.

\begin{figure}
\centerline{\epsfig{figure=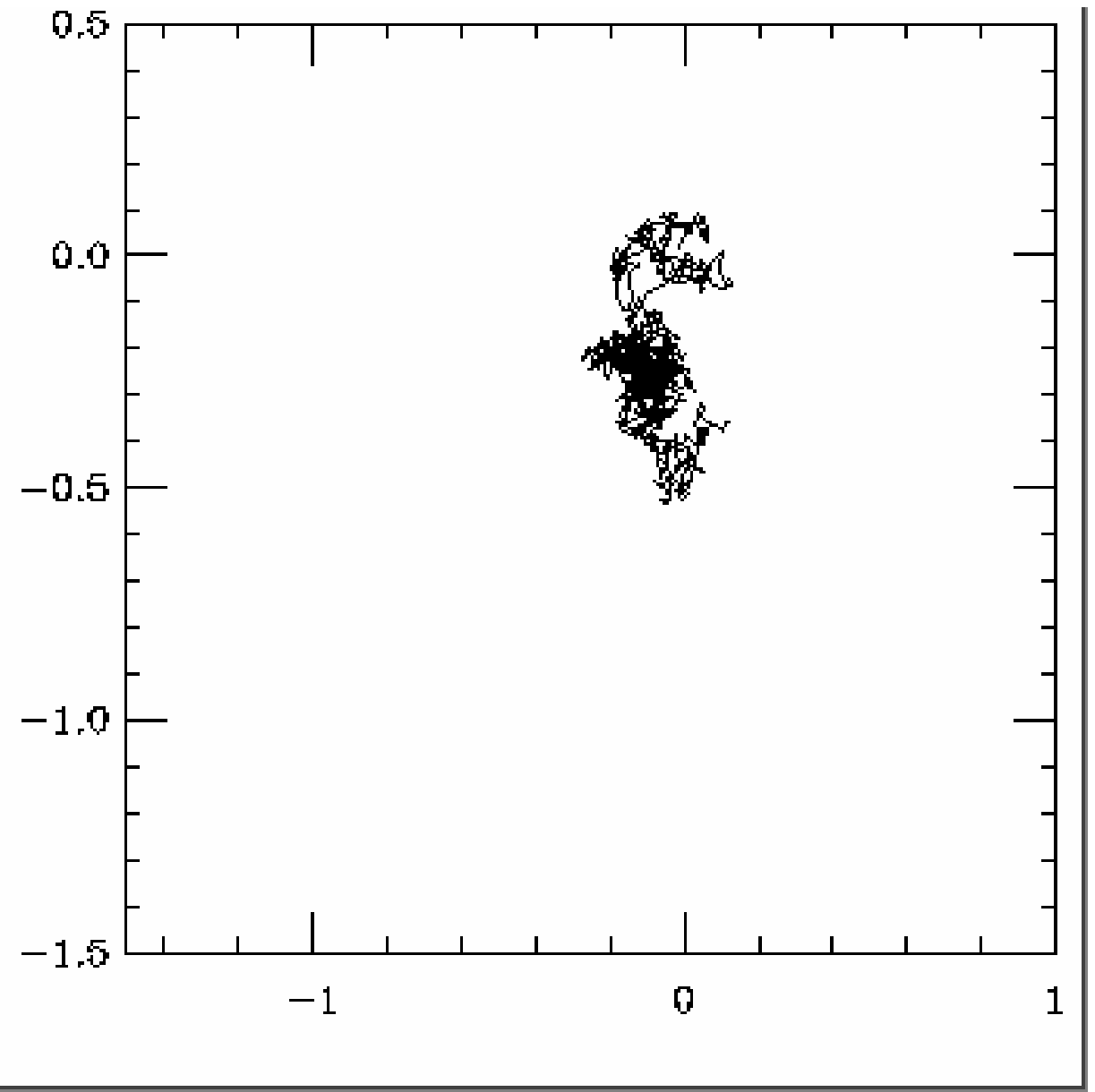,width=5.5cm}}
\centerline{\epsfig{figure=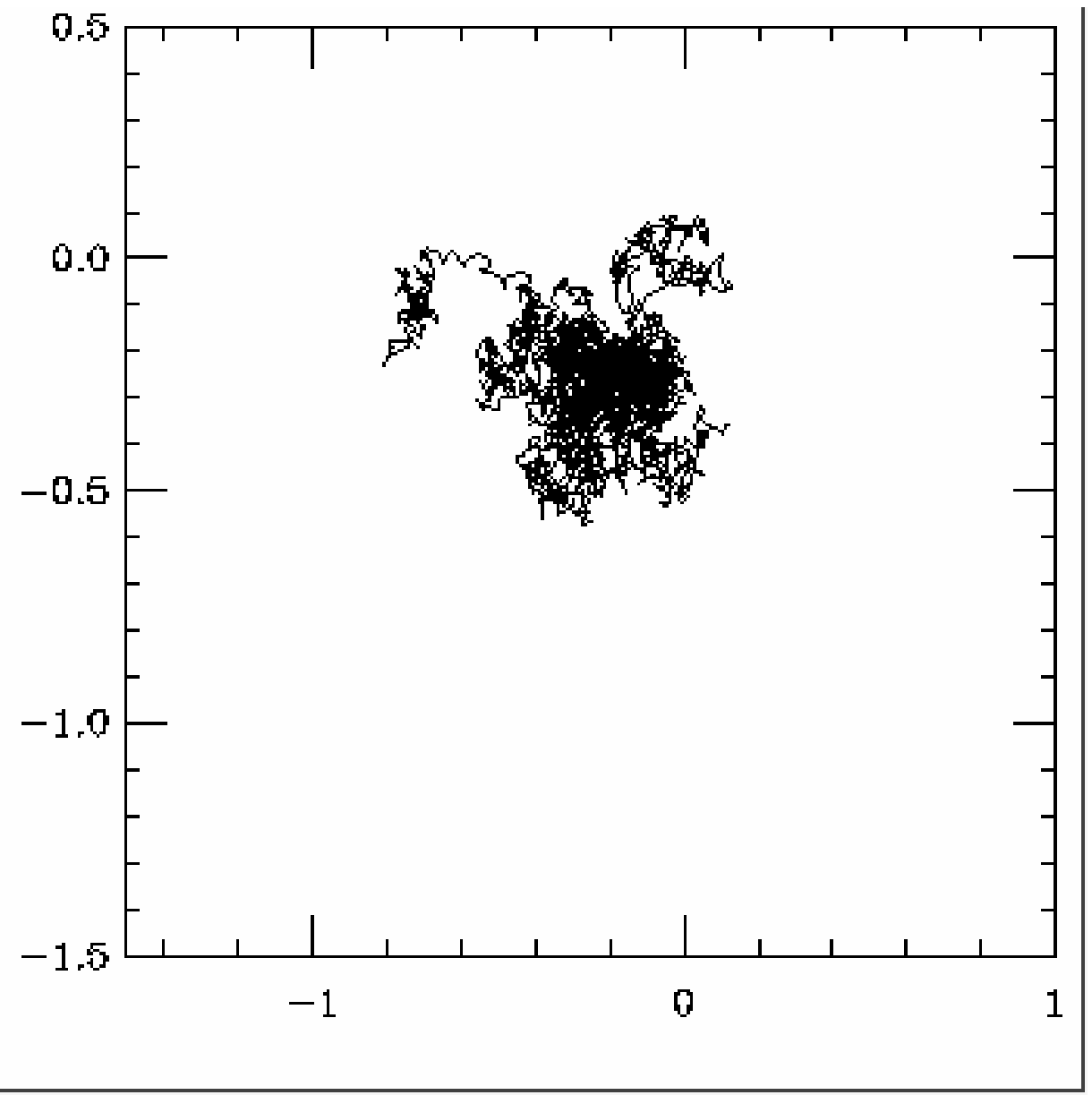,width=5.5cm}}
\centerline{\epsfig{figure=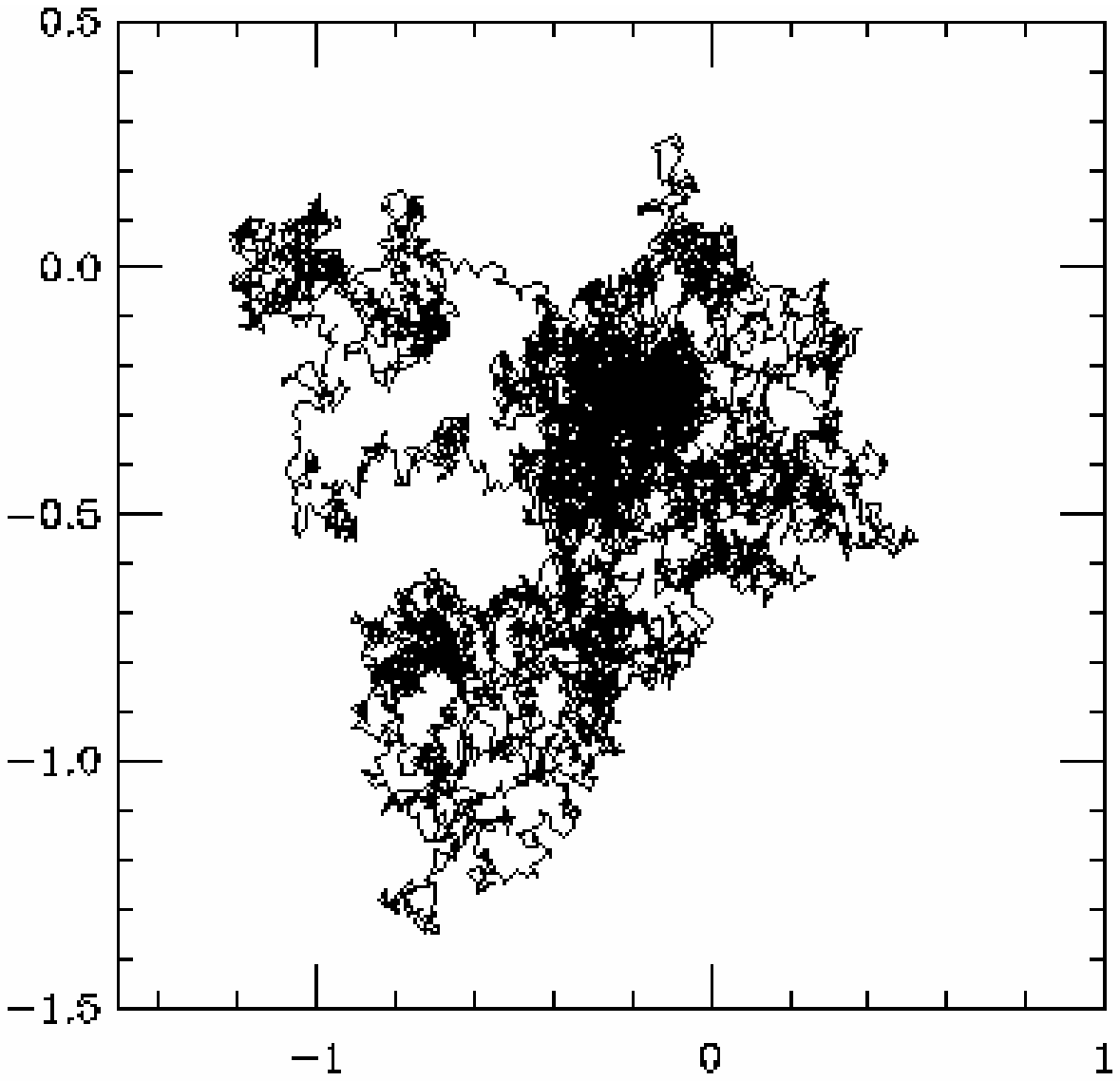,width=5.5cm}}
\caption{
Trajectory of a simulated Brownian particle projected into the $x$-$y$
plane, with $D=0.16$ $\mu$\,m$^{2}$/s.
The $x$ and $y$ axes are marked in microns.
It starts from the origin $(x,y)=(0,0)$ at $t=0$, and the pictures show the
trajectory after 1 sec, 3 sec and 10 sec.
}
\label{brownian-motion-II}
\end{figure}

Langevin's equation was the first example of a {\em stochastic
differential equation\/}--- a differential equation with a random term
$\Lan(t)$ and hence whose solution is, in some sense, a random
function.%
\footnote{
The rigorous mathematical foundation of the theory of
stochastic differential equations was not available until the work of
Ito some 40 years after Langevin's paper.
}
Each solution of the Langevin equation represents a different random
trajectory and, using only rather simple properties of the fluctuating
force $\Lan(t)$, measurable results can be derived.
Figure \ref{brownian-motion-II} shows the trajectory of a Brownian
particle in two dimensions obtained by numerical integration of the
Langevin equation (we shall also study numerical integration of
stochastic differential equations).
It is seen the growth with $t$ of the area covered by the particle,
which corresponds to the increase of $\llangle x^{2}\rrangle -\llangle
x_{0}^{2}\rrangle$ in the one-dimensional case discussed above.

The theory and experiments on Brownian motion during the first two
decades of the XX century, constituted the most important indirect
evidence of the existence of atoms and molecules (which were
unobservable at that time).
This was a strong support for the atomic and molecular theories of
matter, which until the beginning of the century still had strong
opposition by the so-called energeticits.
The experimental verification of the theory of Brownian motion awarded
the 1926 Nobel price to Svedberg and Perrin.
\footnote{
Astonishingly enough, the physical basis of the phenomenon
was already described in the 1st century B.C.E. by Lucretius in {\em
De Rerum Natura\/} (II, 112--141), a didactical poem which constitutes
the most complete account of ancient atomism and Epicureanism.
When observing dust particles dancing in a sunbeam, Lucretius
conjectured that the particles are in such irregular motion since they
are being continuously battered by the invisible blows of restless
atoms.
Although we now know that such dust particles' motion is caused by air
currents, he illustrated the right physics but only with a wrong
example.
Lucretius also extracted the right consequences from the ``observed''
phenomenon, as one that shows macroscopically the effects of the
``invisible atoms'' and hence an indication of their existence.

}

The picture of a Brownian particle immersed in a fluid is typical of a
variety of problems, even when there are no real particles.
For instance, it is the case if there is only a certain (slow or
heavy) degree of freedom that interacts, in a more or less irregular
or random way, with many other (fast or light) degrees of freedom,
which play the role of the bath.
Thus, the general concept of fluctuations describable by
Fokker--Planck and Langevin equations has developed very extensively
in a very wide range of situations.
A great advantage is the necessity of only a few parameters; in the
example of the Brownian particle, essentially the coefficients of the
derivatives in the Kramers--Moyal expansion (allowing in general the
coefficients a $x$ and $t$ dependence)
\begin{equation}
\label{coeffs}
\int_{-\infty}^{\infty}
\Delta\,
\phi(\Delta)
\D\Delta
\;,
\qquad
\int_{-\infty}^{\infty}
\frac{\Delta^{2}}{2}
\phi(\Delta)
\D\Delta
\;,
\ldots
\;.
\end{equation}
It is rare to find a problem (mechanical oscillators, fluctuations in
electrical circuits, chemical reactions, dynamics of dipoles and
spins, escape over metastable barriers, etc.) with cannot be
specified, in at least some degree of approximation, by the
corresponding Fokker--Planck equation, or equivalently, by augmenting
a deterministic differential equation with some fluctuating force or
field, like in Langevin's approach.
In the following sections we shall describe the methods developed for
a systematic and more rigorous study of these equations.

\mynewpage

\section{Stochastic variables}
\markboth{Introduction to the theory of stochastic
processes}{Stochastic variables}


\subsection{Single variable case}

A stochastic or random variable is a quantity $X$, defined by a set of
possible values $\{x\}$ (the ``range", ``sample space", or ``phase space"),
and a probability distribution on this set, $P_{X}(x)$.%
\footnote{
It is advisable to use different notations for the stochastic variable, $X$,
and for the corresponding variable in the probability distribution function,
$x$.
However, one relaxes this convention when no confusion is possible.
Similarly, the subscript $X$ is here and there dropped from the probability
distribution.
	} 
The range can be discrete or continuous, and the probability distribution
is a non-negative function, $P_{X}(x)\geq0$, with $P_{X}(x)\D x$ the
probability that $X\in(x,x+\D x)$.
The probability distribution is normalised in the sense
\[
\int\!\D x\,P_{X}(x)=1
\;,
\]
where the integral extends over the whole range of $X$.

In a discrete range, $\{x_{n}\}$, the probability distribution
consists of a number of delta-type contributions,
$P_{X}(x)=\sum_{n}p_{n}\delta(x-x_{n})$ and the above normalisation
condition reduces to $\sum_{n}p_{n}=1$.
For instance, consider the usual example of casting a die: the range is
$\{x_{n}\}=\{1,2,3,4,5,6\}$ and $p_{n}=1/6$ for each $x_{n}$ (in a honest
die).
Thus, by allowing $\delta$-function singularities in the probability
distribution, one may formally treat the discrete case by the same
expressions as those for the continuous case.


\subsubsection{Averages and moments}

The average of a function $f(X)$ defined on the range of the stochastic
variable $X$, with respect to the probability distribution of this variable,
is defined as
\[
\av{f(X)}=\int\!\D x\,f(x)P_{X}(x)
\;.
\]
The moments of the stochastic variable, $\mu_{m}$, correspond to the
special cases $f(X)=X^{m}$, i.e.,%
\footnote{This definition can formally be extended to $m=0$, with
$\mu_{0}=1$, which expresses the normalisation of $P_{X}(x)$.}
%
\begin{equation}
\label{moments}
\mu_{m}=\av{X^{m}}=\int\!\D x\,x^{m}P_{X}(x)
\;,
\qquad m=1,2,\ldots
\;.
\end{equation}


\subsubsection{Characteristic function}

This useful quantity is defined by the average of $\exp(\iu kX)$,
namely
\begin{equation}
\label{charfunc}
G_{X}(k)=\av{\exp(\iu kX)}=\int\!\D x\,\exp(\iu kx)P_{X}(x)
\;.
\end{equation}
This is merely the Fourier transform of $P_{X}(x)$, and can naturally
be solved for the probability distribution
\[
P_{X}(x)=\frac{1}{2\pi}
\int_{-\infty}^{\infty}\D k\,
\exp(-\iu kx)G_{X}(k)
\;.
\]
The function $G_{X}(k)$ provides an alternative complete
characterisation of the probability distribution.

By expanding the exponential in the integrand of Eq.\ (\ref{charfunc})
and interchanging the order of the resulting series and the integral,
one gets
\begin{equation}
\label{G:expansion}
G_{X}(k)=\sum_{m=0}^{\infty}\frac{(\iu k)^{m}}{m!}\int\!\D x\,x^{m}P_{X}(x)
=\sum_{m=0}^{\infty}\frac{(\iu k)^{m}}{m!}\mu_{m}
\;.
\end{equation}
Therefore, one finds that $G_{X}(k)$ is the {\em moment generating
function}, in the sense that
\begin{equation}
\label{G:generating}
\mu_{m}=(-i)^{m}\eval{\Dpar{^{m}}{k^{m}}G_{X}(k)}{k=0}
\;.
\end{equation}


\subsubsection{Cumulants}

The cumulants, $\kappa_{m}$, are defined as the coefficients of the
expansion of the {\em cumulant function\/} $\ln G_{X}(k)$ in powers of
$\iu k$, that is,
\[
\ln G_{X}(k)=\sum_{m=1}^{\infty}\frac{(\iu k)^{m}}{m!}\kappa_{m}
\;.
\]
Note that, owing to $P_{X}(x)$ is normalised, the $m=0$ term vanishes and
the above series begins at $m=1$.
The explicit relations between the first four cumulants and the corresponding
moments are
\begin{equation}
\label{cumulants:2}
\begin{array}{rcl}
\kappa_{1}
&=&
\mu_{1}
\\[1.ex]
\kappa_{2}
&=&
\mu_{2}-\mu_{1}^{2}
\\[1.ex]
\kappa_{3}
&=&
\mu_{3}-3\mu_{2}\mu_{1}+2\mu_{1}^{3}
\\[1.ex]
\kappa_{4}
&=&
\mu_{4}-4\mu_{3}\mu_{1}
-3\mu_{2}^{2}+12\mu_{2}\mu_{1}^{2}-6\mu_{1}^{4}
\;.
\end{array}
\end{equation}
Thus, the first cumulant is coincident with the first moment (mean) of the
stochastic variable: $\kappa_{1}=\av{X}$; the second cumulant $\kappa_{2}$,
also called the {\em variance\/} and written $\sigma^{2}$, is related to the
first and second moments via
$\sigma^{2}\equiv\kappa_{2}=\av{X^{2}}-\av{X}^{2}$.%
\footnote{
Quantities related to the third- and fourth-order cumulants have also
their own names: {\em skewness}, $\kappa_{3}/\kappa_{2}^{3/2}$, and
{\em kurtosis}, $\kappa_{4}/\kappa_{2}^{2}$.
	} 
We finally mention that there exists a general expression for
$\kappa_{m}$ in terms of the determinant of a $m\times m$ matrix
constructed with the moments $\{\mu_{i}\mid i=1,\ldots,m\}$ (see,
e.g., \cite[p.~18]{risken}):
\begin{equation}
\label{cumulants:general}
\kappa_{m}=(-1)^{m-1}
\left|
\begin{array}{cccccc}
\mu_{1} & 1 & 0 & 0 & 0 & \ldots \\[1ex]
\mu_{2} & \mu_{1} & 1 & 0 & 0 & \ldots \\[1ex]
\mu_{3} & \mu_{2} & {2\choose 1}\mu_{1} & 1 & 0 & \ldots \\[1ex]
\mu_{4} & \mu_{3} & {3\choose 1}\mu_{2} & {3\choose 2}\mu_{1} & 1 & \ldots
\\[1ex]
\mu_{5} & \mu_{4} & {4\choose 1}\mu_{3} & {4\choose 2}\mu_{2} & {4\choose
3}\mu_{1} & \ldots \\[1ex]
\ldots & \ldots & \ldots & \ldots & \ldots & \ldots
\end{array}
\right|_{m}
\;.
\end{equation}
where the ${i\choose k}$ are binomial coefficients.


\subsection{Multivariate probability distributions}

All the above definitions, corresponding to one variable, are readily
extended to higher-dimensional cases.
Consider the $n$-dimensional vector of stochastic variables
$\multi{X}=(X_{1},\ldots,X_{n})$, with a probability distribution
$P_{n}(x_{1},\ldots,x_{n})$.
This distribution is also referred to as the {\em joint probability
distribution\/} and
\[
P_{n}(x_{1},\ldots,x_{n})\D x_{1}\cdots \D x_{n}
\;,
\]
is the probability that $X_{1},\ldots,X_{n}$ have certain values between
$(x_{1},x_{1}+\D x_{1})$,\ldots, $(x_{n},x_{n}+\D x_{n})$.


\paragraph*{Partial distributions.}

One can also consider the probability distribution for some of the
variables.
This can be done in various ways:
\begin{enumerate}
\item
Take a subset of $s<n$ variables $X_{1},\ldots,X_{s}$.
The probability that they have certain values in
$(x_{1},x_{1}+\D x_{1}),\ldots,(x_{s},x_{s}+\D x_{s})$, regardless
of the values of the remaining variables $X_{s+1},\ldots,X_{n}$, is
\[
P_{s}(x_{1},\ldots,x_{s}) =\int\!\D x_{s+1}\cdots \D x_{n}\,
P_{n}(x_{1},\ldots,x_{s},x_{s+1},\ldots,x_{n})
\;,
\]
which is called the {\em marginal distribution\/} for the subset
$X_{1},\ldots,X_{s}$.
Note that from the normalisation of the joint probability it immediately
follows the normalisation of the marginal probability.
\item
Alternatively, one may attribute fixed values to $X_{s+1},\ldots,X_{n}$, and
consider the joint probability of the remaining variables
$X_{1},\ldots,X_{s}$.
This is called the {\em conditional probability}, and it is denoted by
\[
P_{s|n-s}(x_{1},\ldots,x_{s}|x_{s+1},\ldots,x_{n})
\;.
\]
This distribution is constructed in such a way that the total joint
probability $P_{n}(x_{1},\ldots,x_{n})$ is equal to the marginal probability
for $X_{s+1},\ldots,X_{n}$ to have the values $x_{s+1},\ldots,x_{n}$, times
the conditional probability that, this being so, the remaining variables
$X_{1},\ldots,X_{s}$ have the values $(x_{1},\ldots,x_{s})$.
This is Bayes' rule, and can be considered as the {\em definition\/} of
the conditional probability:
\[
P_{n}(x_{1},\ldots,x_{n})
=P_{n-s}(x_{s+1},\ldots,x_{n})
P_{s|n-s}(x_{1},\ldots,x_{s}|x_{s+1},\ldots,x_{n})
\;.
\]
Note that from the normalisation of the joint and marginal probabilities
it follows the normalisation of the conditional probability.
\end{enumerate}


\paragraph*{Characteristic function: moments and cumulants.}

For multivariate probability distributions, the moments are defined by
\[
\av{X_{1}^{m_{1}}\cdots X_{n}^{m_{n}}} =\int\!\D x_{1}\cdots
\D x_{n}\,x_{1}^{m_{1}}\cdots x_{n}^{m_{n}} P(x_{1},\ldots,x_{n})
\;,
\]
while the characteristic (moment generating) function depends on $n$
auxiliary variables $\multi{k}=(k_{1},\ldots,k_{n})$:
\begin{eqnarray}
\label{charfunc:n-dim}
G(\multi{k})
&=&
\av{\exp[\iu (k_{1}X_{1}+\cdots+k_{n}X_{n})]}
\nonumber\\
&=&
\sum_{0}^{\infty}
\frac{(\iu k_{1})^{m_{1}}\cdots(\iu k_{n})^{m_{n}}}{m_{1}!\cdots m_{n}!}
\av{X_{1}^{m_{1}}\cdots X_{n}^{m_{n}}}
\;.
\end{eqnarray}
Similarly, the cumulants of the multivariate distribution, indicated by
double brackets, are defined in terms of the coefficients of the expansion
of $\ln G$ as
\[
\ln G(\multi{k}) =\sum_{0}^{\infty}{^{^{}}}'
\frac{(\iu k_{1})^{m_{1}}\cdots(\iu k_{n})^{m_{n}}}{m_{1}!\cdots m_{n}!}
\cum{X_{1}^{m_{1}}\cdots X_{n}^{m_{n}}}
\;,
\]
where the prime indicates the absence of the term with all the $m_{i}$
simultaneously vanishing (by the normalisation of $P_{n}$).


\paragraph*{Covariance matrix.}

The second-order moments may be combined into a $n$ by $n$ matrix
$\av{X_{i}X_{j}}$.
More relevant is, however, the {\em covariance matrix}, defined by the
second-order cumulants
\[
\cum{X_{i}X_{j}}=\av{X_{i}X_{j}}-\av{X_{i}}\av{X_{j}}
=\av{[X_{i}-\av{X_{i}}][X_{j}-\av{X_{j}}]}
\;.
\]
Each diagonal element is called the {\em variance\/} of the corresponding
variable, while the off-diagonal elements are referred to as the {\em
covariance\/} of the corresponding pair of variables.%
\footnote{Note that $\cum{X_{i}X_{j}}$ is, by construction, a symmetrical
matrix.}


\paragraph*{Statistical independence.}

A relevant concept for multivariate distributions is that of {\em
statistical independence}.
One says that, e.g., two stochastic variables $X_{1}$ and $X_{2}$ are
statistically independent of each other if their joint probability
distribution factorises:
\[
P_{X_{1}X_{2}}(x_{1},x_{2})=P_{X_{1}}(x_{1})P_{X_{2}}(x_{2})
\;.
\]
The statistical independence of $X_{1}$ and $X_{2}$ is also expressed by any
one of the following three equivalent criteria:
\begin{enumerate}
\item All moments factorise:
$\av{X_{1}^{m_{1}}X_{2}^{m_{2}}}=\av{X_{1}^{m_{1}}}\av{X_{2}^{m_{2}}}
\;.$
\item The characteristic function factorises:
$G_{X_{1}X_{2}}(k_{1},k_{2})=G_{X_{1}}(k_{1})G_{X_{2}}(k_{2})
\;.$
\item The cumulants $\cum{X_{1}^{m_{1}}X_{2}^{m_{2}}}$ vanish when both
$m_{1}$ and $m_{2}$ are $\neq0$.
\end{enumerate}
Finally, two variables are called {\em uncorrelated\/} when its {\em
covariance}, $\cum{X_{1}X_{2}}$, is zero, which is a condition weaker
than that of statistical independence.


\subsection{The Gaussian distribution}
\label{gaussian}

This important distribution is defined as
\begin{equation}
\label{gaussian:1D}
P(x)
=\frac{1}{\sqrt{2\pi\sigma^{2}}}
\exp\lrs{-\frac{(x-\mu_{1})^{2}}{2\sigma^{2}}}
\;.
\end{equation}
It is easily seen that $\mu_{1}$ is indeed the average and $\sigma^{2}$
the variance, which justifies the notation.
The corresponding characteristic function is
\begin{equation}
\label{gaussian:1D:charfunct}
G(k)
=\exp(\iu \mu_{1}k-\half \sigma^{2}k^{2})
\;,
\end{equation}
as can be seen from the definition (\ref{charfunc}), by completing the
square in the argument of the total exponential $\iu
kx-(x-\mu_{1})^{2}/2\sigma^{2}$ and using the Gaussian integral
$\int\!\D k\,e^{-\alpha(k-b)^{2}}=\sqrt{\pi/\alpha}$ for complex $b$ as
in the footnote in p.\ \pageref{gaussint}.
Note that the logarithm of this characteristic function comprises
terms up to quadratic in $k$ only.
Therefore, all the cumulants after the second one vanish identically,
which is a property that indeed {\em characterises\/} the Gaussian
distribution.

For completeness, we finally write the Gaussian distribution for $n$
variables $\multi{X}=(X_{1},\ldots,X_{n})$ and the corresponding
characteristic function
\begin{eqnarray*}
P(\multi{x})
&=&
\sqrt{\frac{\det\hat{A}}{(2\pi)^{n}}}
\exp\lrs{
	-\half (\multi{x}-\multi{x}_{0})
	\cdot\hat{A}\cdot(\multi{x}-\multi{x}_{0})
	}
\\
G(\multi{k})
&=&
\exp\lr{ i\multi{x}_{0}\cdot\multi{k}
-\half \multi{k}\cdot\hat{A}^{-1}\cdot\multi{k}
	}
\;.
\end{eqnarray*}
The averages and covariances are given by
$\av{\multi{X}}=\multi{x}_{0}$ and $\cum{X_{i}X_{j}}
=(\hat{A}^{-1})_{ij}$.


\subsection{Transformation of variables}
\label{transformation_variables}

For a given stochastic variable $X$, every related quantity $Y=f(X)$ is
again a stochastic variable.
The probability that $Y$ has a value between $y$ and $y+\Delta y$ is
\[
P_{Y}(y)\Delta y
=\int\limits_{y<f(x)<y+\Delta y}\hspace{-2em}\D x\,P_{X}(x)
\;,
\]
where the integral extends over all intervals of the range of $X$ where the
inequality is obeyed.
Note that one can equivalently {\em define\/} $P_{Y}(y)$ as%
\footnote{
Note also that from Eq.\ (\ref{transformation}), one can
formally write the probability distribution for $Y$ as the following
average [with respect to $P_{X}(x)$ and taking $y$ as a parameter]
\begin{equation}
\label{transformation:2}
P_{Y}(y)=\av{\delta[y-f(X)]}
\;.
\end{equation}
%
	}
\begin{equation}
\label{transformation}
\boxequation{
P_{Y}(y)=\int\!\D x\,\delta[y-f(x)]P_{X}(x)
\;.
            }
\end{equation}
From this expression one can calculate the characteristic function of $Y$:
\begin{eqnarray*}
G_{Y}(k)
&
\stackrel{\text{Eq.\ (\ref{charfunc})}}{=}
&
\int\!\D y\,\exp(\iu ky)P_{Y}(y)
\\
&=&
\int\!\D x\,P_{X}(x)\int\!\D y\,\exp(\iu ky)\delta[y-f(x)]
\\
&=&
\int\!\D x\,P_{X}(x)\exp[\iu kf(x)]
\;,
\end{eqnarray*}
which can finally be written as
\begin{equation}
\label{charfunc:transformation}
\boxequation{
G_{Y}(k)=\av{\exp[\iu kf(X)]}
\;.
}
\end{equation}
As the simplest example consider the linear transformation $Y=\alpha X$.
The above equation then yields $G_{Y}(k)=\av{\exp(\iu k\alpha X)}$, whence
\begin{equation}
\label{charfunc:transformation:linear}
G_{Y}(k)=G_{X}(\alpha k)
\;,
\qquad
(Y=\alpha X)
\;.
\end{equation}


\subsection{Addition of stochastic variables}

The above equations for the transformation of variables remain valid when
$X$ stands for a stochastic variable with $n$ components and $Y$ for one
with $s$ components, where $s$ may or may not be equal to $n$.
For example, let us consider the case of the addition of two stochastic
variables $Y=f(X_{1},X_{2})=X_{1}+X_{2}$, where $s=1$ and $n=2$.
Then, from Eq.\ (\ref{transformation}) one first gets
\begin{eqnarray}
\label{transformation:sum}
P_{Y}(y)
&=&
\int\!\D x_{1}\,
\int\!\D x_{2}\,
\delta[y-(x_{1}+x_{2})]P_{X_{1}X_{2}}(x_{1},x_{2})
\nonumber\\
&=&
\int\!\D x_{1}\,P_{X_{1}X_{2}}(x_{1},y-x_{1})
\;.
\end{eqnarray}


\paragraph*{Properties of the sum of stochastic variables.}

One easily deduces the following three rules concerning the addition of
stochastic variables:
\begin{enumerate}
\item
Regardless of whether $X_{1}$ and $X_{2}$ are independent or not, one
has%
\footnote{
Proof of Eq.\ (\ref{averages:sum}):
\begin{eqnarray*}
\av{Y}
&\equiv&
\int\!\D y\,yP_{Y}(y)
=\int\!\D x_{1}\,\int\!\D x_{2}\,P_{X_{1}X_{2}}(x_{1},x_{2})
\int\!\D y\,y\,\delta[y-(x_{1}+x_{2})]
\\
&=&
\int\!\D x_{1}\,\int\!\D x_{2}\,
P_{X_{1}X_{2}}(x_{1},x_{2})(x_{1}+x_{2})
=\av{X_{1}}+\av{X_{2}}
\;.
\qed
\end{eqnarray*}
	}
\begin{equation}
\label{averages:sum}
\av{Y}=\av{X_{1}}+\av{X_{2}}
\;.
\end{equation}

\item
If $X_{1}$ and $X_{2}$ are {\em uncorrelated}, $\cum{X_{1}X_{2}}=0$, a
similar relation holds for the variances%
\footnote{
Proof of Eq.\ (\ref{variances:sum}):
\[
\av{Y^{2}}
\equiv
\int\!\D y\,y^{2}P_{Y}(y)
=\int\!\D x_{1}\,\int\!\D x_{2}\,
P_{X_{1}X_{2}}(x_{1},x_{2})(x_{1}+x_{2})^{2}
=
\av{X_{1}^{2}}
+\av{X_{2}^{2}}
+2\av{X_{1}X_{2}}
\;.
\]
Therefore
\begin{eqnarray*}
\cum{Y^{2}}
&=&
\av{Y^{2}}-\av{Y}^{2}
=
\underbrace{\av{X_{1}^{2}}-\av{X_{1}}^{2}}_{\cum{X_{1}^{2}}}
+\underbrace{\av{X_{2}^{2}}-\av{X_{2}}^{2}}_{\cum{X_{2}^{2}}}
+2\underbrace{\lr{\av{X_{1}X_{2}}-\av{X_{1}}\av{X_{2}}}}_{\cum{X_{1}X_{2}}}
\end{eqnarray*}
from which the statement follows for uncorrelated variables. \qed
}
\begin{equation}
\label{variances:sum}
\cum{Y^{2}}=\cum{X_{1}^{2}}+\cum{X_{2}^{2}}
\;.
\end{equation}

\item
The characteristic function of $Y=X_{1}+X_{2}$ is%
\footnote{
Proof of Eq.\ (\ref{charfunc:sum}):
\[
G_{Y}(k)
\stackrel{\text{Eq.\ (\ref{charfunc:transformation})}}{=}
\av{\exp[\iu kf(X_{1},X_{2})]}
=
\av{\exp[\iu k(X_{1}+X_{2})]}
\stackrel{\text{Eq.\ (\ref{charfunc:n-dim})}}{=}
G_{X_{1}X_{2}}(k,k)
\;.
\qed
\]
}
\begin{equation}
\label{charfunc:sum}
G_{Y}(k)=G_{X_{1}X_{2}}(k,k)
\;.
\end{equation}
\end{enumerate}
On the other hand, if $X_{1}$ and $X_{2}$ are {\em independent}, Eq.\
(\ref{transformation:sum}) and the factorization of $P_{X_{1}X_{2}}$
and $G_{X_{1}X_{2}}$ yields
\begin{equation}
\label{transformation:sum:independent}
P_{Y}(y)
=\int\!\D x_{1}\,
P_{X_{1}}(x_{1})P_{X_{2}}(y-x_{1})
\;,
\qquad
G_{Y}(k)
=G_{X_{1}}(k)G_{X_{2}}(k)
\;.
\end{equation}
Thus, the probability distribution of the sum of two {\em
independent\/} variables is the {\em convolution\/} of their
individual probability distributions.
Correspondingly, the characteristic function of the sum [which is the
Fourier transform of the probability distribution; see Eq.\
(\ref{charfunc})] is the {\em product\/} of the individual
characteristic functions.


\subsection{Central limit theorem}

As a particular case of transformation of variables, one can also
consider the sum of an arbitrary number of stochastic variables.
Let $X_{1},\ldots,X_{n}$ be a set of $n$ {\em independent\/}
stochastic variables, each having the same probability distribution
$P_{X}(x)$ with zero average and (finite) variance $\sigma^{2}$.
Then, from Eqs.\ (\ref{averages:sum}) and (\ref{variances:sum}) it
follows that their sum $Y=X_{1}+\cdots+X_{n}$ has zero average and
variance $n\sigma^{2}$, which grows linearly with $n$.
On the other hand, the distribution of the arithmetic mean of the
variables, $(X_{1}+\cdots+X_{n})/n$, becomes narrower with increasing
$n$ (variance $\sigma^{2}/n$).
It is therefore more convenient to define a suitable scaled sum
\[
Z=\frac{X_{1}+\cdots+X_{n}}{\sqrt{n}}
\;,
\]
which has variance $\sigma^{2}$ for all $n$.

The {\em central limit theorem\/} states that, even when $P_{X}(x)$ is not
Gaussian, the probability distribution of the $Z$ so-defined tends, as
$n\to\infty$, to a Gaussian distribution with zero mean and variance
$\sigma^{2}$.
This remarkable result is responsible for the important r\^{o}le of the
Gaussian distribution in all fields in which statistics are used and, in
particular, in the equilibrium and non-equilibrium statistical physics.


\paragraph*{Proof of the central limit theorem.}

We begin by expanding the characteristic function of an arbitrary
$P_{X}(x)$ with zero mean as [cf.\ Eq.\ (\ref{G:expansion})]
\begin{equation}
\label{charfunc:expansion}
G_{X}(k)=\int\!\D x\,\exp(\iu kx)P_{X}(x)=1-\half \sigma^{2}k^{2}+\cdots
\;.
\end{equation}
The factorization of the characteristic function of the sum
$Y=X_{1}+\cdots+X_{n}$ of statistically independent variables
[Eq.\ (\ref{transformation:sum:independent})], yields
\[
G_{Y}(k)=\prod_{i=1}^{n}G_{X_{i}}(k)=\lrs{G_{X}(k)}^{n}
\;,
\]
where the last equality follows from the equivalent statistical properties
of the different variables $X_{i}$.
Next, on accounting for $Z=Y/\sqrt{n}$, and using the result
(\ref{charfunc:transformation:linear}) with $\alpha=1/\sqrt{n}$, one has
\begin{eqnarray}
\label{CLT:proof}
G_{Z}(k)
=G_{Y}\lr{\frac{k}{\sqrt{n}}}
=\lrs{G_{X}\lr{\frac{k}{\sqrt{n}}}}^{n}
\simeq\lr{1-\frac{\sigma^{2}k^{2}}{2n}}^{n}
\stackrel{n\to\infty}{\longrightarrow}
\exp\lr{-\half \sigma^{2}k^{2}}
\;,
\quad
\end{eqnarray}
where we have used the definition of the exponential 
$e^{x}={\displaystyle \lim_{n\to\infty}}\lr{1+x/n}^{n}$.
Finally, on comparing the above result with Eqs.\ (\ref{gaussian:1D}),
one gets
\[
P_{Z}(z)=\frac{1}{\sqrt{2\pi\sigma^{2}}}
\exp\lr{-\frac{z^{2}}{2\sigma^{2}}}
\;.\qed
\]

\addtocounter{equation}{2}

\paragraph*{Remarks on the validity of the central limit theorem.}

The derivation of the central limit theorem can be done under more
general conditions.
For instance, it is not necessary that all the cumulants (moments) exist.
However, it is necessary that the moments up to at least second order
exist [or, equivalently, $G_{X}(k)$ being twice differentiable at $k=0$;
see Eq.\ (\ref{G:generating})].
The necessity of this condition is illustrated by the counter-example
provided by the Lorentz--Cauchy distribution:
\[
P(x)=\frac{1}{\pi}\frac{\gamma}{x^{2}+\gamma^{2}}
\;,
\qquad (-\infty<x<\infty)
\;.
\]
It can be shown that, if a set of $n$ independent variables $X_{i}$ have
Lorentz--Cauchy distributions, their sum also has a Lorentz--Cauchy
distribution (see footnote below).
However, for this distribution the conditions for the central limit theorem
to hold are not met, since the integral (\ref{moments}) defining the moments
$\mu_{m}$, does not converge even for $m=1$.%
\footnote{This can also be demonstrated by calculating the corresponding
characteristic function.
To do so, one can use
$\int_{-\infty}^{-\infty}\! d{x}\,\tyfrac{e^{iax}}{(1+x^{2})}=\pi e^{-|a|}$,
which is obtained by computing the residues of the integrand in
the upper (lower) half of the complex plane for $a>0$ ($a<0$).
Thus, one gets
\[
G(k)=\exp(-\gamma|k|)
\;,
\]
which, owing to the presence of the modulus of $k$, is not differentiable at
$k=0$.\qed

We remark in passing that, from $G_{X_{i}}(k)=\exp(-\gamma_{i}|k|)$ and the
second Eq.\ (\ref{transformation:sum:independent}), it follows that the
distribution of the sum of {\em independent\/} Lorentz--Cauchy variables
has a Lorentz--Cauchy distribution (with
$G_{Y}(k)=\exp[-(\sum_{i}\gamma_{i})|k|]$).
	} 

Finally, although the condition of independence of the variables
is important, it can be relaxed to incorporate a sufficiently weak
statistical dependence.


\subsection{Exercise: marginal and conditional probabilities and
moments of a bivariate Gaussian distribution}

To illustrate the definitions given for multivariate distributions,
let us compute them for a simple two-variable Gaussian distribution
\begin{equation}
\label{2Dgauss}
P_{2}(x_{1},x_{2})
=
\sqrt{\frac{1-\lambda^{2}}{(2\pi\sigma^{2})^{2}}}
\exp
\left[
-\frac{1}{2\sigma^{2}}
\left(
x_{1}^{2}-2\lambda\,x_{1}x_{2}+x_{2}^{2}
\right)
\right]
\;,
\end{equation}
where $\lambda$ is a parameter $-1<\lambda<1$, to ensure that the
quadratic form in the exponent is definite positive (the equivalent
condition to assume $\sigma^{2}$ to be positive in the one-variable
Gaussian distribution (\ref{gaussian:1D}).
The normalisation factor can be seen to take this value by direct
integration, or by comparing our distribution with the
multidimensional Gaussian distribution (Sec.\ \ref{gaussian});
here
$\hat{A}
=
\displaystyle
\frac{1}{\sigma^{2}}
\left(
\begin{array}{cc}
1
&
-\lambda
\\[-0.5ex]
-\lambda
&
1
\end{array}
\right)$
so that $\det\hat{A}=(1-\lambda^{2})/\sigma^{4}$.
Finally, if one wishes to fix ideas one can interpret
$P_{2}(x_{1},x_{2})$ as the Boltzmann distribution of two harmonic
oscillators coupled by a potential term $\propto\lambda\,x_{1}x_{2}$.

Let us first rewrite the distribution in a form that will facilitate to do
the integrals by completing once more the
square $-2\lambda\,x_{1}x_{2}+x_{2}^{2}=(x_{2}-\lambda
x_{1})^{2}-\lambda^{2}x_{1}^{2}$
\begin{equation}
\label{2Dgauss:2}
P_{2}(x_{1},x_{2})
=
C
\exp
\left[
-\frac{1-\lambda^{2}}{2\sigma^{2}}
x_{1}^{2}
\right]
\,
\exp
\left[
-\frac{1}{2\sigma^{2}}
(x_{2}-\lambda x_{1})^{2}
\right]
\;.
\end{equation}
and $C=\sqrt{(1-\lambda^{2})/(2\pi\sigma^{2})^{2}}$ is the
normalisation constant.
We can now compute the marginal probability of the individual
variables (for one of them since they are equivalent), defined by
$P_{1}(x_{1}) = \int \D x_{2}\, P_{2}(x_{1},x_{2})$
\[
P_{1}(x_{1})
=
C
\,
e^
{
-\frac{1-\lambda^{2}}{2\sigma^{2}}
x_{1}^{2}
}
\underbrace{
\int\!\D x_{2}\,
e^
{
-(x_{2}-\lambda x_{1})^{2}/2\sigma^{2}
}
}_{\sqrt{\pi/\alpha}\quad\alpha=1/2\sigma^{2}}
\;.
\]
Therefore, recalling the form of $C$, we merely have
\begin{equation}
\label{2Dgauss:marginal}
P_{1}(x_{1})
=
\frac{1}{\sqrt{2\pi\sigma_{\lambda}^{2}}}
\,
\exp
\left(
-\frac{x_{1}^{2}}{2\sigma_{\lambda}^{2}}
\right)
\;,
\qquad
\mbox{with}
\qquad
\sigma_{\lambda}^{2}=\sigma^{2}/(1-\lambda^{2})
\;.
\end{equation}

We see that the marginal distribution depends on $\lambda$, which
results in a modified variance.
To see that $\sigma_{\lambda}^{2}$ is indeed the variance
$\cum{x_{1}^{2}}=\av{x_{1}^{2}}-\av{x_{1}}^{2}$, note that $\llangle
x_{1}^{m}\rrangle$ can be obtained from the marginal distribution only
(this is a {\em general\/} result)
\[
\llangle x_{1}^{m}\rrangle
=
\int\!\D x_{1}
\int\!\D x_{2}\,
x_{1}^{m}
P_{2}(x_{1},x_{2})
=
\int\!\D x_{1}\,
x_{1}^{m}
\!
\underbrace{
\int\!\D x_{2}\,
P_{2}(x_{1},x_{2})
}_{P_{1}(x_{1})}
=
\int\!\D x_{1}\,
x_{1}^{m}
P_{1}(x_{1})
\]
Then inspecting the marginal distribution obtained [Eq.\
(\ref{2Dgauss:marginal})] we get that the first moments vanish and the
variances are indeed equal to $\sigma_{\lambda}^{2}$:
\begin{equation}
\label{2Dgauss:moments}
\boxequation{
\begin{array}{rclcrcl}
\llangle x_{1}\rrangle
&
=
&
0
&
&
\llangle x_{2}\rrangle
&
=
&
0
\\[1.ex]
\llangle x_{1}^{2}\rrangle
&
=
&
\sigma_{\lambda}^{2}
&
&
\llangle x_{2}^{2}\rrangle
&
=
&
\sigma_{\lambda}^{2}
\end{array}
}
\end{equation}
\begin{figure}[!b]
\hspace*{-3.5em}
\begin{tabular}{cc}
\epsfig{figure=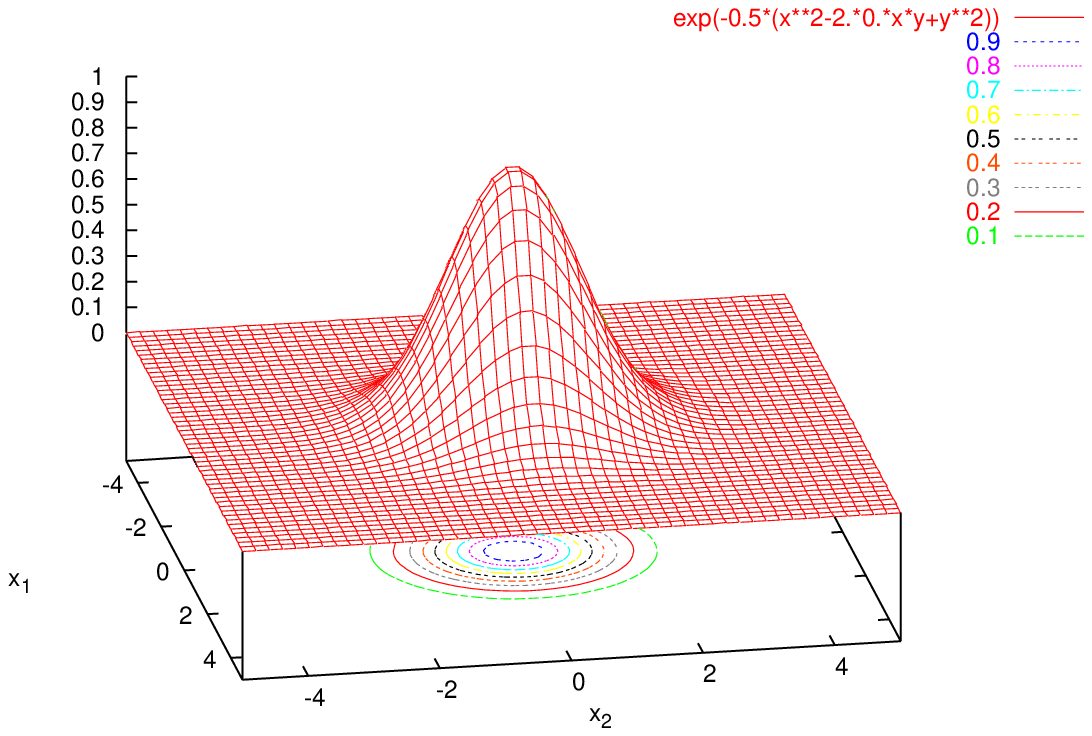,width=7.5cm}
&
\epsfig{figure=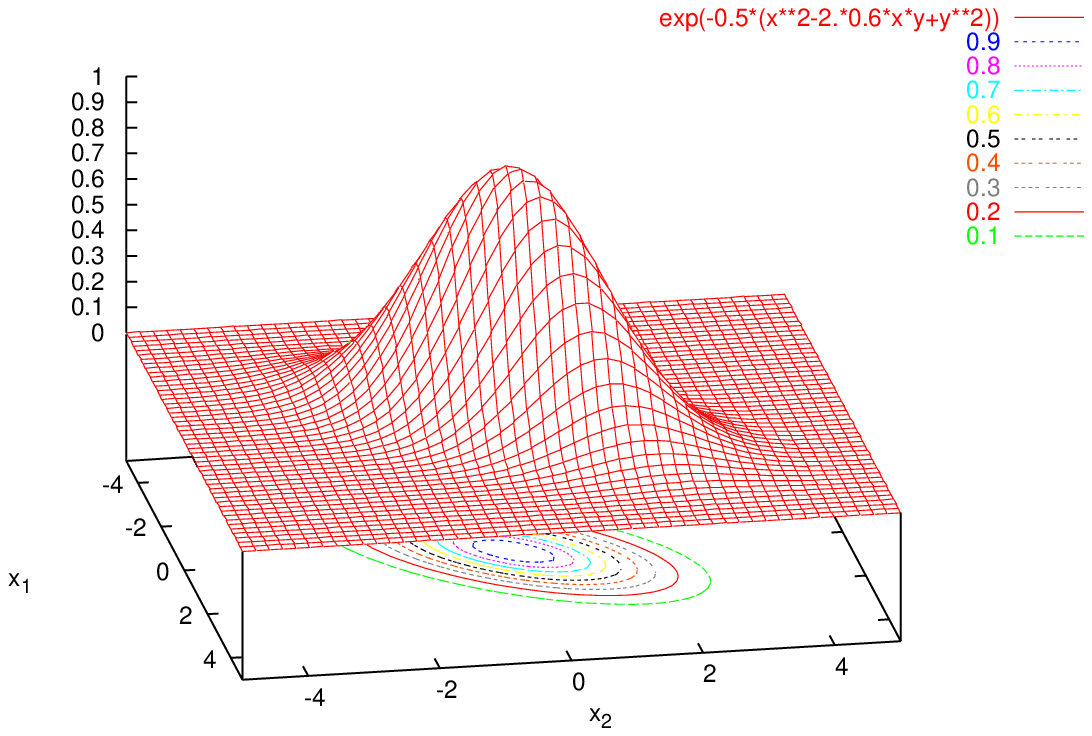,width=7.5cm}
\\
\epsfig{figure=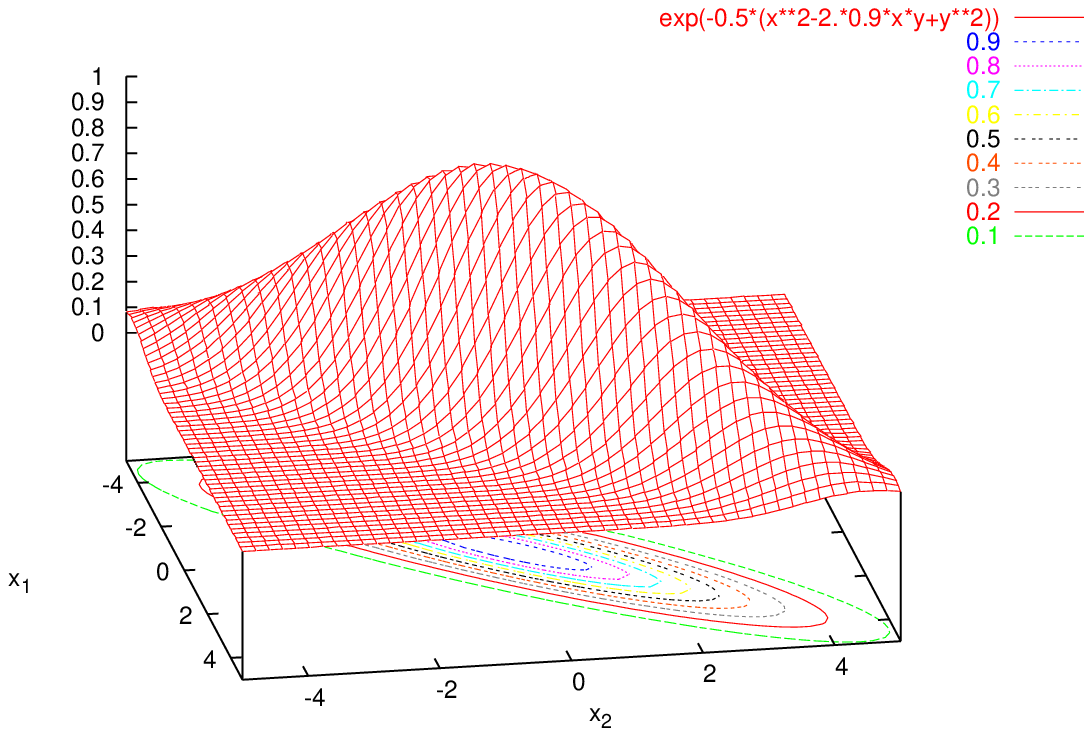,width=7.5cm}
&
\epsfig{figure=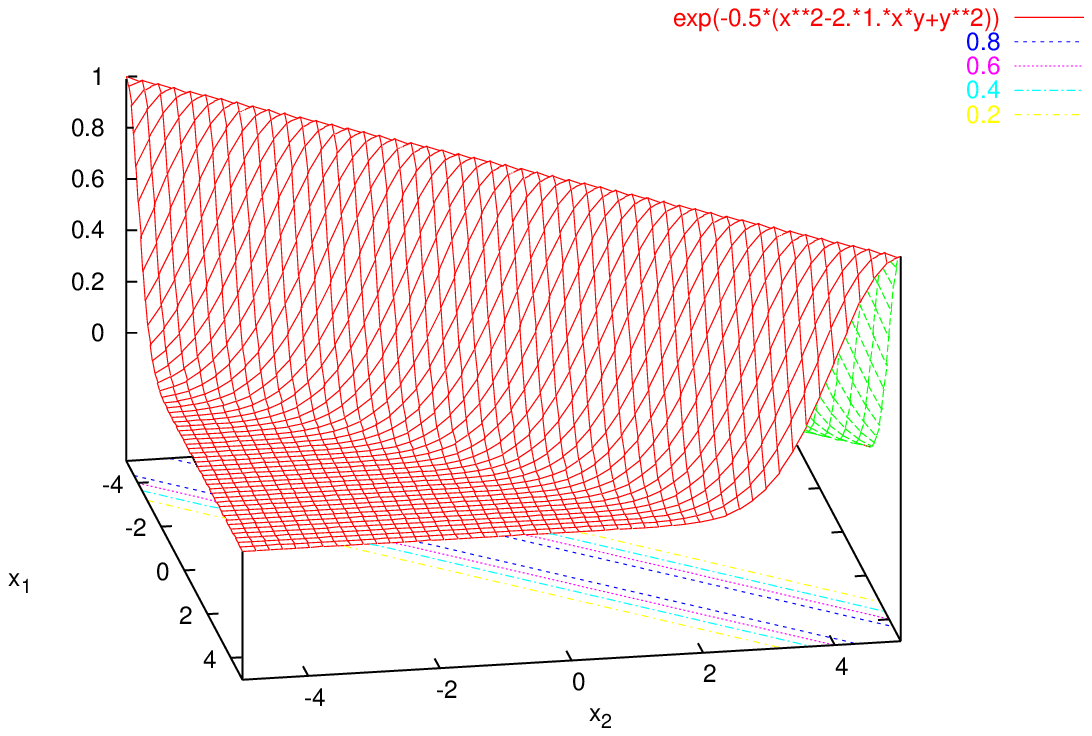,width=7.5cm}
\\
\end{tabular}
\caption{
Gaussian distribution (\ref{2Dgauss}) for $\lambda=0$, $0.6$,
$0.9$ and $1$ (non-normalised).
}
\label{fig:gauss}
\end{figure}

To complete the calculation of the moments up to second order we need
the covariance of $x_{1}$ and $x_{2}$:
$\cum{x_{1}x_{2}}=\av{x_{1}x_{2}}-\av{x_{1}}\av{x_{2}}$ which reduces
to calculate $\llangle x_{1}x_{2}\rrangle$.
This can be obtained using the form (\ref{2Dgauss:2}) for the
distribution
\begin{eqnarray*}
\llangle x_{1}x_{2}\rrangle
&=&
\int\!\D x_{1}
\int\!\D x_{2}\,
x_{1}x_{2}
P_{2}(x_{1},x_{2})
\\[-2.ex]
&=&
C
\int\!\D x_{1}
x_{1}
\exp
\left[
-\frac{1-\lambda^{2}}{2\sigma^{2}}
x_{1}^{2}
\right]
\overbrace{
\int\!\D x_{2}\,
x_{2}
\exp
\left[
-\frac{1}{2\sigma^{2}}
(x_{2}-\lambda x_{1})^{2}
\right]
}^{\lambda x_{1}\sqrt{2\pi\sigma^{2}}}
\\[-0.ex]
&=&
\lambda\sqrt{2\pi\sigma^{2}}
C
\underbrace{
\int\!\D x_{1}
x_{1}^{2}
\exp
\left[
-\frac{1-\lambda^{2}}{2\sigma^{2}}
x_{1}^{2}
\right]
}_{
\sqrt{2\pi\sigma^{2}/(1-\lambda^{2})}
\,
\sigma^{2}/(1-\lambda^{2})
}
\quad
\Rightarrow
\quad
\boxequation{
\llangle x_{1}x_{2}\rrangle
=
\frac{\lambda}{1-\lambda^{2}}\sigma^{2}
}
\end{eqnarray*}
since $C=\sqrt{(1-\lambda^{2})/(2\pi\sigma^{2})^{2}}$.
Its is convenient to compute the normalised covariance $\llangle
x_{1}x_{2}\rrangle / \sqrt{\llangle x_{1}^{2}\rrangle\llangle
x_{2}^{2}\rrangle}$, which is merely given by $\lambda$.
Therefore the parameter $\lambda$ in the distribution (\ref{2Dgauss})
is a measure of how much correlated the variables $x_{1}$ and $x_{2}$
are.
Actually in the limit $\lambda\to0$ the variables are not correlated
at all and the distribution factorises.
In the opposite limit $\lambda\to1$ the variables are maximally
correlated, $\llangle x_{1}x_{2}\rrangle/\sqrt{\llangle
x_{1}^{2}\rrangle\llangle x_{2}^{2}\rrangle}=1$.
The distribution is actually a function of $(x_{1}-x_{2})$, so it is
favoured that $x_{1}$ and $x_{2}$ take similar values (see Fig.\
\ref{fig:gauss})
\begin{equation}
\label{2Dgauss:limits}
P_{2}|_{\lambda=0}
=
\frac{1}{\sqrt{2\pi\sigma^{2}}}
\,
e^
{
-x_{1}^{2}/2\sigma^{2}
}
\frac{1}{\sqrt{2\pi\sigma^{2}}}
\,
e^
{
-x_{2}^{2}/2\sigma^{2}
},
\quad
P_{2}|_{\lambda=1}
\to
e^
{
-(x_{1}-x_{2})^{2}/2\sigma^{2}
}
\;.
\end{equation}
We can now interpret the increase of the variances with $\lambda$: the
correlation between the variables allows them to take arbitrarily
large values, with the only restriction of their difference being
small (Fig.\ \ref{fig:gauss}).

To conclude we can compute the conditional probability by using Bayes
rule $P_{1|1}(x_{1}|x_{2}) = P_{2}(x_{1},x_{2})/P_{1}(x_{2})$ and
Eqs.\ (\ref{2Dgauss}) and (\ref{2Dgauss:marginal})
\begin{eqnarray*}
P_{1|1}(x_{1}|x_{2})
&=&
\frac
{\sqrt{\frac{1-\lambda^{2}}{(2\pi\sigma^{2})^{2}}}
\exp
\left[
-\frac{1}{2\sigma^{2}}
\left(
x_{1}^{2}-2\lambda\,x_{1}x_{2}+x_{2}^{2}
\right)
\right]
}
{\sqrt{\frac{1-\lambda^{2}}{2\pi\sigma^{2}}}
\,
\exp
\left(
-\frac{1-\lambda^{2}}{2\sigma^{2}}
x_{2}^{2}
\right)
}
\\
&=&
\frac{1}{\sqrt{2\pi\sigma^{2}}}
\exp
\left[
-\frac{1}{2\sigma^{2}}
\left(
x_{1}^{2}-2\lambda\,x_{1}x_{2}+[\cross 1-(\cross 1-\lambda^{2})]x_{2}^{2}
\right)
\right]
\;,
\end{eqnarray*}
and hence (recall that here $x_{2}$ is a parameter; the known output
of $X_{2}$)
\begin{equation}
\label{2Dgauss:conditional}
P_{1|1}(x_{1}|x_{2})
=
\frac{1}{\sqrt{2\pi\sigma^{2}}}
\exp
\left[
-\frac{1}{2\sigma^{2}}
\left(
x_{1}-\lambda\,x_{2}
\right)^{2}
\right]
\;.
\end{equation}
Then, at $\lambda=0$ (no correlation) the values taken by $x_{1}$ are
independent of the output of $x_{2}$ while for $\lambda\to1$ they are
centered around those taken by $x_{2}$, and hence strongly conditioned
by them.

\mynewpage

\section{Stochastic processes and Markov processes}
\label{sec:stopro-markov}
\markboth{Introduction to the theory of stochastic processes}{Stochastic
processes and Markov processes}

Once a stochastic variable $X$ has been defined, an infinity of other
stochastic variables derive from it, namely, all the quantities $Y$ defined
as functions of $X$ by some mapping.
These quantities $Y$ may be any kind of mathematical object; in particular,
also functions of an auxiliary variable $t$:
\[
Y_{X}(t)=f(X,t)
\;,
\]
where $t$ could be the time or some other parameter.
Such a quantity $Y_{X}(t)$ is called a {\em stochastic process}.
On inserting for $X$ one of its possible values $x$, one gets an ordinary
function of $t$, $Y_{x}(t)=f(x,t)$, called a {\em sample function\/} or {\em
realisation\/} of the process.
In physical language, one regards the stochastic process as the ``ensemble"
of these sample functions.%
\footnote{As regards the terminology, one also refers to a stochastic
time-dependent quantity as a {\em noise term}.
This name originates from the early days of radio, where the great
number of highly irregular electrical signals occurring either in the
atmosphere, the receiver, or the radio transmitter, certainly sounded like
noise on a radio.}

It is easy to form averages on the basis of the underlying probability
distribution $P_{X}(x)$.
For instance, one can take $n$ values $t_{1},\ldots,t_{n}$, and form the
$n$th moment
\begin{equation}
\label{moments:hierarchy}
\av{Y(t_{1})\cdots Y(t_{n})} =\int\!\D x\,Y_{x}(t_{1})\cdots
Y_{x}(t_{n})P_{X}(x)
\;.
\end{equation}
Of special interest is the {\em auto-correlation function}
\begin{eqnarray*}
\kappa(t_{1},t_{2})\equiv\cum{Y(t_{1})Y(t_{2})}
&=&
\av{Y(t_{1})Y(t_{2})}-\av{Y(t_{1})}\av{Y(t_{2})}
\\
&=&
\bbav{[Y(t_{1})-\av{Y(t_{1})}][Y(t_{2})-\av{Y(t_{2})}]}
\;,
\end{eqnarray*}
which, for $t_{1}=t_{2}=t$, reduces to the time-dependent variance
$\cum{Y(t)^{2}}=\sigma(t)^{2}$.

A stochastic process is called {\em stationary\/} when the moments are not
affected by a shift in time, that is, when
\begin{equation}
\label{Y:stationary}
\av{Y(t_{1}+\tau)\cdots Y(t_{n}+\tau)} =\av{Y(t_{1})\cdots Y(t_{n})}
\;.
\end{equation}
In particular, $\av{Y(t)}$ is then independent of the time, and the
auto-correlation function $\kappa(t_{1},t_{2})$ only depends on the time
difference $|t_{1}-t_{2}|$.
Often there exist a constant $\tau_{c}$ such that
$\kappa(t_{1},t_{2})\simeq0$ for $|t_{1}-t_{2}|>\tau_{c}$; one then calls
$\tau_{c}$ the {\em auto-correlation time\/} of the stationary stochastic
process.

\sloppy
If the stochastic quantity consist of several components $Y_{i}(t)$, the
auto-correlation function is replaced by the correlation matrix
\[
K_{ij}(t_{1},t_{2})=
\cum{Y_{i}(t_{1})Y_{j}(t_{2})}
\;.
\]
The diagonal elements are the {\em auto-correlations\/} and the off-diagonal
elements are the {\em cross-correlations}.
Finally, in case of a zero-average stationary stochastic process, this
equation reduces to
\[
K_{ij}(\tau) =\av{Y_{i}(t)Y_{j}(t+\tau)} =\av{Y_{i}(0)Y_{j}(\tau)}
\;.
\]

\fussy


\subsection{The hierarchy of distribution functions}

A stochastic process $Y_{X}(t)$, defined from a stochastic variable $X$ in
the way described above, leads to a hierarchy of probability distributions.
For instance, the probability distribution for $Y_{X}(t)$ to take the value
$y$ at time $t$ is [cf.\ Eq.\ (\ref{transformation})]
\[
P_{1}(y,t)=\int\!\D x\,\delta[y-\underbrace{Y_{x}(t)}_{f(x,t)}]P_{X}(x)
\;.
\]
Similarly, the joint probability distribution that $Y$ has the value
$y_{1}$ at $t_{1}$, and also the value $y_{2}$ at $t_{2}$, and so on up to
$y_{n}$ at $t_{n}$, is
\begin{equation}
\label{hierarchy:distribution}
P_{n}(y_{1},t_{1};\ldots;y_{n},t_{n})
=\int\!\D x\,
\delta[y_{1}-Y_{x}(t_{1})]\cdots\delta[y_{n}-Y_{x}(t_{n})]P_{X}(x)
\;.
\end{equation}
In this way an infinite hierarchy of probability distributions $P_{n}$,
$n=1,2,\ldots$, is defined.
They allow one the computation of all the averages already introduced, e.g.,%
\footnote{
This result is demonstrated by introducing the definition
(\ref{hierarchy:distribution}) in the right-hand side of Eq.\
(\ref{moments:Y:2}):
\begin{eqnarray*}
\lefteqn{\int\!\D y_{1}\cdots \D y_{n}\, y_{1}\cdots y_{n}
P_{n}(y_{1},t_{1};\ldots;y_{n},t_{n})}
\hspace{3em}
\\
&=&
\int\!\D y_{1}\cdots \D y_{n}\, y_{1}\cdots y_{n}
\int\!\D x\,
\delta[y_{1}-Y_{x}(t_{1})]\cdots\delta[y_{n}-Y_{x}(t_{n})]P_{X}(x)
\\
&=&
\int\!\D x\,Y_{x}(t_{1})\cdots Y_{x}(t_{n})P_{X}(x)
\\[-1.ex]
&
\stackrel{\text{Eq.\ (\ref{moments:hierarchy})}}{=}
&
\av{Y(t_{1})\cdots Y(t_{n})}
\;.
\qed
\end{eqnarray*}
             } 
\begin{equation}
\label{moments:Y:2}
\av{Y(t_{1})\cdots Y(t_{n})}
=\int\!\D y_{1}\cdots \D y_{n}\, y_{1}\cdots y_{n}
P_{n}(y_{1},t_{1};\ldots;y_{n},t_{n})
\;.
\end{equation}
We note in passing that, by means of this result, the definition
(\ref{Y:stationary}) of stationary processes, can be restated in terms of the
dependence of the $P_{n}$ on the time differences alone, namely
\[
P_{n}(y_{1},t_{1}+\tau;\ldots;y_{n},t_{n}+\tau)
=P_{n}(y_{1},t_{1};\ldots;y_{n},t_{n})
\;.
\]
Consequently, a necessary (but not sufficient) condition for the stochastic
process being stationary is that $P_{1}(y_{1})$ does not depend on the time.

Although the right-hand side of Eq.\ (\ref{hierarchy:distribution}) also has a
meaning when some of the times are equal, one regards the $P_{n}$ to be
defined only when all times are different.
The hierarchy of probability distributions $P_{n}$ then obeys the following
consistency conditions:
\begin{enumerate}
\item
$P_{n}\geq0
\;;$
\item
$P_{n}$ is invariant under permutations of two pairs $(y_{i},t_{i})$
and $(y_{j},t_{j})
\;;$
\item\label{consitency:3rd}
$\Intdef{}{}{y_{n}}P_{n}(y_{1},t_{1};\ldots;y_{n},t_{n})
=P_{n-1}(y_{1},t_{1};\ldots;y_{n-1},t_{n-1})
\;;$
\item
$\Intdef{}{}{y_{1}}P_{1}(y_{1},t_{1})=1
\;.$
\end{enumerate}
Inasmuch as the distributions $P_{n}$ enable one to compute all the
averages of the stochastic process [Eq.\ (\ref{moments:Y:2})], they
constitute a complete specification of it.
Conversely, according to a theorem due to Kolmogorov, it is possible to
prove that the inverse is also true, i.e., that any set of functions
obeying the above four consistency conditions determines a stochastic
process $Y(t)$.


\subsection{Gaussian processes}

A stochastic process is called a {\em Gaussian\/} process, if all its
$P_{n}$ are multivariate Gaussian distributions (Sec.\ \ref{gaussian}).
%
In that case, all cumulants beyond $m=2$ vanish and, recalling that
$\cum{Y(t_{1})Y(t_{2})}=\av{Y(t_{1})Y(t_{2})}-\av{Y(t_{1})}\av{Y(t_{2})}$,
one sees that a Gaussian process is fully specified by its average
$\av{Y(t)}$ and its second moment $\av{Y(t_{1})Y(t_{2})}$.
Gaussian stochastic processes are often used as an approximate description
for physical processes, which amounts to assuming that the higher-order
cumulants are negligible.


\subsection{Conditional probabilities}

The notion of conditional probability for multivariate distributions can be
applied to stochastic processes, via the hierarchy of probability
distributions introduced above.
For instance, the conditional probability $P_{1|1}(y_{2},t_{2}|y_{1},t_{1})$
represents the probability that $Y$ takes the value $y_{2}$ at $t_{2}$,
given that its value at $t_{1}$ ``was" $y_{1}$.
It can be constructed as follows: from all sample functions $Y_{x}(t)$ of
the ensemble representing the stochastic process, select those passing
through the point $y_{1}$ at the time $t_{1}$; the fraction of this {\em
sub-ensemble\/} that goes through the gate $(y_{2},y_{2}+\D y_{2})$ at
the time $t_{2}$ is precisely $P_{1|1}(y_{2},t_{2}|y_{1},t_{1})\D y_{2}$.
More generally, one may fix the values of $Y$ at $n$ different times
$t_{1},\ldots,t_{n}$, and ask for the joint probability at $m$ other times
$t_{n+1},\ldots,t_{n+m}$.
This leads to the general definition of the conditional probability
$P_{m|n}$ by Bayes' rule:
\begin{equation}
\label{bayesrule:2}
P_{m|n}(y_{n+1},t_{n+1};\ldots;y_{n+m},t_{n+m}|
y_{1},t_{1};\ldots;y_{n},t_{n})
=
\frac{P_{n+m}(y_{1},t_{1};\ldots;y_{n+m},t_{n+m})}
{P_{n}(y_{1},t_{1};\ldots;y_{n},t_{n})}
\;.
\end{equation}
%
Note that the right-hand side of this equation is well defined in terms of
the probability distributions of the hierarchy $P_{n}$ previously
introduced.
Besides, from their consistency conditions it follows
the normalisation of the $P_{m|n}$.


\subsection{Markov processes}
\label{subsec:markov}

Among the many possible classes of stochastic processes, there is one that
merits a special treatment---the so-called Markov processes.

\begin{sloppypar}
Recall that, for a stochastic process $Y(t)$, the conditional probability
$P_{1|1}(y_{2},t_{2}|y_{1},t_{1})$, is the probability that $Y(t_{2})$
takes the value $y_{2}$, provided $Y(t_{1})$ has taken the value $y_{1}$.
In terms of this quantity one can express $P_{2}$ as
\begin{equation}
\label{bayesrule:particular}
P_{2}(y_{1},t_{1};y_{2},t_{2})
=P_{1}(y_{1},t_{1})P_{1|1}(y_{2},t_{2}|y_{1},t_{1})
\;.
\end{equation}
However, to construct the higher-order $P_{n}$ one needs transition
probabilities $P_{n|m}$ of higher order, e.g.,
$P_{3}(y_{1},t_{1};y_{2},t_{2};y_{3},t_{3})
=
P_{2}(y_{1},t_{1};y_{2},t_{2})
P_{1|2}(y_{3},t_{3}|y_{1},t_{1};y_{2},t_{2})$.
A stochastic process is called a {\em Markov process}, if for any set of
$n$ successive times $t_{1}<t_{2}<\cdots<t_{n}$, one has
\begin{equation}
\label{markov}
P_{1|n-1}(y_{n},t_{n}|y_{1},t_{1};\ldots;y_{n-1},t_{n-1})
=P_{1|1}(y_{n},t_{n}|y_{n-1},t_{n-1})
\;.
\end{equation}
In words: the conditional probability distribution of $y_{n}$ at $t_{n}$,
given the value $y_{n-1}$ at $t_{n-1}$, is uniquely determined, and is
not affected by any knowledge of the values at earlier times.

A Markov process is therefore fully determined by the two distributions
$P_{1}(y,t)$ and $P_{1|1}(y',t'|y,t)$, from which the entire
hierarchy $P_{n}(y_{1},t_{1};\ldots;y_{n},t_{n})$ can be constructed.
For instance, consider $t_{1}<t_{2}<t_{3}$; $P_{3}$ can be written as
\begin{eqnarray}
\label{markov:2}
\hspace{-1.em}
P_{3}(y_{1},t_{1};y_{2},t_{2};y_{3},t_{3})
\hspace{-1.5em}
& &
\stackrel{\text{Eq.\ (\ref{bayesrule:2})}}{=}
P_{2}(y_{1},t_{1};y_{2},t_{2})
P_{1|2}(y_{3},t_{3}|y_{1},t_{1};y_{2},t_{2})
\nonumber\\
& &
\stackrel{\text{Eq.\ (\ref{markov})}}{=}
P_{2}(y_{1},t_{1};y_{2},t_{2})
P_{1|1}(y_{3},t_{3}|y_{2},t_{2})
\nonumber\\
& &
\stackrel{\text{Eq.\ (\ref{bayesrule:particular})}}{=}
P_{1}(y_{1},t_{1})P_{1|1}(y_{2},t_{2}|y_{1},t_{1})
P_{1|1}(y_{3},t_{3}|y_{2},t_{2})
\;.
\quad
\end{eqnarray}
\end{sloppypar}

From now on, we shall only deal with Markov processes.
Then, the only independent conditional probability is $P_{1|1}(y',t'|y,t)$,
so we shall omit the subscript $1|1$ henceforth and call
$P_{1|1}(y',t'|y,t)$ the {\em transition probability}.


\subsection{\ChK\ equation}

Let us now derive an important identity that must be obeyed by the
transition probability of any Markov process.
On integrating Eq.\ (\ref{markov:2}) over $y_{2}$, one obtains
($t_{1}<t_{2}<t_{3}$)
\[
P_{2}(y_{1},t_{1};y_{3},t_{3}) =P_{1}(y_{1},t_{1})
\Intdef{}{}{y_{2}}P(y_{2},t_{2}|y_{1},t_{1}) P(y_{3},t_{3}|y_{2},t_{2})
\;,
\]
where the consistency condition \ref{consitency:3rd} of the hierarchy of
distribution functions $P_{n}$ has been used to write the left-hand side.
Now, on dividing both sides by $P_{1}(y_{1},t_{1})$ and using the
special case (\ref{bayesrule:particular}) of Bayes' rule, one gets
\begin{equation}
\label{chapmankolmogorov}
\boxequation{ P(y_{3},t_{3}|y_{1},t_{1}) =\Intdef{}{}{y_{2}}
P(y_{3},t_{3}|y_{2},t_{2}) P(y_{2},t_{2}|y_{1},t_{1})
\;,
													}
\end{equation}
which is called the \ChK\ equation.
The time ordering is essential: $t_{2}$ must lie between $t_{1}$ and
$t_{3}$ for Eq.\ (\ref{chapmankolmogorov}) to hold.
This is required in order to the starting Eq.\ (\ref{markov:2}) being valid,
specifically, in order to the second equality there being
derivable from the first one by dint of the definition (\ref{markov}) of a
Markov process.

Note finally that, on using Eq.\ (\ref{bayesrule:particular}) one can
rewrite the particular case $P_{1}(y_{2},t_{2})=\int
\D y_{1}\,P_{2}(y_{2},t_{2};y_{1},t_{1})$ of the relation among the
distributions of the hierarchy as
\begin{equation}
\label{consistency}
\boxequation{
P_{1}(y_{2},t_{2})=\Intdef{}{}{y_{1}}
P(y_{2},t_{2}|y_{1},t_{1}) P_{1}(y_{1},t_{1})
\;.
													}
\end{equation}
This is an additional relation involving the two probability distributions
characterising a Markov process.
Reciprocally, any non-negative functions obeying Eqs.\
(\ref{chapmankolmogorov}) and (\ref{consistency}), define a Markov
process uniquely.


\subsection{Examples of Markov processes}


\paragraph*{\WL\ process.}

This stochastic process was originally introduced in order to describe
the behaviour of the {\em position\/} of a free Brownian particle in
one dimension.
On the other hand, it plays a central r\^{o}le in the rigorous
foundation of the stochastic differential equations.
The \WL\ process is defined in the range $-\infty<y<\infty$ and $t>0$
through [cf.\ Eq.\ (\ref{pdf-diffusion})]
\begin{subeqnarray}
\label{wienerlevy}
\slabel{wienerlevy:P1}
P_{1}(y_{1},t_{1})
&=&
\frac{1}{\sqrt{2\pi t_{1}}}
\exp\lr{-\frac{y_{1}^{2}}{2t_{1}}}
\;,
\\
\slabel{wienerlevy:Ptrans}
P(y_{2},t_{2}|y_{1},t_{1})
&=&
\frac{1}{\sqrt{2\pi (t_{2}-t_{1})}}
\exp\lrs{-\frac{(y_{2}-y_{1})^{2}}{2(t_{2}-t_{1})}}
\;,
\quad
(t_{1}<t_{2})
\;.
\qquad
\end{subeqnarray}
This is a non-stationary ($P_{1}$ depends on $t$), Gaussian process.
The second-order moment is%
\begin{eqnarray}
\label{wienerlevy:moment:a}
\av{Y(t_{1})Y(t_{2})}
&=&
\min(t_{1},t_{2})
\;,
\end{eqnarray}
Proof: Let us assume $t_{1}<t_{2}$.
Then, from Eq.\ (\ref{moments:Y:2}) we have
\begin{eqnarray*}
\av{Y(t_{1})Y(t_{2})}
&=&
\Intdef{}{}{y_{1}}\Intdef{}{}{y_{2}}
y_{1}y_{2}
P_{2}(y_{1},t_{1};y_{2},t_{2})
\\
&=&
\Intdef{}{}{y_{1}}
y_{1}
P_{1}(y_{1},t_{1})
\underbrace{
\Intdef{}{}{y_{2}}
y_{2}
P(y_{2},t_{2}|y_{1},t_{1})
}_{y_{1}\;\text{by~Eq.\ (\ref{wienerlevy:Ptrans})}}
=
\underbrace{
\Intdef{}{}{y_{1}}y_{1}^{2}
P_{1}(y_{1},t_{1})
}_{t_{1}~\text{by~Eq.\ (\ref{wienerlevy:P1})}}
\;,
\end{eqnarray*}
where we have used that $t_{1}$ is the time-dependent variance of $P_{1}$.
\qed
%


\paragraph*{\OU\ process.}

This stochastic process was constructed to describe the behaviour of
the {\em velocity\/} of a free Brownian particle in one dimension (see
Sec.\ \ref{subsec:FP-examples}).
It also describes the position of an overdamped particle in an
harmonic potential.
It is defined by ($\Delta t=t_{2}-t_{1}>0$)
\begin{subeqnarray}
\label{ornsteinuhlenbeck}
\slabel{ornsteinuhlenbeck:P1}
P_{1}(y_{1})
&=&
\frac{1}{\sqrt{2\pi }}\exp\lr{-\half y_{1}^{2}}
\;,
\\
\slabel{ornsteinuhlenbeck:Ptrans}
P(y_{2},t_{2}|y_{1},t_{1})
&=&
\frac{1}{\sqrt{2\pi(1-e^{-2\Delta t})}}
\exp\lrs{-\frac{(y_{2}-y_{1}e^{-\Delta t})^{2}}{2(1-e^{-2\Delta t})}}
\;.
\end{subeqnarray}
The \OU\ process is stationary, Gaussian, and Markovian.
According to a theorem due to Doob, it is essentially the only process
with these three properties.
Concerning the Gaussian property, it is clear for $P_{1}(y_{1})$.
For
$P_{2}(y_{2},t_{2};y_{1},t_{1})
=
P_{1}(y_{1})P(y_{2},t_{2}|y_{1},t_{1})$ [Eq.\
(\ref{bayesrule:particular})], we have
\begin{equation}
P_{2}(y_{2},t_{2};y_{1},t_{1})
=
\frac{1}{\sqrt{(2\pi)^{2}(1-e^{-2\Delta t})}}
\exp
\lrs{
-\frac
{y_{1}^{2}-2y_{1}y_{2}e^{-\Delta t}+y_{2}^{2}}
{2(1-e^{-2\Delta t})}
}
\;.
\end{equation}
This expression can be identified with the bivariate Gaussian
distribution (\ref{2Dgauss}) and the following parameters
\[
\lambda
=
e^{-\Delta t}
\;,
\qquad
\sigma^{2}
=
1-e^{-2\Delta t}
\;,
\]
with the particularity that $\sigma^{2}=1-\lambda^{2}$ in this case.
Therefore, we immediately see that the \OU\ process has an exponential
auto-correlation function $\av{Y(t_{1})Y(t_{2})}=e^{-\Delta t}$, since
$\sigma^{2}\lambda/(1-\lambda^{2})=\lambda$ in this case.%
\footnote{
This result can also be obtained by using Eqs.\ (\ref{ornsteinuhlenbeck})
directly:
\begin{eqnarray*}
\kappa(t_{1},t_{2})
&=&
\av{Y(t_{1})Y(t_{2})}-\overbrace{\av{Y(t_{1})}\av{Y(t_{2})}}^{0}
\\
&=&
\int\!\D y_{1}\D y_{2}\,y_{1}y_{2}
\underbrace{P_{2}(y_{1},t_{1};y_{2},t_{2})}%
_{\lefteqn{\scriptstyle P_{1}(y_{1})P(y_{2},t_{2}|y_{1},t_{1})}}
\\[-4ex]
&=&
\Intdef{-\infty}{\infty}{y_{1}}
y_{1}\frac{1}{\sqrt{2\pi}}\exp\lr{-\half y_{1}^{2}}
\underbrace{\Intdef{-\infty}{\infty}{y_{2}}
y_{2}
\frac{1}{\sqrt{2\pi(1-e^{-2\Delta t})}}
\exp\BBs{-\frac{(y_{2}-\overbrace{y_{1}e^{-\Delta t}}^{\mu_{1}})^{2}}
{2\underbrace{(1-e^{-2\Delta t})}_{\sigma^{2}}}}}_{y_{1}e^{-\Delta t}}
\\[-4ex]
&=&e^{-\Delta t}
\underbrace{\Intdef{-\infty}{\infty}{y_{1}}
y_{1}^{2}\frac{1}{\sqrt{2\pi}}\exp\lr{-\half y_{1}^{2}}}_{1}
\;.\qed
\end{eqnarray*}
          } 

The evolution with time of the distribution
$P_{2}(y_{2},t_{2};y_{1},t_{1})$, seen as the velocity of a Brownian
particle, has a clear meaning.
At short times the velocity is strongly correlated with itself: then
$\lambda\sim1$ and the distribution would be like in the lower right
panel of Fig.\ \ref{fig:gauss} [with a shrinked variance
$\sigma^{2}=(1-\lambda^{2})\to0$].
As time elapses $\lambda$ decreases and we pass form one panel to the
previous and, at long times, $\lambda\sim0$ and the velocity has lost
all memory of its value at the initial time due to the collisions and
hence $P_{2}(y_{2},t_{2};y_{1},t_{1})$ is completely uncorrelated.
\enlargethispage*{10.ex}


\paragraph*{Exercise: check by direct integration that the transition
probability (\ref{ornsteinuhlenbeck:Ptrans}) obeys the Chapman--Kolgomorov
equation (\ref{chapmankolmogorov}).}


\mynewpage

\section{The master equation: \KM\
\\
expansion and \FP\ equation}
\label{sec:ME-KM-FP}
\markboth{Introduction to the theory of stochastic processes}{The master
equation: \KM\ expansion and \FP\ equation}

The \ChK\ equation (\ref{chapmankolmogorov}) for Markov processes is not of
much assistance when one searches for solutions of a given problem, because
it is essentially a property of the solution.
However, it can be cast into a more useful form---the {\em master
equation}.


\subsection{The master equation}
\label{subsec:ME}

The master equation is a differential equation for the transition
probability.
Accordingly, in order to derive it, one needs first to ascertain how the
transition probability behaves for short time differences.

Firstly, on inspecting the \ChK\ equation (\ref{chapmankolmogorov}) for
equal time arguments one finds the natural result
\[
P(y_{3},t_{3}|y_{1},t)
=\Intdef{}{}{y_{2}}
P(y_{3},t_{3}|y_{2},t) P(y_{2},t|y_{1},t)
\quad
\Rightarrow
\quad
P(y_{2},t|y_{1},t)=\delta(y_{2}-y_{1})
\;,
\]
which is the zeroth-order term in the short-time behaviour of
$P(y',t'|y,t)$.
Keeping this in mind one {\em adopts\/} the following expression for the
short-time transition probability:
\begin{equation}
\label{shorttimes}
\boxequation{
P(y_{2},t+\Delt|y_{1},t)
=\delta(y_{2}-y_{1})[1-a^{(0)}(y_{1},t)\Delt]
	+W_{t}(y_{2}|y_{1})\Delt+O[(\Delt)^{2}]
\;,
             }
\end{equation}
where $W_{t}(y_{2}|y_{1})$ is interpreted as the {\em transition probability
per unit time\/} from $y_{1}$ to $y_{2}$ at time $t$.
%
%
Then, the coefficient $1-a^{(0)}(y_{1},t)\Delt$ is to be interpreted as the
probability that no ``transition" takes place during $\Delt$.
Indeed, from the normalisation of $P(y_{2},t_{2}|y_{1},t_{1})$
one has:
\[
1
=\Intdef{}{}{y_{2}}P(y_{2},t+\Delt|y_{1},t)
\simeq
1-a^{(0)}(y_{1},t)\Delt
+\Intdef{}{}{y_{2}}W_{t}(y_{2}|y_{1})\Delt
\;.
\]
Therefore, to first order in $\Delt$, one gets%
\footnote{The reason for the notation $a^{(0)}$ will become clear below.}
%
\begin{equation}
\label{a_0&W}
a^{(0)}(y_{1},t)=\Intdef{}{}{y_{2}}W_{t}(y_{2}|y_{1})
\;,
\end{equation}
which substantiates the interpretation mentioned:
$a^{(0)}(y_{1},t)\Delt$ is the total probability of escape from $y_{1}$
in the time interval $(t,t+\Delt)$ and, thus,
$1-a^{(0)}(y_{1},t)\Delt$ is the probability that no transition takes
place during this time.

Now we can derive the differential equation for the transition
probability from the \ChK\ equation (\ref{chapmankolmogorov}).
Insertion of the above short-time expression for the transition
probability in into it yields
\begin{eqnarray*}
P(y_{3},t_{2}+\Delt|y_{1},t_{1})
&=&
\Intdef{}{}{y_{2}}
\overbrace{P(y_{3},t_{2}+\Delt|y_{2},t_{2})}
^{\lefteqn{\qquad\qquad\quad
\sy{\delta(y_{3}-y_{2})[1-a^{(0)}(y_{2},t_{2})\Delt]
  +W_{t_{2}}(y_{3}|y_{2})\Delt}}}
         P(y_{2},t_{2}|y_{1},t_{1})
\\
&\simeq&
[1-a^{(0)}(y_{3},t_{2})\Delt]P(y_{3},t_{2}|y_{1},t_{1})
\\
& &
{}+\Delt\Intdef{}{}{y_{2}}W_{t_{2}}(y_{3}|y_{2})P(y_{2},t_{2}|y_{1},t_{1})
\;.
\end{eqnarray*}
Next, on using Eq.\ (\ref{a_0&W}) to write $a^{(0)}(y_{3},t_{2})$ in terms of
$W_{t_{2}}(y_{2}|y_{3})$, one has
\begin{eqnarray*}
\lefteqn{\frac{1}{\Delt}
        \lrs{P(y_{3},t_{2}+\Delt|y_{1},t_{1})
        -P(y_{3},t_{2}|y_{1},t_{1})}}
\hspace{3em}
\\
& &
\simeq
\Intdef{}{}{y_{2}}\lrs{W_{t_{2}}(y_{3}|y_{2})P(y_{2},t_{2}|y_{1},t_{1})
-W_{t_{2}}(y_{2}|y_{3})P(y_{3},t_{2}|y_{1},t_{1})}
\;,
\end{eqnarray*}
which in the limit $\Delt\to0$ yields, after some changes in notation
($y_{1},t_{1}\to y_{0},t_{0}$, $y_{2},t_{2}\to y',t$, and $y_{3}\to y$), the
{\em master equation}
\begin{equation}
\label{masterequation}
\boxequation{
\Dpar{}{t} P(y,t|y_{0},t_{0})
=\Intdef{}{}{y'}
\lrs{W_{t}(y|y')P(y',t|y_{0},t_{0})
	-W_{t}(y'|y)P(y,t|y_{0},t_{0})}
\;,
            }
\end{equation}
which is an integro-differential equation.

The master equation is a differential form of the \ChK\ equation (and
sometimes it is referred to as such).
Therefore, it is an equation for the transition probability
$P(y,t|y_{0},t_{0})$, but not for $P_{1}(y,t)$.
However, an equation for $P_{1}(y,t)$ can be obtained by using the concept
of ``extraction of a sub-ensemble".
Suppose that $Y(t)$ is a stationary Markov process characterised by
$P_{1}(y)$ and $P(y,t|y_{0},t_{0})$.
Let us define a new, non-stationary Markov process $Y^{\ast}(t)$ for $t\geq
t_{0}$ by setting
\begin{subeqnarray}
\label{subensemble}
\slabel{subensemble:a}
P_{1}^{\ast}(y_{1},t_{1})&=&P(y_{1},t_{1}|y_{0},t_{0})
\;,
\\
\slabel{subensemble:b}
P^{\ast}(y_{2},t_{2}|y_{1},t_{1})&=&P(y_{2},t_{2}|y_{1},t_{1})
\;.
\end{subeqnarray}
This is a sub-ensemble of $Y(t)$ characterised by taking the sharp value
$y_{0}$ at $t_{0}$, since $P_{1}^{\ast}(y_{1},t_{0})=\delta(y_{1}-y_{0})$.
More generally, one may extract a sub-ensemble in which at a given time
$t_{0}$ the values of $Y^{\ast}(t_{0})$ are distributed according to a given
probability distribution $p(y_{0})$:
\begin{equation}
\label{subensemble:a'}
P_{1}^{\ast}(y_{1},t_{1})
=\Intdef{}{}{y_{0}}P(y_{1},t_{1}|y_{0},t_{0})p(y_{0})
\;,
\end{equation}
and $P^{\ast}(y_{2},t_{2}|y_{1},t_{1})$ as in Eq.\ (\ref{subensemble:b}).
Physically, the extraction of a sub-ensemble means that one ``prepares" the
system in a certain non-equilibrium state at $t_{0}$.

By construction, the above $P_{1}^{\ast}(y_{1},t_{1})$ obey
the same differential equation as the transition probability (with
respect its first pair of arguments), that is, $P_{1}^{\ast}(y_{1},t_{1})$
obeys the master equation.
Consequently, we may write, suppressing unessential indices,
\begin{equation}
\label{masterequation:P1}
\Dpar{P(y,t)}{t}
=
\Intdef{}{}{y'}\lrs{W(y|y')P(y',t)-W(y'|y)P(y,t)}
\;.
\end{equation}
If the range of $Y$ is a discrete set of states labelled with $n$, the
equation reduces to
\begin{equation}
\label{masterequation:discrete}
\frac{\D p_{n}(t)}{\D t}
=
\sum_{n'}
\left[
W_{nn'}
p_{n'}(t)
-
W_{n'n}
p_{n}(t)
\right]
\;.
\end{equation}
In this form the meaning becomes clear: {\em the master equation is a
balance (gain--loss) equation for the probability of each state}.
The first term is the ``gain" due to ``transitions" from other
``states" $n'$ to $n$, and the second term is the ``loss" due to
``transitions" into other configurations.
Remember that $W_{n'n}\geq0$ and that the term with $n=n'$ does not
contribute.

Owing to $W(y|y')\Delt$ is the transition probability in a short time
interval $\Delt$, it can be computed, for the system under study, by means of
any available method valid for short times, e.g., by Dirac's time-dependent
perturbation theory leading to the ``golden rule".
Then, the master equation serves to determine the time evolution of
the system over long time periods, at the expense of {\em assuming\/} the
Markov property.

The master equation can readily be extended to the case of a
multi-component Markov process $Y_{i}(t)$, $i=1$, $2,\ldots$, $N$, on
noting that the \ChK\ equation (\ref{chapmankolmogorov}) is valid as it stands
by merely replacing $y$ by $\multi{y}=(y_{1},\cdots,y_{N})$.
Then, manipulations similar as those leading to Eq.\
(\ref{masterequation:P1}) yield the multivariate counterpart of the
master equation
\begin{equation}
\label{masterequation:P1:n-dim}
\boxequation{
\Dpar{P(\multi{y},t)}{t}
=\Intdef{}{}{\multi{y}'}\lrs{W(\multi{y}|\multi{y}')P(\multi{y}',t)
-W(\multi{y}'|\multi{y})P(\multi{y},t)}
\;.
            }
\end{equation}


\paragraph*{Example: the decay process.}

Let us consider an typical example of master equation describing a
decay process, in which $p_{n}(t)$ determines the probability of
having at time $t$, $n$ surviving ``emitters'' (radioactive nuclei,
excited atoms emitting photons, etc.).
The transition probability in a short interval is
\[
W_{n,n'}
\Delta t
=
\left\{
\begin{array}{cl}
0
&
\mbox{for}
\quad
n>n'
\\
\gamma
n'
\Delta t
&
\mbox{for}
\quad
n=n'-1
\\
\Order(\Delta t)^{2}
&
\mbox{for}
\quad
n<n'-1
\end{array}
\right.
\]
That is, there are not transitions to a state with more emitters (they
can only decay; reabsortion is negligible), and the decay probability
of more that one decay in $\Delta t$ is of higher order in $\Delta t$.
The decay parameter $\gamma$ can be computed with quantum mechanical
techniques.
The corresponding master equation is Eq.\
(\ref{masterequation:discrete}) with $W_{n,n'}=\gamma
n'\delta_{n,n'-1}$
\[
\frac{\D p_{n}(t)}{\D t}
=
W_{n,n+1}
\,
p_{n+1}(t)
-
W_{n-1,n}
\,
p_{n}(t)
\;.
\]
and hence
\begin{equation}
\label{masterequation:discrete:decay}
\frac{\D p_{n}(t)}{\D t}
=
\gamma
(n+1)
\,
p_{n+1}(t)
-
\gamma n
\,
p_{n}(t)
\;.
\end{equation}

Without finding the complete solution for $p_{n}(t)$, we can derive
the equation for the average number of surviving emitters $\llangle
N\rrangle(t)=\sum_{n=0}^{\infty}n\,p_{n}(t)$
\begin{eqnarray*}
\sum_{n=0}^{\infty}
n(\D p_{n}/\D t)
&=&
\gamma
\overbrace{
\sum_{n=0}^{\infty}
n(n+1)
p_{n+1}
}^{k=n+1}
-
\gamma
\overbrace{
\sum_{n=0}^{\infty}
n^{2}
p_{n}
}^{n=0\to1}
\\[-1.ex]
&=&
\gamma
\sum_{k=1}^{\infty}
[(\cross k-1)k-\cross k^{2}]
p_{k}
=
-
\gamma
\overbrace{
\sum_{k=1}^{\infty}
k
p_{k}
}^{\llangle N\rrangle}
\;.
\end{eqnarray*}
Therefore the differential equation for $\llangle N\rrangle$ and its
solution are:
\begin{equation}
\label{average:decay}
\frac{\D}{\D t}
\llangle N\rrangle
=
-
\gamma
\llangle N\rrangle
\;,
\quad
\Rightarrow
\quad
\boxequation{
\llangle N\rrangle(t)
=
n_{0}e^{-\gamma t}
\;.
}
\end{equation}


\subsection{The \KM\ expansion and the \FP\ equation}
\label{subsec:KM-FP}

The \KM\ expansion of the master equation casts this
integro-dif\-fer\-en\-tial equation into the form of a differential equation
of infinite order.
It is therefore not easier to handle but, under certain conditions, one may
break off after a suitable number of terms.
When this is done after the second-order terms one gets a partial
differential equation of second order for $P(y,t)$ called the \FP\ equation.

Let us first express the transition probability $W$ as a function of the
size $r$ of the jump from one configuration $y'$ to another one $y$, and of
the starting point $y'$:
\begin{equation}
\label{Wy'r}
W(y|y')=W(y';r)
\;,\qquad r=y-y'
\;.
\end{equation}
The master equation (\ref{masterequation:P1}) then reads,
\begin{equation}
\label{masterequation:P1:jumps}
\Dpar{P(y,t)}{t}
	=\Intdef{}{}{r}W(y-r;r)P(y-r,t)-P(y,t)\Intdef{}{}{r}W(y;-r)
\;,
\end{equation}
where the sign change associated with the change of variables $y'\to
r=y-y'$, is absorbed in the boundaries (integration limits), by considering
a symmetrical integration interval extending from $-\infty$ to $\infty$:
\[
\Intdef{-\infty}{\infty}{y'}f(y')
=-\Intdef{y+\infty}{y-\infty}{r}f(y-r)
=-\Intdef{\infty}{-\infty}{r}f(y-r)
=\Intdef{-\infty}{\infty}{r}f(y-r)
\;.
\]
Moreover, since finite integration limits would incorporate an additional
dependence on $y$, we shall restrict our attention to problems to which the
boundary is irrelevant.

Let us now assume that the changes on $y$ occur via small jumps, i.e.,
that $W(y';r)$ is a sharply peaked function of $r$ but varies slowly
enough with $y'$.
A second assumption is that $P(y,t)$ itself also varies slowly with $y$.
It is then possible to deal with the shift from $y$ to $y-r$ in the
first integral in Eq.\ (\ref{masterequation:P1:jumps}) by means of a
Taylor expansion:
\begin{eqnarray*}
\Dpar{P(y,t)}{t}
&=&
\Intdef{}{}{r}W(y;r)P(y,t) +\sum_{m=1}^{\infty}
\frac{(-1)^{m}}{m!}\Intdef{}{}{r}r^{m}
\Dpar{^{m}}{y^{m}}[W(y;r)P(y,t)]
\\
& &
{}-P(y,t)\Intdef{}{}{r}W(y;-r)
\\
&=&
\sum_{m=1}^{\infty}
\frac{(-1)^{m}}{m!}
\Dpar{^{m}}{y^{m}}
\lrc{\lrs{\Intdef{}{}{r}r^{m}W(y;r)}
     P(y,t)}
\;,
\end{eqnarray*}
where we have used that the first and third terms on the first right-hand
side cancel each other.%
\footnote{This can be shown upon interchanging $-r$ by $r$ and
absorbing the sign change in the integration limits, as discussed above.}
Note that the dependence of $W(y;r)$ on its second argument $r$ is fully
kept; an expansion with respect to it, is not useful as $W$ varies
rapidly with $r$.
Finally, on introducing the {\em jump moments}
\begin{equation}
\label{jumpmoments}
a^{(m)}(y,t)=\Intdef{}{}{r}r^{m}W(y;r)
\;,
\end{equation}
one gets the {\em \KM\ expansion of the master equation}:
\begin{equation}
\label{kramersmoyal}
\boxequation{
\Dpar{P(y,t)}{t}=
\sum_{m=1}^{\infty}\frac{(-1)^{m}}{m!}
\Dpar{^{m}}{y^{m}}\lrs{a^{(m)}(y,t)P(y,t)}
\;.
													}
\end{equation}

Formally, Eq.\ (\ref{kramersmoyal}) is identical with the master
equation and is therefore not easier to deal with, but it suggest that
one may break off after a suitable number of terms.
For instance, there could be situations where, for $m>2$,
$a^{(m)}(y,t)$ is identically zero or negligible.
In this case one is left with
\begin{equation}
\label{fokkerplanck}
\boxequation{
\Dpar{P(y,t)}{t}
=
-\Dpar{}{y}\lrs{a^{(1)}(y,t)P(y,t)}
+\frac{1}{2}\Dpar{^{2}}{y^{2}}\lrs{a^{(2)}(y,t)P(y,t)}
\;,
			}
\end{equation}
which is the celebrated {\em \FP\ equation}.
The first term is called the {\em drift\/} or {\em transport\/} term and the
second one the {\em diffusion\/} term, while $a^{(1)}(y,t)$ and
$a^{(2)}(y,t)$ are the drift and diffusion ``coefficients".

It is worth recalling that, being derived from the master equation, the
\KM\ expansion, and the \FP\ equation as a special case of it, involve the
transition probability $P(y,t|y_{0},t_{0})$ of the Markov stochastic
process, not its one-time probability distribution $P_{1}(y,t)$.
However, they also apply to the $P_{1}^{\ast}(y,t)$ of every subprocess
that can be extracted from a Markov stochastic process by imposing an
initial condition [see Eqs.\ (\ref{subensemble}) and
(\ref{subensemble:a'})].


\subsection{The jump moments}

The transition probability per unit time $W(y'|y)$ enters
in the definition (\ref{jumpmoments}) of the jump moments.
Therefore, in order to calculate $a^{(m)}(y,t)$, we must use the
relation (\ref{shorttimes}) between $W(y'|y)$ and the transition
probability for short time differences.

Firstly, from Eq.\ (\ref{Wy'r}) one sees that
$W(y';r)=W(y|y')$ with $y=y'+r$.
Accordingly, one can write
\[
W(y;r)=W(y'|y)
\;,\qquad y'=y+r
\;.
\]
On inserting this expression in Eq.\ (\ref{jumpmoments}) one can write
the jump moments as%
\footnote{
This equation makes clear the notation employed.
The quantity $a^{(0)}(y,t)$ [Eq.\ (\ref{a_0&W})], which was introduced
in Eq.\ (\ref{shorttimes}), is indeed the $m=0$ jump moment.
%
  } 
%
\begin{equation}
\label{jumpmoments:2}
\boxequation{
a^{(m)}(y,t)=\Intdef{}{}{y'}(y'-y)^{m}W(y'|y)
\;.
             }
\end{equation}
In order to calculate the jumps moments we introduce the quantity
\[
{\cal A}^{(m)}(y;\tau,t)=\Intdef{}{}{y'}(y'-y)^{m}P(y',t+\tau|y,t)
\;,
\qquad
(m\geq1)
\;,
\]
which is the average of $[Y(t+\tau)-Y(t)]^{m}$ with sharp initial value
$Y(t)=y$ (conditional average).
Then, by using the short-time transition probability
(\ref{shorttimes}), one can write
\begin{eqnarray*}
{\cal A}^{(m)}(y;\tau,t)
&=&
\Intdef{}{}{y'}(y'-y)^{m}
\lrc{\delta(y'-y)[1-a^{(0)}(y,t)\tau]+W(y'|y)\tau+{\cal O}(\tau^{2})}
\\
&=&
\tau\Intdef{}{}{y'}(y'-y)^{m}W(y'|y)+{\cal O}(\tau^{2})
\\
&=&a^{(m)}(y,t)\tau+{\cal O}(\tau^{2})
\;,
\qquad
(m\geq1)
\;,
\end{eqnarray*}
where the integral involving the first term in the short-time transition
probability vanishes due to the presence of the Dirac delta.
Therefore, one can calculate the jump moments from the derivatives of the
conditional averages as follows
\[
a^{(m)}(y,t)
=\eval{\frac{\partial}{\partial\tau}{\cal A}^{(m)}(y;\tau,t)}{\tau=0}
\;.
\]
Finally, on writing
\[
{\cal A}^{(m)}(y;\Delt,t)
=\Intdef{}{}{y'}(y'-y)^{m}P(y',t+\Delt|y,t)
=\eval{\BBav{[Y(t+\Delt)-Y(t)]^{m}}}{Y(t)=y}
\;,
\]
one can alternatively express the jump moments as
\begin{equation}
\label{jumpmoments:derivative}
\boxequation{
	a^{(m)}(y,t)
=\lim_{\Delt\to0}\frac{1}{\Delt}
\eval{\BBav{[Y(t+\Delt)-Y(t)]^{m}}}{Y(t)=y}
\;.
			}
\end{equation}
In Sec.\ \ref{langevin} below, which is devoted to the Langevin equation,
we shall calculate the corresponding jump moments in terms of the
short-time conditional averages by means of this formula.


\subsection{Expressions for the multivariate case}

The above formulae can be extended to the case of a multi-component
Markov process $Y_{i}(t)$, $i=1$, $2,\ldots$, $N$.
Concerning the {\em \KM\ expansion\/} one only needs to use the multivariate
Taylor expansion to get
\begin{equation}
\label{kramersmoyal:n-dim}
\frac{\partial P}{\partial t}=
\sum_{m=1}^{\infty}\frac{(-1)^{m}}{m!}
\sum_{j_{1}\ldots j_{m}}
\frac{\partial^{m}} {\partial y_{j_{1}}\cdots\partial y_{j_{m}}}
\lrs{a_{j_{1},\ldots,j_{m}}^{(m)}(\multi{y},t)P}
\;,
\end{equation}
while the {\em \FP\ equation\/} is then given by
\begin{equation}
\label{fokkerplanck:n-dim}
\boxequation{
\frac{\partial P}{\partial t}= -\sum_{i}\frac{\partial}{\partial y_{i}}
\lrs{a_{i}^{(1)}(\multi{y},t)P}
+\frac{1}{2}\sum_{ij}
\frac{\partial^{2}} {\partial y_{i}\partial y_{j}}
\lrs{a_{ij}^{(2)}(\multi{y},t)P}
\;.
	}
\end{equation}
In these equations, the jump moments are given by the natural generalisation
of Eq.\ (\ref{jumpmoments:2}), namely%
%
%
\begin{equation}
\label{jumpmoments:n-dim:def}
a_{j_{1},\ldots,j_{m}}^{(m)}(\multi{y},t)
=\Intdef{}{}{\multi{y}'}
(y_{j_{1}}'-y_{j_{1}})\cdots(y_{j_{m}}'-y_{j_{m}})
W(\multi{y}'|\multi{y})
\;,
\end{equation}
and can be calculated by means of the corresponding generalisation of
Eq.\ (\ref{jumpmoments:derivative}):
\begin{equation}
\label{jumpmoments:derivative:n-dim}
\boxequation{
a_{j_{1},\ldots,j_{m}}^{(m)}(\multi{y},t)
=\eval{\lim_{\Delt\to0}\frac{1}{\Delt}
		\bigg\langle
		\prod_{\mu=1}^{m}[Y_{j_{\mu}}(t+\Delt)-Y_{j_{\mu}}(t)]
		\bigg\rangle
		}{Y_{k}(t)=y_{k}}
,			}
\end{equation}
that is, by means of the derivative of the corresponding conditional
average.


\subsection{Examples of \FP\ equations}
\label{subsec:FP-examples}


\paragraph*{Diffusion equation for the position.}

In Einstein's explanation of Brownian motion he arrived at an equation
of the form [see Eq.\ (\ref{fokkerplanck:einstein})]
\begin{equation}
\label{smoluchowski:free}
\boxequation{
\frac{\partial P}{\partial t}
=
D
\frac{\partial^{2} P}{\partial x^{2}}
\;.
}
\end{equation}
Comparing with the \FP\ equation (\ref{fokkerplanck}), we see that in
this case $a^{(1)}(x,t)\equiv 0$, since no forces act on the particle
and hence the net drift is zero.
Similarly $a^{(2)}(x,t)=2D$, which is independent of space and time.
This is because the properties of the surrounding medium are
homogeneous [otherwise $D=D(x)$].
The solution of this equation for $P(x,t=0)=\delta(x)$ was Eq.\
(\ref{pdf-diffusion}), which corresponds to the \WL\ process
(\ref{wienerlevy}).

This equation is a special case of the {\em Smoluchowski equation\/}
for a particle with large damping coefficient $\gamma$ (overdamped
particle), the special case corresponding to no forces acting on the
particle.


\paragraph*{Diffusion equation in phase space $(x,v)$.}

The true diffusion equation of a free Brownian particle is
\begin{equation}
\label{KK:free}
\boxequation{
\frac{\partial P}{\partial t}
=
-v
\frac{\partial P}{\partial x}
+
\gamma
\left(
\frac{\partial }{\partial v}
v
+
\frac{\kT}{m}
\frac{\partial^{2}}{\partial v^{2}}
\right)
P
\;.
}
\end{equation}
This equation is the no potential limit of the {\em Klein--Kramers
equation\/} for a particle with an arbitrary damping coefficient
$\gamma$.
From this equation one can obtain the diffusion equation
(\ref{smoluchowski:free}) using {\em singular perturbation
theory}, as the leading term in a expansion in powers of $1/\gamma$.
Alternatively, we shall give a proof of this in the context of the
Langevin equations corresponding to these Fokker--Planck equations.%
\footnote{
We shall see that the Langevin equation $m\ddot{x} = -\vis\,
\dot{x} + \Lan(t)$ [Eq.\ (\ref{langevin:original})] leads to Eq.\
(\ref{KK:free}), while the overdamped approximation $\vis\, \dot{x}
\simeq \Lan(t)$ corresponds to Eq.\ (\ref{smoluchowski:free}).
}

We have stated without proof that the \OU\ process describes the time
evolution of the transition probability of the velocity of a free
Brownian particle.
We shall demonstrate this, by solving the equation for the marginal
distribution for $v$ obtained from (\ref{KK:free}).
The marginal probability is $P_{V}(v,t)=\int\!\D x\,P(x,v,t)$.
Integrating Eq.\ (\ref{KK:free}) over $x$, using $\int\!\D
x\,\partial_{x}P(x,v,t)=0$, since $P(x=\pm\infty,v,t)=0$, we find
\begin{equation}
\label{KK:free:marginal}
\frac{\partial P_{V}}{\partial t}
=
\gamma
\left(
\frac{\partial }{\partial v}
v
+
\frac{\kT}{m}
\frac{\partial^{2}}{\partial v^{2}}
\right)
P_{V}
\;.
\end{equation}
We will see that this equation also describes the position of an
overdamped particle in an harmonic potential.
Thus, let us find the solution of the generic equation
\begin{equation}
\label{FP:OU}
\boxequation{
\tau
\partial_{t}
P
=
\partial_{y}
(yP)
+
D
\partial_{y}^{2}
P
\;.
}
\end{equation}
Introducing the characteristic function (\ref{charfunc}) (so we are
solving by the Fourier transform method)
\[
G(k,t)
=\int\!\D y\,e^{\iu ky}P(y,t)
\;,
\qquad
P(y,t)
=
\frac{1}{2\pi}
\int\D k\,e^{-\iu ky}G(k,t)
\;,
\]
the second order partial differential equation (\ref{FP:OU})
transforms into a first order one
\begin{equation}
\label{FP:OU:char}
\tau
\partial_{t}
G
+
k
\partial_{k}
G
=
-
D
k^{2}
G
\;,
\end{equation}
which can be solved by means of the method of characteristics.%
\footnote{
In brief, if we have a differential equation of the form
\[
P
\frac{\partial f}{\partial x}
+
Q
\frac{\partial f}{\partial y}
=
R
\;,
\]
and $u(x,y,f)=a$ and $v(x,y,f)=b$ are two solutions of the subsidiary
system
\[
\frac{\D x}{P}
=
\frac{\D y}{Q}
=
\frac{\D f}{R}
\;,
\]
the general solution of the original equation is an arbitrary function
of $u$ and $v$, $h(u,v)=0$.
}

In this case the subsidiary system is
\[
\frac{\D t}{\tau}
=
\frac{\D k}{k}
=
-
\frac{\D G}{D k^{2}G}
\;.
\]
Two integrals are easily obtained considering the systems $t,k$ and
$k,G$:
\begin{eqnarray*}
\frac{\D t}{\tau}
&=&
\frac{\D k}{k}
\quad
\rightarrow
\quad
k
=
a\,e^{t/\tau}
\quad
\rightarrow
\quad
u=ke^{-t/\tau}=a
\\
-D
k\D k
&=&
\D G/G
\;\;
\rightarrow
\;\;
-\half D
k^{2}
=
\ln G
+{\rm c}
\;\;
\rightarrow
\;\;
v
=
e^{-\half D k^{2}}
G
=b
\end{eqnarray*}
Then, the solution $h(u,v)=0$ can be solved for $v$ as $v=\phi(u)$ with
still an arbitrary function $\phi$, leading the desired general
solution of Eq.\ (\ref{FP:OU:char})
\begin{equation}
\label{FP:OU:char:gensol}
G
=
e^{-\half D k^{2}}
\phi(ke^{-t/\tau})
\;,
\end{equation}
by means of the methods of characteristics.

The solution for sharp initial values $P(y,t=0)=\delta(y-y_{0})$ leads
to $G(k,t=0)=\exp(\iu ky_{0})$, from which we get the functional form
of $\phi$: $\phi(k)
=
\exp(\iu ky_{0}+\half D k^{2})$.
Therefore, one finally obtains for $G(k,t)$
\begin{equation}
\label{FP:OU:char:parsol}
G(k,t)
=
\exp
\left[
\iu y_{0}e^{-t/\tau}\,k
-
\half
D(1-e^{-2t/\tau})
\,
k^{2}
\right]
\;,
\end{equation}
which is the characteristic function of a Gaussian distribution [see
Eq.\ (\ref{gaussian:1D:charfunct})], with $\mu_{1}=y_{0}e^{-t/\tau}$
and $\sigma^{2}=D(1-e^{-2t/\tau})$.
Therefore, the probability distribution solving Eq.\ (\ref{FP:OU}) is
\begin{equation}
\label{ornsteinuhlenbeck:Ptrans:2}
\boxequation{
P(y,t|y_{0},0)
=
\frac{1}{\sqrt{2\pi D(1-e^{-2t/\tau})}}
\exp
\lrs{
-\frac
{(y-y_{0}e^{-t/\tau})^{2}}
{2D(1-e^{-2t/\tau})}
}
\;.
}
\end{equation}
which, as stated, is the transition probability of the \OU\ process [Eq.\
(\ref{ornsteinuhlenbeck:Ptrans})].\qed

Note finally that the parameters for the original equation for $P_{V}$
[Eq.\ (\ref{KK:free:marginal})], are simply $\mu_{1}=v_{0}e^{-t/\tau}$ and
$\sigma^{2}=(\kT/m)(1-e^{-2t/\tau})$.
Thus, at long times we have $P_{V}\propto\exp(-\half mv^{2}/\kT)$
which is simply the statistical mechanical equilibrium Boltzmann
distribution for free particles.


\paragraph*{Diffusion equation for a dipole.}

The diffusion equation for a dipole moment $p$ in an electric field $E$ is
(neglecting inertial effects)
\begin{equation}
\label{debye}
\boxequation{
\zeta
\frac{\partial P}{\partial t}
=
\frac{1}{\sin\vartheta}
\frac{\partial }{\partial\vartheta}
\left[
\sin\vartheta
\left(
\kT
\frac{\partial P}{\partial\vartheta}
+
pE
\sin\vartheta
P
\right)
\right]
\;.
}
\end{equation}
This equation was introduced by Debye in the 1920's, and constitutes
the first example of rotational Brownian motion.
$\zeta$ is the viscosity coefficient (the equivalent to $\gamma$ in
translational problems).
It is easily seen that $P_{0}\propto\exp(pE\cos\vartheta/\kT)$ is the
stationary solution of Eq.\ (\ref{debye}), which leads to the famous
result for the average dipole moment of an assembly of dipoles,
$\llangle\cos\vartheta\rrangle=\coth\alpha-1/\alpha$ with $\alpha=pE/\kT$
and to {\em Curie's paramagnetic law\/} at low fields.
However, Eq.\ (\ref{debye}) also governs non-equilibrium situations,
and in particular the time evolution between different equilibrium
states.

\mynewpage

\section{The Langevin equation}
\label{langevin}
\markboth{Introduction to the theory of stochastic processes}{The
Langevin equation}


\subsection{Langevin equation for one variable}

The Langevin equation for one variable is a ``differential equation"
of the form [cf.\ Eq.\ (\ref{langevin:original})]
\begin{equation}
\label{langevinequation}
\boxequation{
\frac{\D y}{\D t}=\drift(y,t)+\diff(y,t)\Lan(t)
\;,
}
\end{equation}
where $\Lan(t)$ is a given stochastic process.
The choice for $\Lan(t)$ that renders $y(t)$%
\footnote{Hereafter, we use the same symbol for the stochastic process
$Y(t)$ and its realisations $y(t)$.}
a Markov process is that of the Langevin ``process" (white noise),
which is Gaussian and its statistical properties are
%
%
\begin{subeqnarray}
\label{langevin:moments}
\slabel{langevin:1st_moment}
\av{\Lan(t)}
&=&0
\;,
\\
\slabel{langevin:2nd_moment}
\llangle
\Lan(t_{1})\Lan(t_{2})
\rrangle
&=&2D\delta(t_{1}-t_{2})
\;.
\end{subeqnarray}
Since Eq.\ (\ref{langevinequation}) is a {\em first-order\/}
differential equation, for each sample function (realisation) of
$\Lan(t)$, it determines $y(t)$ uniquely when $y(t_{0})$ is given.
In addition, the values of the fluctuating term at different times are
statistically independent, due to the delta-correlated nature of
$\Lan(t)$.
Therefore, the values of $\Lan(t)$ at previous times, say $t'<t_{0}$, cannot
influence the conditional probabilities at times $t>t_{0}$.
From these arguments it follows the Markovian character of the
solution of the Langevin equation (\ref{langevinequation}).

The terms $\drift(y,t)$ and $\diff(y,t)\Lan(t)$ are often referred to as
the {\em drift\/} ({\em transport}) and {\em diffusion\/} terms,
respectively.
Due to the presence of $\Lan(t)$, Eq.\ (\ref{langevinequation}) is a
{\em stochastic differential equation}, that is, a differential
equation comprising random terms with given stochastic properties.
To solve a Langevin equation then means to determine the statistical
properties of the process $y(t)$.

%

Finally, the higher-order moments of $\Lan(t)$ are obtained from the
second order ones (\ref{langevin:moments}), by assuming relations like
those of the multivariate Gaussian case, i.e., all odd moments of
$\Lan(t)$ vanish and, e.g.,
\begin{eqnarray}
\label{wick:langevin}
\av{\Lan(t_{1})\Lan(t_{2})\Lan(t_{3})\Lan(t_{4})}
=(2D)^{2}
&\big[&
\av{\Lan(t_{1})\Lan(t_{2})}\av{\Lan(t_{3})\Lan(t_{4})}
\nonumber\\
& &
{}+\av{\Lan(t_{1})\Lan(t_{3})}\av{\Lan(t_{2})\Lan(t_{4})}
\nonumber\\
& &
{}+\av{\Lan(t_{1})\Lan(t_{4})}\av{\Lan(t_{2})\Lan(t_{3})}\big]
\;.
\end{eqnarray}
To check this result, we shall demonstrate a slightly more general
result known as Novikov theorem.


\subsubsection{Novikov theorem and Wick formula}

Novikov theorem states that for a multivariate Gaussian distribution
(Sec.\ \ref{gaussian}) with zero mean
\begin{equation}
\label{gaussian:P:centered}
P(\multi{x})
=
\sqrt{\frac{\det\hat{A}}{(2\pi)^{n}}}
\exp
\left(
-\half
\multi{x}
\cdot
\hat{A}
\cdot
\,
\multi{x}
\right)
\;,
\end{equation}
the averages of the type $\llangle x_{i}f(\multi{x})\rrangle$, can be
obtained as
\begin{equation}
\label{novikov}
\boxequation{
\llangle x_{i}f(\multi{x})\rrangle
=
\sum_{m}
\llangle x_{i}x_{m}\rrangle
\llangle
\frac{\partial f}{\partial x_{m}}
\rrangle
\;.
}
\end{equation}
Applying this result to $f(\multi{x})=x_{j}x_{k}x_{\ell}$ and using
$\partial x_{i}/\partial x_{m}=\delta_{im}$, we have
\[
\llangle x_{i}x_{j}x_{k}x_{\ell}\rrangle
=
\sum_{m}
\llangle x_{i}x_{m}\rrangle
\Big\langle
\underbrace{
\delta_{jm}x_{k}x_{\ell}
+
x_{j}\delta_{km}x_{\ell}
+
x_{j}x_{k}\delta_{\ell m}
}_{\partial f/\partial x_{m}}
\Big\rangle
\;.
\]
Therefore, using the Kronecker's delta to do the sum, we get Wick's formula
\begin{equation}
\label{wick}
\llangle
x_{i}x_{j}x_{k}x_{\ell}
\rrangle
=
\llangle
x_{i}x_{j}
\rrangle
\llangle
x_{k}x_{\ell}
\rrangle
+
\llangle
x_{i}x_{k}
\rrangle
\llangle
x_{j}x_{\ell}
\rrangle
+
\llangle
x_{i}x_{\ell}
\rrangle
\llangle
x_{j}x_{k}
\rrangle
\;.
\end{equation}
Equation (\ref{wick:langevin}) then follows because $\Lan(t)$ is
assumed to be a Gaussian process, which {\em by definition\/} means
that the $n$ times probability distribution
$P_{n}(\Lan_{1},t_{1};\ldots;\Lan_{n},t_{n})$ is a multivariate
Gaussian distribution.


\paragraph*{Proof of Novikov theorem.}

We shall demonstrate this theorem in three simple steps:

\noindent
(1) If we denote by $E(\multi{x})=\half\sum_{ij}x_{i}A_{ij}x_{j}$
minus the exponent in the Gaussian distribution
(\ref{gaussian:P:centered}), we have
\begin{eqnarray*}
\frac{\partial E}{\partial x_{m}}
&=&
\frac{1}{2}
\frac{\partial }{\partial x_{m}}
\sum_{ij}x_{i}A_{ij}x_{j}
=
\half
\sum_{ij}(\delta_{im}A_{ij}x_{j}+x_{i}A_{ij}\delta_{jm})
\\
&=&
\half
\sum_{j}A_{mj}x_{j}
+
\half
\sum_{i}x_{i}A_{im}
\\[-1.ex]
\mbox{\small [$A_{ij}$ sym.]}
\quad
&=&
\sum_{j}A_{mj}x_{j}
\quad
\Rightarrow
\quad
\boxequation{
x_{i}
=
\sum_{m}(A^{-1})_{im}
\frac{\partial E}{\partial x_{m}}
}
\;.
\end{eqnarray*}

\noindent
(2) Using the definition of average for $\llangle
x_{i}f(\multi{x})\rrangle$, inserting the above expression for $x_{i}$
and integrating by parts, we have
\begin{eqnarray*}
\llangle x_{i}f(\multi{x})\rrangle
&=&
{\rm C}
\int\!\D \multi{x}\,
x_{i}f(\multi{x})
e^{-E}
\\[-2.ex]
&=&
{\rm C}
\sum_{m}(A^{-1})_{im}
\int\!\D \multi{x}\,
f(\multi{x})
\overbrace{
\frac{\partial E}{\partial x_{m}}
e^{-E}
}^{-\partial_{x_{m}}e^{-E}}
\\
&=&
\sum_{m}(A^{-1})_{im}
{\rm C}
\int\!\D \multi{x}\,
\frac{\partial f}{\partial x_{m}}
e^{-E}
\quad
\Rightarrow
\quad
\boxequation{
\llangle x_{i}f(\multi{x})\rrangle
=
\sum_{m}
(A^{-1})_{im}
\llangle
\frac{\partial f}{\partial x_{m}}
\rrangle
}
\end{eqnarray*}

\noindent
(3) Finally, we demonstrate that $\llangle
X_{i}X_{j}\rrangle = (\hat{A}^{-1})_{ij}$ (a particular case of the
result $\cum{X_{i}X_{j}}=(\hat{A}^{-1})_{ij}$ given without proof in
Sec.\ \ref{gaussian}).
Indeed, using the above result for $f=x_{j}$ and $\partial
x_{j}/\partial x_{m}=\delta_{jm}$, we have $\llangle
x_{i}x_{j}\rrangle = \sum_{m} (A^{-1})_{im} \delta_{jm} =
(A^{-1})_{ij}$.
Insertion of this in the above result completes the proof of Novikov's
theorem.


\subsection{The \KM\ coefficients for the Langevin equation}

Since the solution of the Langevin equation is a Markov process, it
obeys a master equation, which may be written in the \KM\ form
(\ref{kramersmoyal}).
Let us calculate the successive coefficients
(\ref{jumpmoments:derivative}) occurring in that expansion.
We first cast the differential equation (\ref{langevinequation}) into the form
of an integral equation
\begin{equation}
\label{langevinequation:int}
y(t+\Delt)-y =\Intdef{t}{t+\Delt}{t_{1}}\drift[y(t_{1}),t_{1}]
+\Intdef{t}{t+\Delt}{t_{1}}\diff[y(t_{1}),t_{1}]\Lan(t_{1})
\;,
\end{equation}
where $y$ stands for the initial value $y(t)$.
On expanding according to
\begin{eqnarray*}
\drift[y(t_{1}),t_{1}]
&=&
\drift(y,t_{1})+\drift'(y,t_{1})[y(t_{1})-y]+\cdots
\;,
\\
\diff[y(t_{1}),t_{1}]
&=&
\diff(y,t_{1})+\diff'(y,t_{1})[y(t_{1})-y]+\cdots
\;,
\end{eqnarray*}
where the prime denotes partial derivative with respect to $y$ evaluated at
the initial point:
\[
\drift'(y,t)\equiv\eval{\Dpar{\drift}{y}}{y}
\qquad
\diff'(y,t)\equiv\eval{\Dpar{\diff}{y}}{y}
\;,
\]
one gets
\begin{eqnarray}
\label{langevinequation:int:2}
y(t+\Delt)-y&=&
\Intdef{t}{t+\Delt}{t_{1}}\drift(y,t_{1})
\nonumber\\
& &
{}+\Intdef{t}{t+\Delt}{t_{1}}\drift'(y,t_{1})[y(t_{1})-y]+\cdots
\nonumber\\
& &
{}+\Intdef{t}{t+\Delt}{t_{1}}\diff(y,t_{1})\Lan(t_{1})
\nonumber\\
& &
{}+\Intdef{t}{t+\Delt}{t_{1}}\diff'(y,t_{1})[y(t_{1})-y]\Lan(t_{1})+\cdots
\;.
\end{eqnarray}
For $y(t_{1})-y$ in the above integrands we iterate
Eq.\ (\ref{langevinequation:int:2}) to get
\begin{eqnarray}
\label{langevinequation:int:3}
y(t+\Delt)\!-\!y
&=&
\Intdef{t}{t+\Delt}{t_{1}}\drift(y,t_{1})
\nonumber\\
& &
{}+\Intdef{t}{t+\Delt}{t_{1}}\drift'(y,t_{1})
\Intdef{t}{t_{1}}{t_{2}}\drift(y,t_{2})
\nonumber\\
& &
{}+\Intdef{t}{t+\Delt}{t_{1}}\drift'(y,t_{1})
\Intdef{t}{t_{1}}{t_{2}}\diff(y,t_{2})\Lan(t_{2}) +\cdots
\nonumber\\
& &
{}+\Intdef{t}{t+\Delt}{t_{1}}\diff(y,t_{1})\Lan(t_{1})
\nonumber\\
& &
{}+\Intdef{t}{t+\Delt}{t_{1}}\diff'(y,t_{1})\Lan(t_{1})
\Intdef{t}{t_{1}}{t_{2}}\drift(y,t_{2})
\nonumber\\
& &
{}+\Intdef{t}{t+\Delt}{t_{1}}\diff'(y,t_{1})\Lan(t_{1})
\Intdef{t}{t_{1}}{t_{2}}\diff(y,t_{2})\Lan(t_{2})+\cdots
\;.
\end{eqnarray}
If we take the average of this equation for fixed $y=y(t)$, by using the
statistical properties (\ref{langevin:moments}), we obtain the conditional
average required to get $a^{(1)}(y,t)$
\begin{eqnarray*}
\label{langevinequation:int:av}
\av{y(t+\Delt)-y}
&=&
\Intdef{t}{t+\Delt}{t_{1}}\drift(y,t_{1})
+\Intdef{t}{t+\Delt}{t_{1}}\drift'(y,t_{1})
\Intdef{t}{t_{1}}{t_{2}}\drift(y,t_{2})
\\
& &
{}+2D\Intdef{t}{t+\Delt}{t_{1}}\diff'(y,t_{1})
\Intdef{t}{t_{1}}{t_{2}}\diff(y,t_{2})\delta(t_{2}-t_{1})+\cdots
\;.
\end{eqnarray*}
Next, on using for the Dirac delta the result
$\Intdef{t_{0}}{t_{1}}{t}
\delta(t-t_{0})f(t)=\half f(t_{0})$, we obtain
\begin{equation}
\label{G&delta}
\Intdef{t}{t_{1}}{t_{2}}\diff(y,t_{2})\delta(t_{2}-t_{1})
=\half \diff(y,t_{1})
\;.
\end{equation}
Finally, on considering that $a^{(1)}
=
{\displaystyle\lim_{\Delt\to0}}
\frac{1}{\Delt}\av{y(t+\Delt)-y}|_{y(t)=y}$, for
the calculation of which only terms through order $\Delt$ need to be retained,
one finally gets
\[
a^{(1)}(y,t)=\drift(y,t)+D\,\diff(y,t)\Dpar{\diff(y,t)}{y}
\;.
\]

\sloppy
Other integrals not written down in the above formulae do not contribute in
the limit $\Delt\to0$.
This can be seen as follows: each Langevin fluctuating term on the
right-hand side of Eq.\ (\ref{langevinequation:int:3}), is accompanied
by an integral.
The lowest-order terms are written in that expression, whereas higher-order
terms can be of two types:
(i) Integrals of the form of, e.g.,
\[
\av{
		\Intdef{t}{t+\Delt}{t_{1}}\cdots\Lan(t_{1})
		\Intdef{t}{t_{1}}{t_{2}}\cdots\Lan(t_{2})
		\Intdef{t}{t_{2}}{t_{3}}\cdots\Lan(t_{3})
		\Intdef{t}{t_{3}}{t_{4}}\cdots\Lan(t_{4})
		}
\;,
\]
which can only give a contribution proportional to $(\Delt)^{2}$, as it is
seen by using the splitting of
$\av{\Lan(t_{1})\Lan(t_{2})\Lan(t_{3})\Lan(t_{4})}$
in sum of products of the form
$\av{\Lan(t_{i})\Lan(t_{j})}\av{\Lan(t_{k})\Lan(t_{\ell})}$
[Eq.\ (\ref{wick:langevin})].
(ii) Integrals containing no Langevin terms, which are proportional to
$(\Delt)^{n}$, where $n$ is the number of simple integrals.
Both types of terms clearly vanish when dividing by $\Delt$ and taking the
limit $\Delt\to0$.

\fussy

On using the same type of arguments to identify some vanishing
integrals one can compute the second coefficient in the \KM\
expansion,
$a^{(2)}
=
{\displaystyle\lim_{\Delt\to0}}\frac{1}{\Delt}\av{[y(t+\Delt)-y]^{2}}|_{y(t)
=
y}$,
obtaining
\[
a^{(2)}(y,t)
=\lim_{\Delt\to0}\frac{1}{\Delt}
\Intdef{t}{t+\Delt}{t_{1}}\diff(y,t_{1})
\Intdef{t}{t+\Delt}{t_{2}}\diff(y,t_{2})2D\delta(t_{1}-t_{2})
=2D\,\diff^{2}(y,t)
\;,
\]
whereas all the coefficients $a^{(m)}$ vanish for $m\geq3$.
Thus, on collecting all these results one can finally write
\begin{equation}
\label{a_nu:langevin}
\boxeqnarray{22.em}{
a^{(1)}(y,t)
&=&
\drift(y,t)+D\,\diff(y,t)\Dpar{\diff(y,t)}{y}
\;,
\\
a^{(2)}(y,t)&=&2D\,\diff^{2}(y,t)
\;,
\\[1ex] a^{(m)}(y,t)&=&0
\;,
\qquad\mbox{for}\quad m\geq3
\;.
}
\end{equation}


\subsection{\FP\ equation for the Langevin equation}

From Eq.\ (\ref{a_nu:langevin}) it follows that, for the Markov
stochastic process determined by the Langevin equation
(\ref{langevinequation}) with Gaussian $\delta$-correlated $\xi(t)$, the
\KM\ expansion includes up to second-order terms.
Therefore, the distribution of probability obeys a \FP\ equation [Eq.\
(\ref{fokkerplanck})], which in terms of the above jump moments is
explicitly given by
\begin{equation}
\label{fokkerplanck:langevin}
\boxequation{
\frac{\partial P}{\partial t}=
-\Dpar{}{y}\lrc{\lrs{\drift(y,t)+D\,\diff(y,t)\Dpar{\diff(y,t)}{y}}P}
+D\Dpar{^{2}}{y^{2}}\lrs{\diff^{2}(y,t)P} .
	}
\end{equation}
Note that, along with the deterministic drift $\drift(y,t)$,
$a^{(1)}(y,t)$ contains a term, $D\diff(y,t)\diff'(y,t)$, which is called
the {\em noise-induced drift}.
%
This equation is very important, since it allows one to construct the
Fokker--Planck equation directly in terms of the coefficients
appearing in the equation of motion.
In some cases, it can even be done by simply inspection of that
equation.


\subsubsection{Multivariate case}

The stochastic differential (Langevin) equation for a multi-component
process $\multi{y}=(y_{1},\cdots,y_{N})$ has the form
\begin{equation}
\label{langevinequation:n-dim}
\boxequation{
\frac{\D y_{i}}{\D t}
=
\drift_{i}(\multi{y},t)
+
\sum_{k}\diff_{ik}(\multi{y},t)\Lan_{k}(t)
\;,
}
\end{equation}
where the $\Lan_{k}(t)$ are $N_{\text{L}}$ white-noise terms.%
\footnote{The number of Langevin sources, $N_{\text{L}}$, does
not need to be equal to the number of equations.
For example, the sum in $k$ in Eq.\ (\ref{langevinequation:n-dim}) can
even have one term, $N_{\text{L}}=1$ ---the case of ``scalar noise".
} 
The statistical properties of the $\Lan_{k}(t)$ are
\begin{subeqnarray}
\label{langevin:n-dim}
\slabel{langevin:n-dim:1st_moment}
\llangle
\Lan_{k}(t)
\rrangle
&=&
0
\;,
\\
\slabel{langevin:n-dim:2nd_moment}
\llangle
\Lan_{k}(t_{1})\Lan_{\ell}(t_{2})
\rrangle
&=&
2D\delta_{k\ell}\delta(t_{1}-t_{2})
\;.
\end{subeqnarray}
Again, the higher-order moments are obtained from these ones, on assuming
relations like those of the (multivariate) Gaussian case.

The successive coefficients (\ref{jumpmoments:derivative:n-dim})
occurring in the \KM\ expansion (\ref{kramersmoyal:n-dim}) can be
calculated by using arguments entirely analogous to those employed
above to identify some vanishing integrals.
On doing so, one gets the following generalisation of Eqs.\
(\ref{a_nu:langevin}) in the multivariate case:
\begin{eqnarray}
\label{a_nu:langevin:n-dim}
a_{i}^{(1)}(\multi{y},t)
&=&
\drift_{i}(\multi{y},t)
+D\sum_{jk}\,\diff_{jk}(\multi{y},t)
\frac{\partial\diff_{ik}(\multi{y},t)}{\partial y_{j}}
\;,
\nonumber\\
a_{ij}^{(2)}(\multi{y},t)
&=&2D\sum_{k}\,\diff_{ik}(\multi{y},t)\diff_{jk}(\multi{y},t)
\;,
\\
a_{j_{1},\ldots,j_{m}}^{(m)}(\multi{y},t)
&=&0
\;,
\qquad\mbox{for}\quad m\geq3
\nonumber
\;.
\end{eqnarray}
Again, for the Markov stochastic process defined by the set
(\ref{langevinequation:n-dim}) of Langevin equations, the \KM\
expansion of the master equation includes up to second-order terms, so
that the probability distribution obeys a \FP\ equation
\begin{equation}
\label{fokkerplanck:langevin:n-dim}
\boxeqnarray{30.em}{
\frac{\partial P}{\partial t}
&=&
-\sum_{i}
\frac{\partial}{\partial y_{i}}
\bigg\{
\bigg[
\drift_{i}(\multi{y},t)
+
D\sum_{jk}\,\diff_{jk}(\multi{y},t)
\frac{\partial\diff_{ik}(\multi{y},t)}{\partial y_{j}}
\bigg]
P
\bigg\}
\\
& &
{}+
D
\sum_{ij}
\frac{\partial^{2}}{\partial y_{i}\partial y_{j}}
\bigg\{
\bigg[
\sum_{k}\,\diff_{ik}(\multi{y},t)\diff_{jk}(\multi{y},t)
\bigg]
P
\bigg\}
\;,
		}
\end{equation}
which is entirely determined by the coefficients of the Langevin equation.


\subsection{Examples of Langevin equations and derivation of their
\FP\ equations}


\subsubsection{
Diffusion in phase-space: Klein--Kramers equation
}

Let us consider the following generalisation of the original Langevin
equation (\ref{langevin:original}), in order to account for the
presence of an external potential $U(x,t)$ (e.g., gravity in the
Brownian motion problem)
\begin{equation}
\label{langevin:newton}
\boxequation{
m
\frac{\D^{2}x}{\D t^{2}}
=
-\vis\,
\frac{\D x}{\D t}
-\frac{\partial U}{\partial x}
+
m\Lan(t)
\;.
}
\end{equation}
This is simply Newton equation augmented by the fluctuating force [for
convenience we have extracted $m$ from $\xi(t)$].

Let us divide by the mass, introduce $V=U/m$ and the notation
$V'=\partial V/\partial x$, and write (\ref{langevin:newton}) as a
pair of first-order differential equations
\begin{eqnarray}
\frac{\D x}{\D t}
&=&
v
\\
\frac{\D v}{\D t}
&=&
-(\gamma\,v+V')
+
\Lan(t)
\;.
\end{eqnarray}
Then, comparing with the multivariate Langevin equation
(\ref{langevinequation:n-dim}), we identify $\xi_{x}(t)\equiv0$ and
$\xi_{v}(t)=\xi(t)$, as well as
\[
\begin{array}{lcll}
\drift_{x}
=
v
&
&
\diff_{xx}\equiv0
&
\diff_{xv}\equiv0
\\
\drift_{v}
=
-(\gamma\,v+V')
&
&
\diff_{vx}\equiv0
&
\diff_{vv}=1
\end{array}
\]
Inserting these results in the general \FP\ equation
(\ref{fokkerplanck:langevin:n-dim}), one gets (note
$\partial_{j}\diff_{ik}\equiv0$)
\[
\frac{\partial P}{\partial t}
=
\left[
-\frac{\partial }{\partial x}
v
-\frac{\partial }{\partial v}
[-(\gamma\,v+V')]
+
\Dcoeff
\frac{\partial^{2}}{\partial v^{2}}
\right]
P
\;.
\]
Gathering the Hamiltonian terms and identifying
$\Dcoeff/\gamma=\kT/m$, we finally find the famous {\em Klein--Kramers
equation\/}
\begin{equation}
\label{KK}
\boxequation{
\frac{\partial P}{\partial t}
=
-v
\frac{\partial P}{\partial x}
+
V'
\frac{\partial P}{\partial v}
+
\gamma
\left(
\frac{\partial }{\partial v}
v
+
\frac{\kT}{m}
\frac{\partial^{2}}{\partial v^{2}}
\right)
P
\;.
}
\end{equation}
The result $\Dcoeff/\gamma=\kT/m$ comes from inserting the Boltzmann
distribution $P_{0}\propto\exp[-(\half mv^{2}+mV)/\kT]$ and finding
the conditions for it to be a stationary solution.
This is equivalent to P. Langevin recourse to the equipartition
theorem (Sec.\ \ref{langevin:explanation}) to find $\llangle
mv^{2}\rrangle=\kT$.
Note finally that in the absence of potential, the Klein--Kramers
equation leads to the equation for free diffusion (\ref{KK:free}),
with solution (for the marginal distribution $P_{V}(v,t)=\int\!\D
x\,P(x,v,t)$) given by the \OU\ process
(\ref{ornsteinuhlenbeck:Ptrans:2}).


\subsubsection{
Overdamped particle: Smoluchowski equation
}

Let us consider the overdamped limit of the Newton--Langevin equation
(\ref{langevin:newton})
\begin{equation}
\label{langevin:newton:overdamped}
\boxequation{
\frac{\D x}{\D t}
=
-V'/\gamma
+
\Lan(t)/\gamma
\;.
}
\end{equation}
Comparing with the univariate Langevin equation
(\ref{langevinequation}), we identify
\[
\drift
=
-V'/\gamma
\;,
\qquad
B\equiv1/\gamma
\;.
\]
Inserting these results in the \FP\ equation
(\ref{fokkerplanck:langevin}), one gets (putting again
$\Dcoeff/\gamma=\kT/m$) the {\em Smoluchowski equation}
\begin{equation}
\label{smoluchowski}
\boxequation{
\frac{\partial P}{\partial t}
=
\left(
\frac{1}{\gamma}
\frac{\partial }{\partial x}
V'
+
\frac{\kT}{m\gamma}
\frac{\partial^{2}}{\partial x^{2}}
\right)
P
\;.
}
\end{equation}
The result $\Dcoeff/\gamma=\kT/m$ can also be obtained on inserting
the marginal Boltzmann distribution $P_{0}\propto\exp[-mV(x)/\kT]$ and
finding the conditions for it to be a stationary solution.

In the absence of potential, the Smoluchowski equation leads to the
equation for free diffusion (\ref{smoluchowski:free}) or Einstein's
Eq.\ (\ref{fokkerplanck:einstein}), with solution given by the \WL\
process [Eqs.\ (\ref{pdf-diffusion}) or (\ref{wienerlevy})].
Note also that in an harmonic potential
$V(x)=\half\omega_{0}^{2}x^{2}$, the equation is equivalent to Eq.\
(\ref{FP:OU}), with parameters
$\tau=\gamma/\omega_{0}^{2}$, $D=\kT/m\omega_{0}^{2}$, whose
solution is the \OU\ process (\ref{ornsteinuhlenbeck:Ptrans:2}).
Therefore we can at once write for the overdamped harmonic oscillator
\begin{equation}
\label{ornsteinuhlenbeck:Ptrans:HO}
P(x,t)
=
\sqrt{\frac{m\omega_{0}^{2}}{2\pi\kT(1-e^{-2t/\tau})}}
\exp
\lrs{
-\frac
{m\omega_{0}^{2}(x-x_{0}e^{-t/\tau})^{2}}
{2\kT(1-e^{-2t/\tau})}
}
\;,
\qquad
\tau=\gamma/\omega_{0}^{2}
\;.
\end{equation}
Thus, at long times we have $P\propto\exp(-\half
m\omega_{0}^{2}x^{2}/\kT)$ which is simply the statistical mechanical
equilibrium Boltzmann distribution for the harmonic oscillator.
In addition Eq.\ (\ref{ornsteinuhlenbeck:Ptrans:HO}) tells us how the
relaxation to the equilibrium state proceeds.

\renewcommand{\mm}{s}
\renewcommand{\eff}{}
\renewcommand{\bfl}{\vec{\xi}}

\subsubsection{
Equations for a classical spin (dipole)
}

For classical spins, the Langevin equation is the {\em stochastic
Landau--Lifshitz equation\/}, which for instance describes the
dynamics of the magnetic moment of a magnetic nanoparticle.
Written in simplified units (fields in frequency units), it
reads
%
\begin{equation}
\label{stolleq:mod}
\boxequation
{
\frac{\D\m}{\D t}
=
\m
\vecpro
\left[\Beff+\bfl(t)\right]
-\lambda
\m
\vecpro
\left(\m\vecpro\Beff\right)
\;,
}
\end{equation}
Here, $\Beff=-\partial\Hs/\partial\m$ is the effective field associated
with the Hamiltonian of the spin $\Hs(\s)$ (the equivalent to
$F=-\partial U/\partial x$ in mechanical problems), and the double vector
product is the damping term, which rotates $\m$ towards the potential
minima (preserving its length).
The stochastic properties of the components of $\bfl(t)$ are the
usual ones [Eq.\ (\ref{langevin:n-dim})], but $\bfl(t)$ is now
interpreted as a fluctuating field.
Finally, the damping coefficient $\lambda$ measures the relative
importance of the relaxation and precession terms.

The stochastic Landau--Lifshitz equation (\ref{stolleq:mod}), can be
cast into the form of the general system of Langevin equations
(\ref{langevinequation:n-dim}), by identifying
\begin{eqnarray}
\label{Fi:ll:mod}
\drift_{i}
&=&
\sum_{jk}
\epsilon_{ijk}\,
\mj
B_{\eff k}
+
\lambda
\sum_{k}
(\mm^{2}\delta_{ik}-\mi\mk)
B_{\eff k}
\;,
\\
\label{Gik:ll:mod}
\diff_{ik}
&=&
\sum_{j}\epsilon_{ijk}\mj
\;.
\end{eqnarray}
where $\epsilon_{ijk}$ is the antisymmetrical unit tensor of rank
three (Levi-Civita symbol)%
\footnote{
This tensor is defined as the tensor antisymmetrical in all three
indices with $\epsilon_{xyz}=1$.
Therefore, one can write the vector product of $\vec{A}$ and $\vec{B}$ as
$\big(\vec{A}\vecpro\vec{B}\big)_{i}
=\sum_{jk}\epsilon_{ijk}A_{j}B_{k}$.
In addition, one has the useful contraction property
\begin{eqnarray}
\label{contraction:1}
\sum_{k}\epsilon_{ijk}\epsilon_{i'j'k}
&=&
\delta_{ii'}\delta_{jj'}-\delta_{ij'}\delta_{ji'}
\\
\label{contraction:2}
\sum_{jk}\epsilon_{ijk}\epsilon_{i'jk}
&=&
2\delta_{ii'}
\;,
\qquad
\sum_{ijk}\epsilon_{ijk}\epsilon_{ijk}
=
6
\;.
\end{eqnarray}
where the last two are obtained by repeated contraction of the first one.
}
and we have expanded the triple vector products
$-\m\vecpro(\m\vecpro\Beff)$ by using the rule 
$\vec{a}\vecpro\big(\vec{b}\vecpro\vec{c}\big)
=
\vec{b}\big(\vec{a}\cdot\vec{c}\big)
-\vec{c}\big(\vec{a}\cdot\vec{b}\big)$
(``BAC-CAB" rule).
%

To calculate the {\em noise-induced\/} drift coefficient of the
Fokker--Planck equation [the term accompanying $\drift_{i}$ in Eq.\
(\ref{fokkerplanck:langevin:n-dim})] we need the derivative of the
diffusion ``coefficient":
\begin{equation}
\label{Gij:derivative:ll:mod}
\frac{\partial\diff_{ik}}{\partial\mj}
=
\epsilon_{ijk}
\;.
\end{equation}
On multiplying by Eq.\ (\ref{Gik:ll:mod}) for $\diff_{ik}$, summing, and
using the second contraction property (\ref{contraction:2}) for the
$\epsilon_{ijk}$, one finds
\[
\sum_{jk}\diff_{jk}\frac{\partial\diff_{ik}}{\partial\mj}
=
\sum_{\ell}
\overbrace{
\Big(
\sum_{jk}\epsilon_{j\ell k}\epsilon_{ijk}
\Big)
}^{-2\delta_{i\ell}}
\ml
\;,
\quad
\Rightarrow
\quad
\boxequation
{
\Dcoeff
\sum_{jk}\,
\diff_{jk}
\frac{\partial\diff_{ik}}{\partial\mj}
=
-2\Dcoeff\mi
}
\]
Let us compute now the coefficient in the diffusion term
$\sum_{k}\diff_{ik}\diff_{jk}$:
\begin{eqnarray*}
\label{GikGjk:ll}
\sum_{k}\diff_{ik}\diff_{jk}
&=&
\sum_{k}
\Big(
\sum_{r}
\epsilon_{irk}\mr
\Big)
\Big(
\sum_{s}
\epsilon_{jsk}\ms
\Big)
=
\sum_{r,s}
\mr\ms
\overbrace{
\sum_{k}
\epsilon_{irk}
\epsilon_{jsk}
}^{\delta_{ij}\delta_{rs}-\delta_{is}\delta_{rj}}
\\
&=&
\mm^{2}\delta_{ij}-\mi\mj
\;,
\quad
\Rightarrow
\quad
\sum_{k}\diff_{ik}\diff_{jk}
=
\mm^{2}\delta_{ij}-\mi\mj
\;,
\end{eqnarray*}
where we have employed the contraction rule (\ref{contraction:1}).

Therefore the Langevin equation associated to the stochastic
Landau--Lifshitz equation (\ref{stolleq:mod}), reads
\begin{eqnarray*}
\frac{\partial P}{\partial t}
&=&
-\sum_{i}\Dpar{}{\mi}
\bgbgc{
\bgbgs{
\sum_{jk}
\epsilon_{ijk}\,
\mj
B_{\eff k}
+
\lambda
\sum_{k}
(\mm^{2}\delta_{ik}-\mi\mk)
B_{\eff k}
-
2\Dcoeff\mi
}
P
}
\\
& &
{}+
D\sum_{ij}\Dpar{^{2}}{\mi\partial\mj}
\bgbgs{
\bgbg{
\mm^{2}\delta_{ij}-\mi\mj
}
P
}
\;.
\end{eqnarray*}
Taking the $\mj$-derivative in the last term by
using
$\sum_{j}\partial_{j}(\mm^{2}\delta_{ij}-\mi\mj)
=
\sum_{j}(2\mj\delta_{ij}-\delta_{ij}\mj-\mi\delta_{jj})
=
2\si-\si-3\si$,
we obtain $\Dcoeff\sum_{i}\partial_{i} [-2\mi
P+\sum_{j}(\mm^{2}\delta_{ij}-\mi\mj)\partial_{j}P]$.
The first term cancels $-2\Dcoeff\mi$ in the drift and the
2nd can be combined with
$\sum_{k}(\mm^{2}\delta_{ik}-\mi\mk)B_{\eff k}$.
Finally, returning to a vector notation
\begin{equation}
\boxequation{
\frac{\partial P}{\partial t}
=
-\Dpar{}{\m}
\cdot
\Bigg\{
\m
\vecpro
\B
-
\lambda
\,
\m
\vecpro
\Bigg[
\m\vecpro
\bigg(
\B
-
\kT
\Dpar{}{\m}
\bigg)
\Bigg]
\Bigg\}
P
\;,
}
\end{equation}
where $(\partial/\partial\m)\cdot\vec{J} =\sum_{i} (\partial
J_{i}/\partial\mi)$ (divergence) and, by analogy with the mechanical
problems, we have set $\Dcoeff=\lambda\kT$.
This equation can be seen as the rotational counterpart of the
Klein--Kramers equation.

The electric case corresponds to the precession term dominated by the
damping term (a sort of Smoluchowski equation)
\begin{equation}
\frac{\partial P}{\partial t}
=
\lambda
\Dpar{}{\vec{p}}
\cdot
\Bigg\{
\,
\vec{p}
\vecpro
\Bigg[
\vec{p}\vecpro
\bigg(
\vec{E}
-
\kT
\Dpar{}{\vec{p}}
\bigg)
\Bigg]
\Bigg\}
P
\;.
\end{equation}
Then, introducing spherical coordinates $(\vartheta,\varphi)$ and
assuming the Hamiltonian to be axially symmetric, the equation above
reduces to
\begin{equation}
\label{debye:II}
\frac{1}{\lambda}
\frac{\partial P}{\partial t}
=
\frac{1}{\sin\vartheta}
\frac{\partial {}}{\partial\vartheta}
\left[
\sin\vartheta
\left(
\frac{\partial\Hs}{\partial\vartheta}P
+
\kT
\frac{\partial\W}{\partial\vartheta}
\right)
\right]
\;.
\end{equation}
This equation corresponds to Eq.\ (\ref{debye}) by identifying
$1/\lambda\to\zeta$ and $\Hs=-pE\cos\vartheta$, which is the
Hamiltonian of the dipole in an external field.

Let us solve Eq.\ (\ref{debye}) for a time dependent field $E(t)$ with
$E(t<0)=\Delta E$, while $E(t>0)=0$.
That is, in the distant past the system was equilibrated in the
presence of a small field $\Delta E$ which at $t=0$ is removed, and we
seek for the time evolution of $P(\vartheta,t)$.
For $t<0$ it is easily seen that $P_{0}=N\exp(p\Delta
E\cos\vartheta/\kT)$ is the stationary solution of Eq.\
(\ref{debye}), since $\partial
P_{0}/\partial\vartheta=-(p\Delta E\sin\vartheta/\kT)P_{0}$.
Then, introducing the notation $\alpha=p\Delta E/\kT$, and using that
$\alpha\ll1$, we have
\begin{equation}
\label{P0}
P_{0}(\vartheta)
\simeq
N
(1+\alpha\cos\vartheta)
\;,
\quad
(t<0)
\;.
\end{equation}
For $t>0$ the field is removed and for the solution we use the ansatz
\begin{equation}
\label{P1}
P(\vartheta,t)
=
N
\left[
1+\alpha g(t)\cos\vartheta
\right]
\;,
\quad
(t>0)
\;,
\end{equation}
with $g(t)$ a function to be determined by inserting this $P$ in the
Fokker--Planck equation (\ref{debye}) with $E(t>0)=0$:
\[
\zeta
N
\alpha
\frac{\D g}{\D t}
\cos\vartheta
=
-\kT
N
\alpha\,g
\frac{1}{\sin\vartheta}
\frac{\partial }{\partial\vartheta}
\left(
\sin^{2}\!\vartheta
\right)
\;.
\]
Therefore, defining the {\em Debye relaxation time}
\begin{equation}
\label{tau:debye}
\tD
=
\zeta/2\kT
\;,
\end{equation}
we have
\vspace{-1.ex}
\begin{equation}
\label{debye:g}
\D g/\D t
=
-g/\tD
\qquad
\Longrightarrow
\qquad
g(t)
=
\overbrace{g(0)}^{1}
e^{-t/\tD}
\;.
\end{equation}
The initial condition $g(0)=1$, comes from the matching of the
distributions (\ref{P0}) and (\ref{P1}) at $t=0$.
Since the normalisation constant follows from
$1=\int_{0}^{\pi}\D\vartheta\sin\vartheta P_{0}=2N$, we finally have
\begin{equation}
\label{P1:final}
\boxequation{
P(\vartheta,t)
=
\frac{1}{2}
\left[
1+
\frac{p\Delta E}{\kT}
e^{-t/\tD}
\,
\cos\vartheta
\right]
\;.
}
\end{equation}
It is easy to compute now the average dipole moment along the field
direction $\llangle
p\cos\vartheta\rrangle=\int_{0}^{\pi}\D\vartheta\sin\vartheta
P(\vartheta,t)p\cos\vartheta$, by the change of variables $z=\cos\vartheta$
\begin{equation}
\label{response:t:debye}
\llangle
p\cos\vartheta
\rrangle
=
\frac{p}{2}
\int_{-1}^{1}\D z
\left[
z+
\frac{p\Delta E}{\kT}
e^{-t/\tD}
\,
z^{2}
\right]
\quad
\Rightarrow
\quad
\boxequation{
\llangle
p\cos\vartheta
\rrangle
=
\frac{p^{2}\Delta E}{3\kT}
e^{-t/\tD}
\;.
}
\end{equation}
This result goes from the {\em Curie law\/} for the linear response of a
paramagnet $p^{2}\Delta E/3\kT$ in the initial equilibrium regime, to
zero at long times, corresponding to the final equilibrium state in the
absence of field.
The solution of the Fokker--Planck equation (\ref{P1:final}) provides
also a complete description the intermediate non-equilibrium regime.

\mynewpage

\section[Linear response theory, dynamical susceptibilities,\\
and relaxation times (Kramers' theory)]
{Linear response theory, dynamical
susceptibilities, and relaxation times (Kramers' theory)}
\markboth{Introduction to the theory of stochastic processes}
{Linear response theory, dynamical susceptibilities, and relaxation
times}

In this section we shall consider some general results that can be
obtained for the response of a very general class of dynamical systems
(and in particular those having a ``Fokker--Planck dynamics''),
results which hold when the external perturbation is weak enough.
In the context of this {\em linear response theory}, due to Kubo, the
definition of dynamical {\em response functions\/} appears naturally
(in the time and frequency domain), and sometimes, associated with
them, quantities characterising the lifetime of certain states---the
{relaxation times\/}.

\renewcommand{\fp}{}


\subsection{Linear response theory}
\label{LRT}

Let us consider a system governed by an evolution equation of the type
\begin{equation}
\label{fokkerplanck:generic}
\boxequation{
\partial_{t}P
=
\LFP
P
\;,
}
\end{equation}
where $\LFP$ is a linear operator (linear in its action on $P$), and
$P$ characterises the system.
The natural example in this context is the Fokker--Planck operator
\begin{equation}
\label{fokkerplanck:operator}
\LFP
=
-\partial_{y}
\left(
a^{(1)}\,\cdot
\right)
+
\half
\partial_{y}^{2}
\left(
a^{(2)}\,\cdot
\right)
\end{equation}
Well, let us apply an external perturbation, and separate the part of
the \FP\ operator accounting for the coupling with the perturbation,
and the unperturbed part, which we assume to have a stationary
solution $P_{0}$:
\[
\LFP
=
\LFPo
+
\LFPext(t)
\;,
\qquad
\LFPo
P_{0}
=
0
\;.
\]
These conditions are quite general.
For instance $\LFP$ could be the Liouville operator of a mechanical
system, $P_{0}$ the equilibrium Boltzmann distribution, and $\LFPext$
the part corresponding to the external potential.
In the Fokker--Planck case $\LFPext$ could be $V_{\rm
ext}'\partial_{v}$ in the Klein--Kramers equation or
$\gamma^{-1}\partial_{x}(V_{\rm ext}'\cdot)$ in the Smoluchowski
equation.
However, the external perturbation does not need to be a force or a field.
For instance, external modulations of the system parameters, like the
bath temperature $\Delta T(t)$ are also possible; then in the
Smoluchowski equation we will have $\LFPext\propto\Delta
T\partial_{x}^{2}$.

If the perturbation is weak enough, we can write the deviation from
the stationary state as $P=P_{0}+p$, and the evolution equation would
lead to first order to
\[
\underbrace{
\partial_{t}P_{0}
}_{0}
+
\partial_{t}p
=
\left[
\LFPo
+
\LFPext(t)
\right]
\left(
P_{0}
+
p
\right)
\simeq
\underbrace{
\LFPo
P_{0}
}_{0}
+
\LFPo
p
+
\LFPext(t)
P_{0}
\;,
\]
(we have disregarded $\LFPext\,p$).
The resulting equation can be solved formally%
\footnote{
Indeed, taking the $t$ derivative of the presumed solution, we have
\[
\partial_{t}
p
=
\underbrace{
\LFPext(t)
P_{0}
}_{\mbox{\scriptsize integrand at $\tp=t$}}
+
\underbrace{
\LFPo
\int_{-\infty}^{t}\!\D \tp\,
e^{(t-\tp)\LFPo}
\LFPext(\tp)
P_{0}
}_{\LFPo p}
=
\LFPext(t)
P_{0}
+
\LFPo p
\;
\qed
\]
}
\begin{equation}
\label{solution:p}
\partial_{t}p
=
\LFPo
p
+
\LFPext(t)
P_{0}
\quad
\rightarrow
\quad
\boxequation{
p
=
\int_{-\infty}^{t}\!\D \tp\,
e^{(t-\tp)\LFPo}
\LFPext(\tp)
P_{0}
\;.
}
\end{equation}
This equation gives the formal solution to the problem of time
evolution in the presence of a weak time-dependent perturbation.
{\bf ToDo, warn on the order of the operators}


\subsection{Response functions}


\subsubsection{Time domain}

Let us consider any function $c(y)$ of the variables of the system
$y$, and compute the variation of its average with respect to the
unperturbed state
\begin{equation}
\Delta C(t)
\equiv
\llangle c\rrangle(t)
-
\llangle c\rrangle_{0}
\;.
\end{equation}
To this end we extract the time dependent part of
$\LFPext(y,t)=\LFPext(y)F(t)$ (factorisation is the common case) and
use the solution (\ref{solution:p})
\begin{eqnarray*}
\Delta C(t)
&=&
\int\!\D y\,
c(y)P(y,t)
-
\int\!\D y\,
c(y)P_{0}(y)
\\
&=&
\int\!\D y\,
c(y)\,p(y,t)
\\
&=&
\int_{-\infty}^{t}\!\D \tp\,
\Big[
\int\!\D y\,
c
\,
e^{(t-\tp)\LFPo}
\LFPext
P_{0}
\Big]
F(\tp)
\;.
\end{eqnarray*}
Then, introducing the response function for the quantity $c$
\begin{equation}
\label{response-function}
R_{c}(t)
=
\left\{
\begin{array}{cl}
\displaystyle
\int\!\D y\,
c(y)
\,
e^{t\LFPo}
\LFPext
P_{0}(y)
&
t>0
\\
0
&
t<0
\end{array}
\right.
\;,
\end{equation}
we can write the response $\Delta C(t)$ simply as
\begin{equation}
\label{response:t}
\boxequation{
\Delta C(t)
=
\int_{-\infty}^{\infty}\!\D \tp\,
R_{c}(t-\tp)
F(\tp)
\;.
}
\end{equation}

The following linear response functions are used [$\Theta(t)$ is the
step function $\Theta(t<0)=0$ and $\Theta(t>0)=1$]:
\begin{equation}
F(t)
=
\left\{
\begin{array}{cc}
\delta(t)
&
\mbox{pulse response function $\Delta C_{\rm p}(t)$}
\\
\Theta(t)
&
\mbox{excitation function $\Delta C_{\rm e}(t)$}
\\
\Theta(-t)
&
\mbox{relaxation function $\Delta C_{\rm r}(t)$}
\end{array}
\right.
\end{equation}
Let us consider the last one, also called after-effect function, which
corresponds to switch a constant excitation off at $t=0$
\[
\Delta C_{\rm r}(t)
=
\int_{-\infty}^{\infty}\!\D \tp\,
R_{c}(t-\tp)
\Theta(-\tp)
=
\int_{-\infty}^{0}\!\D \tp\,
R_{c}(
\underbrace{t-\tp}_{\tp'}
)
=
\int_{t}^{\infty}\!\D \tp'\,
R_{c}(\tp')
\;.
\]
Thus, we see that $R_{c}(t)$ can be obtained as the derivative of
$\Delta C_{\rm r}(t)$
\begin{eqnarray}
\label{relaxation}
\Delta C_{\rm r}(t)
=
\int_{t}^{\infty}\!\D \tp\,
R_{c}(\tp)
\qquad
\Longrightarrow
\qquad
\boxequation{
R_{c}(t)
=
-
\frac{\D\,}{\D t}
\Delta
C_{\rm r}
\;.
}
\end{eqnarray}


\subsubsection{Frequency domain}

Introducing now the Fourier transforms of $\Delta C(t)$, $F(t)$, and
$R_{c}(t)$
\[
\Delta \tilde{C}(\omega)
=
\int\!\D t\,
e^{-\iu\omega t}
\Delta C(t)
,
\quad
\tilde{F}(\omega)
=
\int\!\D t\,
e^{-\iu\omega t}
F(t)
,
\quad
\chi_{c}(\omega)
=
\int\!\D t\,
e^{-\iu\omega t}
R_{c}(t)
\;,
\]
the convolution in Eq.\ (\ref{response:t}) relating those quantities,
reduces to a simple product
\begin{equation}
\label{response:w}
\boxequation{
\Delta \tilde{C}(\omega)
=
\chi_{c}(\omega)
\tilde{F}(\omega)
\;.
}
\end{equation}
The complex quantity $\chi_{c}(\omega)$ is known as the
susceptibility.
It can be seen that corresponds to the usual definition: if we excite
with a perturbation $F(t)=e^{\iu\omega t}$, the corresponding response
function $\Delta C(t)$ oscillates in the stationary state with
$e^{\iu\omega t}$ with proportionality coefficient $\chi_{c}(\omega)$
\[
\Delta C(t)
=
\int_{-\infty}^{\infty}\!\D \tp\,
R_{c}(t-\tp)
e^{\iu\omega \tp}
=
\underbrace{
\bigg[
\int_{-\infty}^{\infty}\!\D \tp\,
R_{c}(t-\tp)
e^{-\iu\omega (t-\tp)}
\bigg]
}_{\chi_{c}(\omega)}
e^{\iu\omega t}
\;.
\]

To conclude, we shall demonstrate a famous linear response relation
between the dynamic susceptibility and the relaxation (after-effect)
function.
From the definition of $\chi_{c}(\omega)$, using that $R_{c}(t<0)=0$
and its relation with the relaxation function, we have
\vspace*{-1.ex}
\begin{eqnarray*}
\chi_{c}(\omega)
&\stackrel{\rm def.}{=}&
\int_{-\infty}^{\infty}\!\D t\,
e^{-\iu\omega t}
R_{c}(t)
=
\int_{0}^{\infty}\!\D t\,
e^{-\iu\omega t}
\overbrace{
R_{c}(t)
}^{
-
\D(\Delta C_{\rm r})/\D t
}
\\
&=&
-\Delta C_{\rm r}(t)
e^{-\iu\omega t}
\bigg|_{0}^{\infty}
+
\int_{0}^{\infty}\!\D t\,
\Delta C_{\rm r}(t)
\frac{\D\,}{\D t}
e^{-\iu\omega t}
\\
&=&
\Delta C_{\rm r}(0)
-\iu\omega
\int_{0}^{\infty}\!\D t\,
\Delta C_{\rm r}(t)
e^{-\iu\omega t}
\;.
\end{eqnarray*}
Then, extracting the static (thermal-equilibrium) susceptibility
$\chi_{c}^{\eq}=\chi_{c}(0)=\Delta C_{\rm r}(0)$, we have
\begin{equation}
\label{LRT:chi}
\boxequation{
\chi_{c}(\omega)
=
\chi_{c}^{\eq}
\left[
1
-\iu\omega
\int_{0}^{\infty}\!\D t\,
\frac{\Delta C_{\rm r}(t)}{\Delta C_{\rm r}(0)}
e^{-\iu\omega t}
\right]
\;,
}
\end{equation}
which gives the dynamical susceptibility in terms of the normalised
relaxation function $\Delta C_{\rm r}(t)/\Delta C_{\rm r}(0)$.


\paragraph*{Example: Debye relaxation.}

\begin{figure}
\centerline{\epsfig{figure=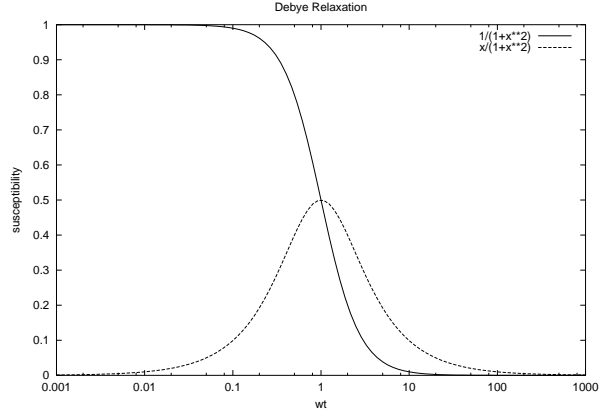,width=8.cm}}
\caption{
Real part (full line) and imaginary part (dashed line) of
the Debye formula (\ref{chi:debye}) for the dynamic susceptibility
(normalised).
}
\label{fig:debye}
\end{figure}

We can immediately apply these linear response results to the example
of the dynamics of the electrical dipole [Eq.\ (\ref{debye})].
There, we calculated the time evolution of the average of the field
projection of the dipole, Eq.\ (\ref{response:t:debye}).
Thus, in this case
\[
c
=
p\cos\vartheta
\;,
\qquad
\Delta P_{\rm r}(t)
=
\frac{p^{2}}{3\kT}
e^{-t/\tD}
\;,
\]
and hence $\chi_{p}^{\eq}=(p^{2}/3\kT)$ (Curie law).
Therefore, we have $\Delta P_{\rm r}(t)/\Delta P_{\rm
r}(0)=e^{-t/\tD}$, so that 
\vspace*{-2.ex}
\begin{equation}
\label{chi:debye}
\chi_{p}(\omega)
=
\frac{p^{2}}{3\kT}
\bigg[
1
-\iu\omega
\overbrace{
\int_{0}^{\infty}\!\D t\,
e^{-(\iu\omega+1/\tD)t}
}^{
1/(\iu\omega+1/\tD)
}
\bigg]
\quad
\Rightarrow
\quad
\boxequation{
\chi_{p}(\omega)
=
\frac{p^{2}}{3\kT}
\frac{1}{1+\iu\omega\tD}
\;.
}
\end{equation}
which is the famous formula displaying ``Debye'' relaxation (see
Fig. \ref{fig:debye}).


\subsection{Relaxation times}

When we considered various examples of Fokker--Planck equations, we
obtained the solution for the Smoluchowski equation of an harmonic
oscillator
\[
P(x,t)
=
\sqrt{\frac{m\omega_{0}^{2}}{2\pi\kT(1-e^{-2t/\tau})}}
\exp
\lrs{
-\frac
{m\omega_{0}^{2}(x-x_{0}e^{-t/\tau})^{2}}
{2\kT(1-e^{-2t/\tau})}
}
\;,
\qquad
\tau=\gamma/\omega_{0}^{2}
\;.
\]
In this equation we see that the time scale for the relaxation to the
equilibrium state $P_{0}\propto\exp(-\half m\omega_{0}^{2}x^{2}/\kT)$,
is given by $\tau=\gamma/\omega_{0}^{2}$.
This quantity is the {\em relaxation time}.
In this problem it depends on the system parameters $\gamma$ and
$\omega_{0}$, but it is independent of the temperature.

Now, in the example of the dielectric dipole, a natural relaxation
time has also appeared $\tD=\zeta/2\kT$ [Eq.\ (\ref{tau:debye})].
In this problem the relaxation time depends on $T$, which is the
common case, however, the dependence is not very strong.

It is very common to find expressions for different relaxation times
that depend exponentially on $T$ (Arrhenius law), which is the generic
behaviour when to establish the equilibrium potential barriers need to
be overcome.
This was not the case of the previous examples, and such dependence
was absent.
We shall solve now a simple problem with potential barriers to see how
the exponential dependence arises (the theoretical study of this
problems was initiated by Kramers in 1940 to study the relaxation rate
of chemical reactions).
{\bf ToDo, warn on low temperature assumption}

\begin{figure}
\centerline{\epsfig{figure=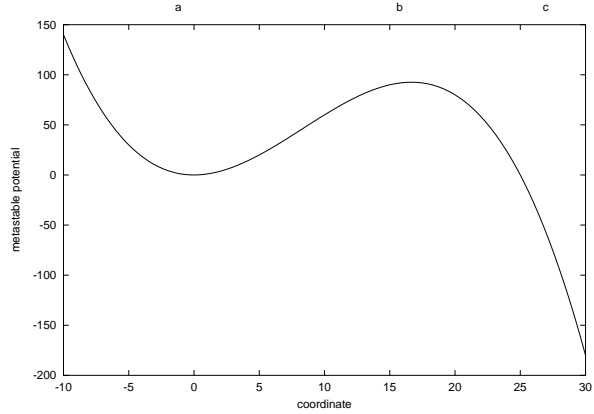,width=8.cm}}
\caption{
Sketch of a metastable potential for the Kramers calculation of the
relaxation time. The point $a$ is at the potential minimum, $b$ at the
maximum, and $c$ is a point at the right-side of the barrier.
}
\label{kramers}
\end{figure}

To simplify the calculation let us consider an overdamped particle,
described by Smoluchowski equation
\begin{equation}
\label{smoluchowski:J}
\frac{\partial P}{\partial t}
=
\frac{1}{\gamma}
\left(
\frac{\partial }{\partial x}
V'
+
\frac{\kT}{m}
\frac{\partial^{2}}{\partial x^{2}}
\right)
P
=
-\frac{\partial J}{\partial x}
\;.
\end{equation}
where the last equality defines the current of probability $J$, and
the metastable potential is depicted in Fig.\ \ref{kramers}.
At very low temperatures the probability of escape from the metastable
minimum is very low (zero at $T=0$; deterministic system).
Therefore the flux of particles over the barrier is very slow, and we
can solve the problem as if it were stationary.
Then the expression for $J$,
\begin{equation}
\label{J:ode}
-\gamma J
=
V'P
+
\frac{\kT}{m}
\frac{\partial P}{\partial x}
\;,
\end{equation}
is assumed to be independent of $x$ and the differential equation for $P$
can be integrated ($U=mV$, $D=\kT/m\gamma$)
\begin{equation}
\label{P:J}
P
=
e^{-U/\kT}
\bigg[
C_{1}
-
\frac{J}{D}
\int_{c}^{x}\!\D x'\,
e^{U(x')/\kT}
\bigg]
\;,
\end{equation}
were $c$ is an arbitrary point.
The integration constant is $C_{1}=P(c)e^{U(c)/\kT}$.
If we choose $c$ well outside the barrier region $c\to\infty$, we have
$P(c)\simeq0$, and we find for the current $J$
\begin{equation}
\label{J:P}
J
=
-D
\frac
{P(x)e^{U(x)/\kT}}
{\displaystyle\int_{c}^{x}\!\D x'\,
e^{U(x')/\kT}}
\;.
\end{equation}
Since $\partial J/\partial x\simeq0$ we can choose $x$ at will; we
set $x=a$ (the metastable minimum), so the integral in the denominator
covers the entire maximum.
The main contribution to that integral comes from a small region about
the maximum $x=b$, so we expand there $U(x)=U_{b}-\half
m\wb^{2}(x-b)^{2}$, being $m\wb^{2}=U''(b)$.
The integration limits can be shifted to $\pm\infty$, so the resulting
Gaussian integral leads
\begin{equation}
\label{P:J:2}
J
=
D
\frac
{P(a)e^{U_{a}/\kT}}
{e^{U_{b}/\kT}\sqrt{2\pi\kT/m\wb^{2}}}
=
D
P(a)
\sqrt{\frac{m\wb^{2}}{2\pi\kT}}
e^{-(U_{b}-U_{a})/\kT}
\;.
\end{equation}
To compute $P(a)$ we use the following argument.
The fraction of particles close to the potential minimum $N_{a}$ can be
obtained integrating $P(x)$ in an interval around $a$, with the
distribution approximated as $P=C_{2}e^{-U(x)/\kT}$ with
$U(x)=U_{a}+\half m\wa^{2}(x-a)^{2}$, where
$m\wa^{2}=U''(a)$.
Then for the particles in the well we have
$N_{a}=C_{2}e^{-U_{a}/\kT}\sqrt{2\pi\kT/m\wa^{2}}$, so that
$P(a)/N_{a}=\sqrt{m\wa^{2}/2\pi\kT}$.

The relaxation rate is defined as $J/N_{a}$ (so the number of
particles in the well $N_{a}$, times the escape rate, gives the flux
$J$ leaving the well).
Then, introducing the above expression for $P(a)/N_{a}$ into Eq.\
(\ref{P:J:2}) divided by $N_{a}$, and using $D/\kT=1/m\gamma$ we finally
have
\begin{equation}
\label{tau:kramers:od}
\boxequation{
\frac{1}{\tau}
=
\frac{\wa\wb}{2\pi\gamma}
e^{-(U_{b}-U_{a})/\kT}
\;.
}
\end{equation}
This formula has the typical exponential dependence on the barrier
height over the temperature.
Although the calculation in other cases (intermediate to weak damping)
is much more elaborated, this exponential dependence always appears.

\enlargethispage*{3.ex}

\mynewpage

\section{Methods for solving Langevin and \FP\ equations
(mostly numerical)}
\label{numerical:methods}
\markboth{Introduction to the theory of stochastic processes}{OJO}

In general the Fokker--Planck or Langevin equations cannot be solved
analytically.
In some cases one can use approximate methods, in others numerical
methods are preferable.
Here we shall discuss some of these methods.


\subsection{Solving Langevin equations by numerical integration}
\label{schemes}

We shall start with methods to integrate the Langevin equations
numerically.
These are the counterpart of the known methods for the
deterministic differential equations.


\subsubsection{The Euler scheme}

In order to integrate the system of Langevin equations
\[
\frac{\D y_{i}}{\D t}
=
\drift_{i}(\multi{y},t)
+
\sum_{k}\diff_{ik}(\multi{y},t)\Lan_{k}(t)
\;,
\qquad
\av{\Lan_{k}(t)}
=
0
\;,
\quad
\av{\Lan_{k}(t)\Lan_{\ell}(\tp)}
=
2D\delta_{k\ell}\delta(t-\tp)
\]
starting at $t=t_{0}$ with the values $\multi{y}_{0}$, to the time
$t_{0}+T$, one first divides the time interval $[t_{0},t_{0}+T]$ into
$N_{\text{s}}$ time steps of length $\Delt$, i.e.,
$t_{n}=t_{0}+n\Delt$.
The stochastic variables at a later time $\multi{y}(t_{n+1})$,
are calculated in terms of $\multi{y}(t_{n})$ according to
\begin{equation}
\label{euler:schemes}
\boxequation{
y_{i}(t_{n+1})=y_{i}(t_{n})+a_{i}^{(1)}[\multi{y}(t_{n}),t_{n}]\Delt
+\sum_{k}\diff_{ik}[\multi{y}(t_{n}),t_{n}]\DelW_{kn}
\;,
            }
\end{equation}
where
$a_{i}^{(1)}=\drift_{i}+D\sum_{jk}\,\diff_{jk}\partial_{j}\diff_{ik}$
is the first jump moment, $\DelW_{kn}$, $k=1,\ldots,N_{\text{L}}$ (the
number of Langevin sources), $n=1,\ldots,N_{\text{s}}$, are
independent
%
%
%
Gaussian numbers with zero mean and variance $2D\Delt$, i.e.,
\begin{equation}
\label{Wcorr}
\av{\DelW_{kn}}=0
\;,
\qquad
\av{\DelW_{kn}\DelW_{k'n'}}=(2D\Delt)\delta_{kk'}\delta_{nn'}
\;.
\end{equation}
The recursive algorithm (\ref{euler:schemes}) is called the Euler
scheme, in analogy with the Euler method to integrate deterministic
differential equations.
By construction, for $\Delt\to0$, the above recursive scheme, leads to the
correct \KM\ coefficients.%
\footnote{
Let us prove this in the simple one-variable case.
Then
\begin{equation}
\label{euler:schemes:1D}
y(t+\Delt)
=
y(t)
+a^{(1)}(y,t)
\Delt
+
\diff(y,t)
\DelW
\;.
\end{equation}
To obtain the \KM\ coefficients, we average this equation for fixed
initial values $y(t)$ (conditional average).
To do so, one can use $\av{\DelW}=0$ and
$\av{\DelW^{2}}=2D\Delt$, to get
\[
\llangle
\diff\DelW
\rrangle
=
0
\;,
\qquad
\blangle
a^{(1)}\Delt
\diff\DelW
\brangle
=
0
\;,
\qquad
\llangle
\diff\DelW
\diff\DelW
\rrangle
=
2D\diff^{2}
\Delt
\;.
\]
Therefore, one obtains
\begin{eqnarray*}
\llangle
y(t+\Delt)
-
y(t)
\rrangle
&=&
a^{(1)}\Delt
\\
\blangle
\left[
y(t+\Delt)
-
y(t)
\right]^{2}
\brangle
&=&
(a^{(1)})^{2}\Delt^{2}
+
2
a^{(1)}\Delt\diff
\llangle
\DelW
\rrangle
+
\diff^{2}
\llangle
\DelW^{2}
\rrangle
=
2D
\diff^{2}
\Delt
+{\cal O}[(\Delt)^{2}]
\;,
\end{eqnarray*}
which lead to the \KM\ coefficients (\ref{a_nu:langevin}) via Eq.\
(\ref{jumpmoments:derivative}).\qed
}


\subsubsection{Stochastic Heun scheme}

This is a higher-order scheme for the numerical integration of the
Langevin equations given by (a sort of Runge--Kutta scheme)
\begin{eqnarray}
\label{heun:scheme}
\boxeqnarray{30.em}{
y_{i}(t+\Delt)
=
y_{i}(t)
&+&
\half 
\lrc{\drift_{i}[\tilde{\multi{y}},t+\Delt]
+\drift_{i}[\multi{y}(t),t]}\Delt
\\
&+&
\half 
\sum_{k}
\lrc{\diff_{ik}[\tilde{\multi{y}},t+\Delt]
+\diff_{ik}[\multi{y}(t),t]}\DelW_{k}
\;,
 			}
\end{eqnarray}
with Euler-type supporting values,
\begin{equation}
\label{euler:support}
\tilde{y}_{i}=y_{i}(t)+\drift_{i}[\multi{y}(t),t]\Delt
+\sum_{k}\diff_{ik}[\multi{y}(t),t]\DelW_{k}
\;.
\end{equation}
Note that if one uses this support value as the numerical integration
algorithm [by identifying $y_{i}(t+\Delt)=\tilde{y}_{i}$], the result
does not agree with the ordinary Euler scheme if
$\partial\diff_{ik}/\partial y_{j}\neq0$ (or equivalently if
$a_{i}^{(1)}\neq\drift_{i}$).
%
%
%



%
The Euler scheme only requires the evaluation of $\drift_{i}$ and
$\diff_{ik}$ at one point per time step, while the Heun scheme
requires two, increasing the computational effort.
Nevertheless, the Heun scheme substitutes the derivatives of
$\diff_{ik}$ by the evaluation of $\diff_{ik}$ at different points.
Besides, it treats the deterministic part of the differential
equations with a second-order accuracy in $\Delt$, being numerically
more stable.
Thus, the computational advantage of the Euler scheme, may disappear
if it needs to be implemented with a smaller integration step
($\Delt$) to avoid numerical instabilities.


\subsubsection{Gaussian random numbers}

The {\em Gaussian\/} random numbers required to simulate the variables
$\DelW$, can be constructed from {\em uniformly\/} distributed ones by
means of the Box--Muller algorithm (see, e.g., Ref.\
\cite[p.~280]{recipes}).
This method is based on the following property: if $r_{1}$ and $r_{2}$ are
random numbers uniformly distributed in the interval $(0,1)$ (as those
pseudo-random ones provided by a computer), the transformation
\begin{equation}
\label{boxmuller}
w_{1}=\sqrt{-2\ln(r_{1})}\cos(2\pi r_{2})
\;,
\quad
w_{2}=\sqrt{-2\ln(r_{1})}\sin(2\pi r_{2})
\;,
\end{equation}
outputs $w_{1}$ and $w_{2}$, which are Gaussian-distributed independent
random numbers of zero mean and variance unity.
Then, if one needs Gaussian numbers with variance $\sigma^{2}$, these
are immediately obtained by multiplying the above $w_{i}$ by $\sigma$
(e.g., $\sigma=\sqrt{2D\Delt}$ in the Langevin equations).

%


\subsubsection{Example I: Brownian particle}

The Langevin equations for a particle subjected to fluctuations and
dissipation evolving in a potential $V(x)$ are
\begin{equation}
\label{langevin:hamilton}
\left\{
\begin{array}{rcl}
\D x/\D t
&=&
v
\\
\D v/\D t
&=&
-V'
-\gamma\,v
+
\Lan(t)
\;,
\qquad
\llangle
\Lan(t)\Lan(\tp)
\rrangle
=
2D
\delta(t-\tp)
\end{array}
\right.
\end{equation}
with $D=\gamma(\kT/m)$.
For the potential we consider that of a constant force field $F$ plus
a periodic substrate potential of the form
\begin{equation}
\label{Vratchet}
V(x)
=
-d
\left[
\sin x
+
(\rat/2)
\sin 2x
\right]
-Fx
\;.
\end{equation}

For the familiar case $\rat=0$, we have the cosine potential and
the Langevin equations describe a variety of systems:

(i) Non-linear pendulum:
\begin{equation}
\label{pendulum}
\ddot{\phi}
+
\gamma\,\dot{\phi}
+
(g/\ell)
\sin\phi
=
N
+
\Lan(t)
\;.
\end{equation}
In this case we have $x=\phi$ (the angle of the pendulum with respect
to the vertical direction), $d=g/\ell$ (gravity over length of the
pendulum), $F=N$ (external torque), and $D=\gamma(\kT/m\ell)$.

(ii) Josephson junction (RCSJ model):
\begin{figure}[!b]
\centerline{\epsfig{figure=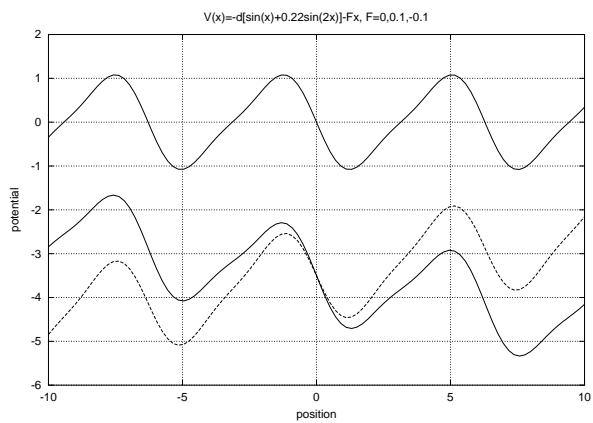,width=8.cm}}
\caption{
Ratchet potential (\ref{Vratchet}), with asymmetry parameter
$\rat=0.44$, at zero force and $F/d=\pm0.15$ (displaced for clarity).
}
\label{fig:Vratchet}
\end{figure}
%
\begin{equation}
\label{JJ}
C
\ddot{\varphi}
+
\frac{1}{R}
\dot{\varphi}
+
\frac{2e}{\hbar}
I_{\rm c}
\sin\varphi
=
\frac{2e}{\hbar}
\left[
I
+
\Lan(t)
\right]
\;.
\end{equation}
Here $x=\varphi$ (the phase across the junction), $\gamma=1/RC$,
$d=2eI_{\rm c}/\hbar C$ (essentially the critical current),
$F=2eI/\hbar C$ (external current), and $D=\kT/R$.

(iii) Others: superionic conductors, phase-locked loops, etc.

When $\rat\neq0$ in Eq.\ (\ref{Vratchet}), $V$ is called a ratchet
potential, where it is more difficult to surmount the barrier to the
left than to the right (like a saw tooth; see Fig.\
\ref{fig:Vratchet}).
Ratchet potentials have been used to model directional motion in
diverse systems, one of them the molecular motors in the cell.

\begin{figure}[!t]
\centerline{\epsfig{figure=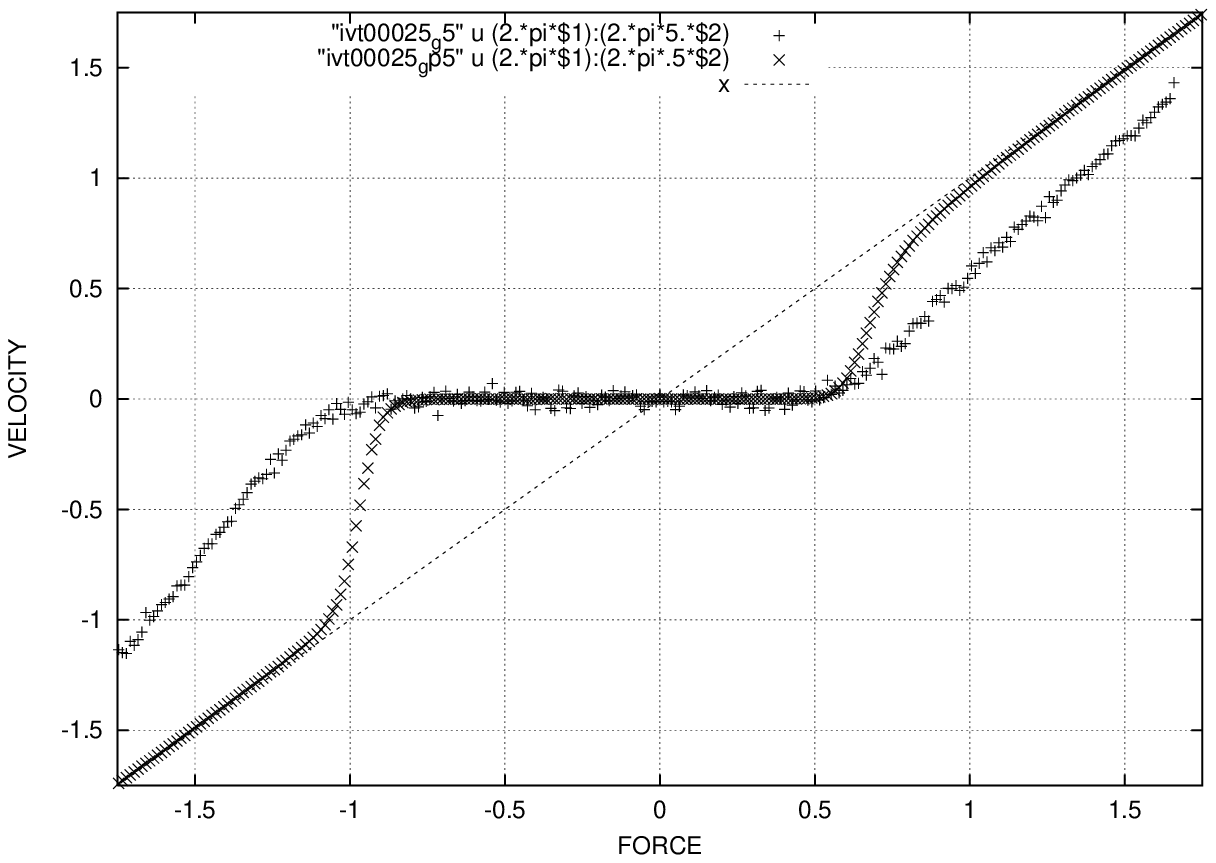,width=8.cm}}
\caption{
$\gamma\llangle v\rrangle$ vs. $F$ for a particle in a ratchet
potential at $T/d=0.1$ with $\gamma=0.5$ and $5$.
The dotted line shows the limit curve $\gamma\llangle v\rrangle=F$.
Results obtained by numerical integration of the Langevin equation
with a Runge--Kutta-like 4th order algorithm for a system of 1000
independent particles.
}
\label{depinning}
\medskip
\centerline{\epsfig{figure=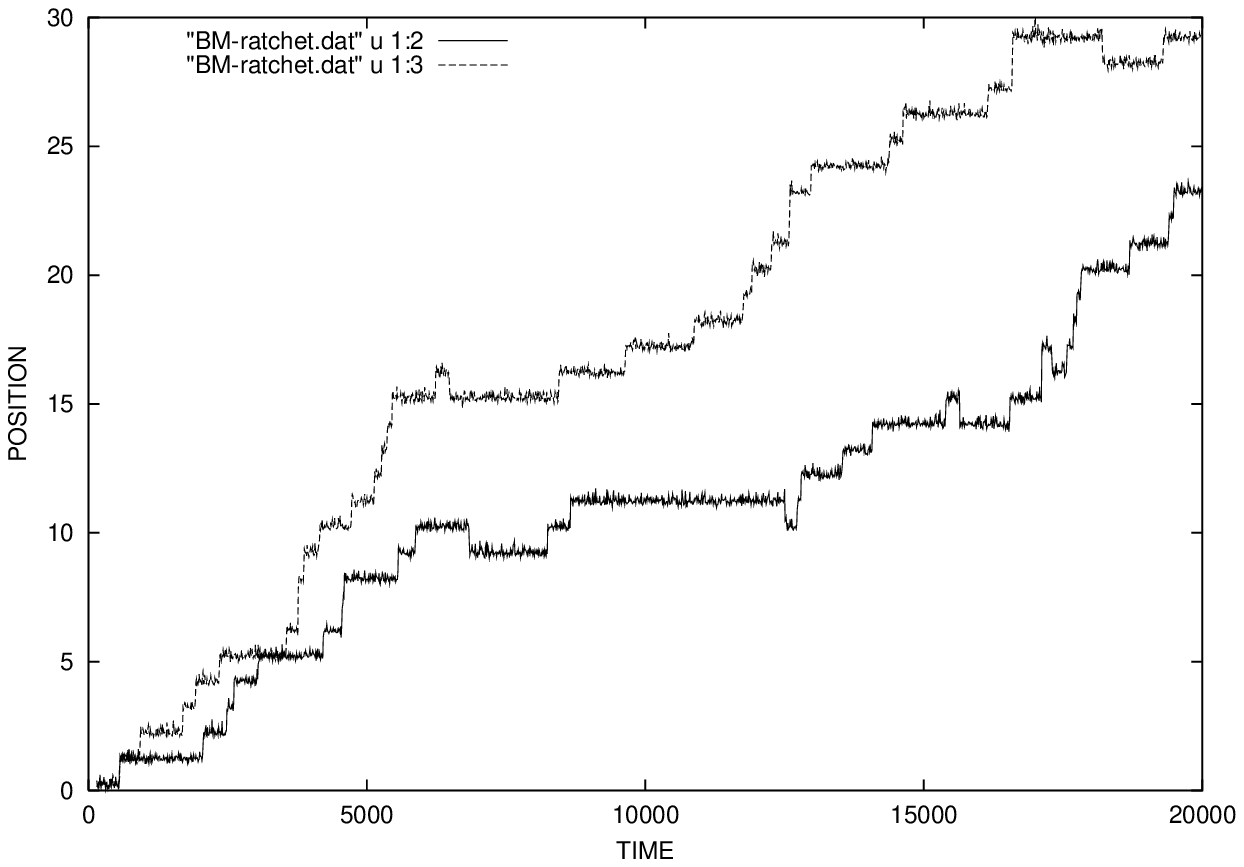,width=8.cm}}
\caption{
Time evolution of the position of two independent Brownian particles
in a ratchet potential corresponding to Fig.\ \ref{depinning}.
The particles are in a weak force $F/d\sim0.03$ so their random-walk is
biased in the force direction.
The other parameters are $T/d\sim0.1$, $\gamma=0.5$.
%
}
\label{BM-ratchet}
\end{figure}

If Fig.\ \ref{depinning} we show the average velocity vs.\ force for a
system of independent particles in a ratchet potential, obtained by
numerical integration of the Langevin equation
(\ref{langevin:hamilton}) with a fourth order algorithm.
It is seen that the depinning (transition to a state with non-zero
velocity), occurs at lower forces to the right that to the left, as
can be expected from the form of the potential.
It is also seen that for lower damping, the transition to the running
state is quite sharp, and the curve quickly goes to the limit
velocity curve $\gamma\llangle v\rrangle=F$.
The reason is that for high damping, if the particle has crossed the
barrier, it will not necessarily pass to a running state, but can be
trapped in the next well, while the weakly damped particle has more
chances to travel, at least a number of wells.

The smooth graphs in Fig.\ \ref{depinning} are obtained averaging the
results for 1000 particles.
The individual trajectories of the particles, however, are quite
irregular.
In Fig.\ \ref{BM-ratchet}, the trajectories of two of them are shown.
It is seen that to the overall trend of advancing in the direction of
the force, there are superimposed Brownian fluctuations (biased random
walk), and indeed we see that the particles can be trapped in the
wells for some time and even to return to the previous well.
\begin{figure}[!b]
\centerline{\epsfig{figure=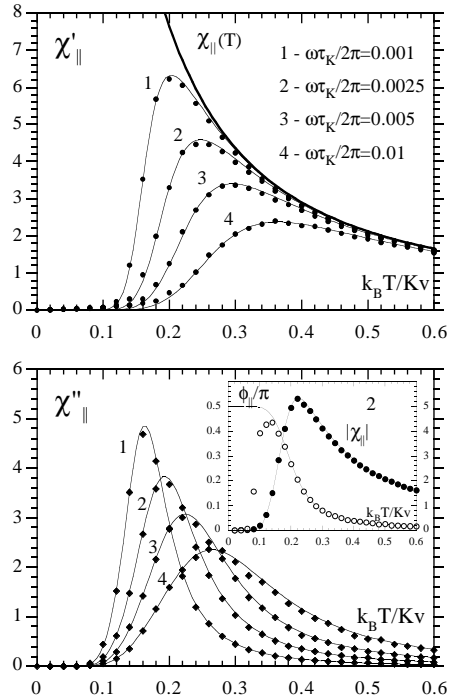,width=8.cm}}
\caption{ Symbols: $\chi(\omega,T)$ vs.\ $T$ obtained by numerical
integration of the stochastic Landau--Lifshitz equation with the Heun
scheme (\ref{heun:scheme}).
Thin lines: simple Debye approximation.
Thick line: thermal-equilibrium susceptibility.
}
\label{chipara:plot}
\end{figure}


\subsubsection{Example II: Brownian spins and dipoles.}

The Langevin equation for a spin subjected to fluctuations and
dissipation is the Landau--Lifshitz equation
\begin{equation}
\label{stolleq:mod:II}
\frac{\D\m}{\D t}
=
\m
\vecpro
\left[\Beff+\bfl(t)\right]
-\lambda
\m
\vecpro
\left(\m\vecpro\Beff\right)
\;,
\quad
\llangle
\Lan_{k}(t)\Lan_{\ell}(\tp)
\rrangle
=
2D
\delta_{k\ell}
\delta(t-\tp)
\;,
\end{equation}
with $\Dcoeff=\lambda\kT/\mm????$ [cf.\ $\Dcoeff=\gamma(\kT/m)$ for
the particle] and $\Beff=-\partial\Hs/\partial\m$ (cf.\ $-\partial
V/\partial x$).
For the Hamiltonian we choose
\begin{equation}
\label{U}
\Hs(\m)
=
-\dU\mz^{2}
-\m
\cdot
\B_{0}
\;,
\qquad
\Beff
=
2\dU\mz\hat{z}
+
\B_{0}
\;,
\end{equation}
with the term of coupling with the external field $\B_{0}$ and a {\em
magnetic anisotropy\/} term (magnetocrystalline, magnetostatic, etc.),
which favours orientation of the spin along $\pm\hat{z}$.
The anisotropy and the field play the role of the substrate potential
and the force for the particle problem.
\begin{figure}[!t]
\centerline{\epsfig{figure=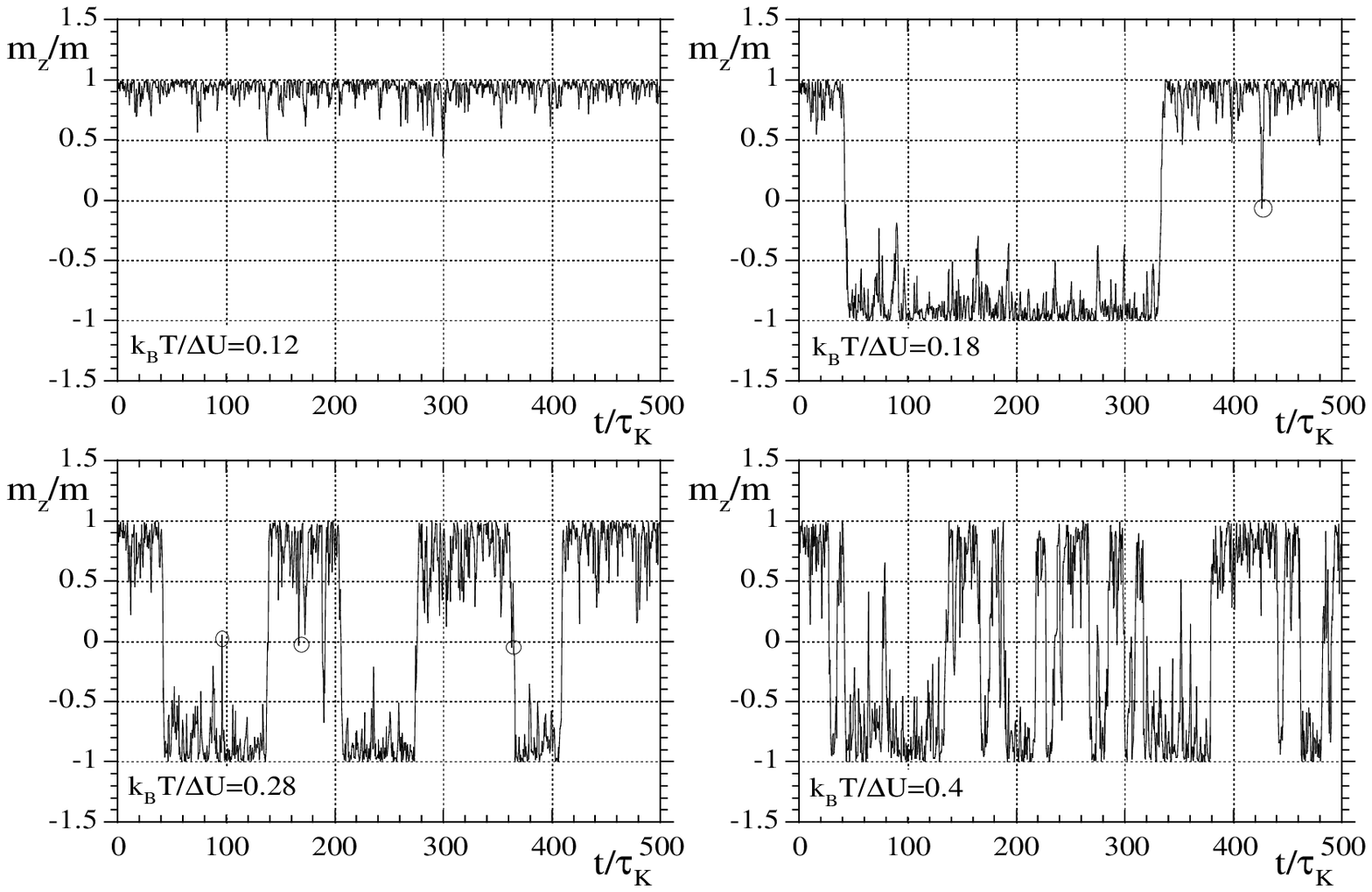,width=13.cm}}
\caption{
Projection of $\m(t)$ onto the anisotropy axis for
$\Hs=-\dU\mz^{2}$ in zero field, for various temperatures.
}
\label{jumps:plot}
\end{figure}

The stochastic Landau--Lifshitz equation describes a variety of
systems
\begin{enumerate}
\item[(i)] Magnetic systems: magnetic nanoparticles and approximately,
magnetic molecular clusters.

\item[(ii)]
Electric systems (taking formally the limit $\lambda\gg1$): permanent
dipole molecules, nematic liquid crystals, and relaxor ferroelectrics.
\end{enumerate}

Figure \ref{chipara:plot} displays the results for the dynamical
susceptibility vs.\ the temperature for an ensemble of 1000 spins with
parallel anisotropy axes for $\lambda=0.1$.
At low temperatures, the relaxation time obeys $\tau\gg2\pi/\omega$,
so that the probability of over-barrier rotations is almost zero.
The spins rotate close to the bottom of the potential wells (see the
panel $\kT/\dU=0.12$ of Fig.\ \ref{jumps:plot}) and the response is
small.
As $T$ is increased the spins can depart from the minima (panel
$\kT/\dU=0.18$) and rotate over the barrier.
However, since the jump mechanism is not efficient enough, the
response lags behind the probing field, and an appreciable imaginary
component of the response arises.
If $T$ is further increased, the very thermal agitation, which up to
this point was responsible for the growth of the response, reaches a
level that simultaneously disturbs the alignment of the magnetic
moments in the field direction.
The response has then a maximum and starts to decrease.
Finally, at still higher temperatures the jumps are so frequent (panel
$\kT/\dU=0.4$) that the spins quickly redistribute according to the
conditions set by the instantaneous field and the response tends to
the thermal equilibrium one.
%

\renewcommand{\fp}{\scriptscriptstyle {\rm FP}}

\subsection{Methods for the \FP\ equation}

There are several methods to solve \FP\ equations not amenable for an
analytical treatment.
Among them we can mention the method of the eigenvalues, which consist
in finding the eigenvalues of the \FP\ operator occurring in the
dynamical equation $\partial_{t}P=\LFP P$
However, the operator $\LFP$ is not in general Hermitian.
If the stationary distribution is known $P_{0}$, and $\LFP$ fulfills
the {\em detailed balance\/} condition, it can be seen that the
transformed operator $\LFPb=P_{0}^{-1/2}\LFP P_{0}^{1/2}$ is
Hermitian, so the problem can be reduced to an ordinary eigenvalue
problem.
\begin{equation}
\LFPb
p_{n}
=
-\lambda_{n}
p_{n}
\;,
\end{equation}
from which we will have
$P(y,t)=\sum_{n}a_{n}p_{n}(y)e^{-\lambda_{n}t}$.
However, along with relying on the knowledge of $P_{0}$ and the
detailed balance condition, the method has the drawback that the
eigenvalue problem may be difficult to solve (as happens for the
Schr\"odinger equation).
Concerning approximate methods, we can mention the {\em small noise
expansion\/} (also for Langevin equations) and the method of {\em
adiabatic elimination of fast variables}, which reduces the
dimensionality of the problem.

In what follows we shall discuss an alternative method, the {\em
continued fraction method\/}, which is a special case of the expansion
into complete sets approach.
The method, although limited in principle to problems with a few
variables, it is {\em exact\/} and in general {\em very efficient},
and, in addition, illustrates more general techniques used in
non-equilibrium statistical mechanics.
We shall discuss the method with two important examples: the
Klein--Kramers equation and the Fokker--Planck equation for the Debye
dipole.


\subsubsection{Solution of the Klein--Kramers equation by
continued--fractions}
\label{CF:KK}

The Klein--Kramers equation (\ref{KK}) can be written in the compact
form as
\begin{equation}
\label{KK:Lrev:Lirr}
\partial_{t}
P
=
\Big(
\underbrace{
\Ls_{\rm rev}
+
\Ls_{\rm irr}
}_{\LFP}
\Big)
P
\;,
\qquad
\left\{
\begin{array}{rcl}
\Ls_{\rm rev}
&=&
-v\,\partial_{x}
+
V'\partial_{v}
=
\pbra
\Hs
\,,
\cdot
\pket_{\rm PB}
\\[1.ex]
\Ls_{\rm irr}
&=&
\gamma
\,
\partial_{v}
\left[
v
+
(\kT/m)
\partial_{v}
\right]
\end{array}
\right.
\end{equation}


\paragraph*{Scaled quantities.}

To simplify the notation, we introduce a thermal rescaling of the
velocity, time, damping, and potential
\begin{equation}
\label{rescaling}
\bar{v}
=
v/\sqrt{\kT/m}
\;,
\quad
\bar{t}
=
t
\times
\sqrt{\kT/m}
\;,
\quad
\bar{\gamma}
=
\gamma/\sqrt{\kT/m}
\;,
\quad
U
=
mV/\kT
\;.
\end{equation}
Well, for the $v$, $t$, and $\gamma$ so defined we shall not keep the
bars and we shall simply write the Klein--Kramers equation as
\begin{equation}
\label{KK:Lrev:Lirr:scaled}
\boxequation{
\partial_{t}
P
=
\left(
\Ls_{\rm rev}
+
\Ls_{\rm irr}
\right)
P
\;,
\qquad
\left\{
\begin{array}{rcl}
\Ls_{\rm rev}
&=&
-v\,\partial_{x}
+
U'\partial_{v}
\\
\Ls_{\rm irr}
&=&
\gamma
\,
\partial_{v}
\left(
v
+
\partial_{v}
\right)
\end{array}
\right.
\;.
}
\end{equation}


\paragraph*{Expansion in an orthonormal basis of $v$.}

We can expand the probability distribution $P(x,v,t)$ in an
orthonormal basis $\psi_{n}(v)$ (to be specified below) as follows
\begin{equation}
\label{P:KK:expansion}
P
=
W
\sum_{n}
c_{n}(x,t)
\psi_{n}(v)
\;,
\end{equation}
where $W=W(x,v)$ is some function we extract for later convenience.
Due to the orthonormality of the $\psi_{n}(v)$, we can write
$c_{n}=\int\!\D v\,\psi_{n}P/W$, and hence
\[
\mbox{OBS: partial derivatives}
\dot{c}_{n}
=
\int\!\D v\,
\psi_{n}
W^{-1}
\underbrace{\partial_{t}P}_{\LFP P}
=
\int\!\D v\,
\psi_{n}
\left(
W^{-1}
\LFP
W
\right)
\underbrace{P/W}_{\sum_{m}c_{m}\psi_{m}}
\;.
\]
Then, we can compactly write
\begin{equation}
\label{cn:Qnm:Lbar}
\boxequation{
\dot{c}_{n}
=
\sum_{m}
\hat{Q}_{n,m}
c_{m}
\;,
\qquad
\hat{Q}_{n,m}
=
\int\!\D v\,
\psi_{n}
\LFPb
\psi_{m}
\quad
\LFPb
=
W^{-1}
\LFP
W
\;.
}
\end{equation}
Then, the equations for the $c_{n}$ are like a system of linear
equations with the matrix elements of $\LFPb$.
However, the $\hat{Q}_{n,m}$ still contain operators over $x$.
Besides, it seems that all the $c_{m}$, contribute to the equation for
$\dot{c}_{n}$.

As for the function $W$, by analogy with the transformation
$P_{0}^{-1/2}\LFP P_{0}^{1/2}$ to simplify the problem, we try it at
least for the $v$ dependent part.
Then, since $P_{0}\propto\exp(-v^{2}/2)$ in our units%
\footnote{
Undoing the thermal rescaling (\ref{rescaling}), we have
$\exp(-v^{2}/2)\to\exp(-mv^{2}/2\kT)$.
} 
we put $W\propto\exp(-v^{2}/4)$.
For the $x$ dependent part, one sets $W\propto\exp(-\epspot U)$,
which for $\epspot=1/2$ corresponds to the $x$-dependent part of
$P_{0}^{1/2}$ (if given by the Boltzmann distribution).
Thus, from these considerations we set
$W\propto\exp[-(v^{2}/4+\epspot U)]$.


\paragraph*{Calculation of $\LFPb=W^{-1}\LFP W$.}

Since $\LFP=\Ls_{\rm rev}+\Ls_{\rm irr}$, we have to calculate
$\Lsb_{\rm rev}=W^{-1}\Ls_{\rm rev}W$
and
$\Lsb_{\rm irr}=W^{-1}\Ls_{\rm irr}W$.

For $\Lsb_{\rm irr}$ we have
\[
\gamma^{-1}
\Lsb_{\rm irr}
f
=
e^{v^{2}/4}
\partial_{v}
\left(
v
+
\partial_{v}
\right)
e^{-v^{2}/4}
f
=
-
\left(
-\partial_{v}^{2}
+
\case{1}{4}
v^{2}
-
\case{1}{2}
\right)
f
\;,
\]
which has the structure of (minus) the Hamilton operator of the
harmonic oscillator in quantum mechanics.
Hence, we introduce creation and annihilation operators $b^{+}$ and
$b$, defined as
\begin{equation}
\label{creation:annihilation}
\left\{
\begin{array}{lcr}
b
&=&
\partial_{v}
+
\half v
\\[0.5ex]
b^{+}
&=&
-
\partial_{v}
+
\half v
\end{array}
\right.
\qquad
\Longrightarrow
\qquad
\big[
b
\,,
b^{+}
\big]
=
1
\;.
\end{equation}
Since the $1/2$ in $\Lsb_{\rm irr}$ cancels the ``zero-point
fluctuations'', we finally have
\begin{equation}
\label{Lbarirr}
\boxequation{
\Lsb_{\rm irr}
=
-\gamma
b^{+}b
\;.
}
\end{equation}
Because of this, it seems natural to choose the $\psi_{n}(v)$ as the
Hermite functions
\begin{equation}
\label{hermite}
\psi_{n}(v)
=
\frac{1}{\sqrt{n!(2\pi)^{1/2}}}
\,
e^{-v^{2}/4}
H_{n}(v)
\;.
\end{equation}

For $\Lsb_{\rm rev}$, since
$\Ls_{\rm rev}
=
-v\,\partial_{x}
+
U'\partial_{v}$,
we need $\overline{\partial_{v}}$ and $\overline{\partial_{x}}$
\begin{eqnarray*}
\overline{\partial_{v}}
f
&=&
e^{v^{2}/4}
\partial_{v}
\left(
e^{-v^{2}/4}
f
\right)
\;,
\qquad
\Rightarrow
\qquad
\overline{\partial_{v}}
=
-b^{+}
\\
\overline{\partial_{x}}
f
&=&
e^{\epspot U}
\partial_{x}
\left(
e^{-\epspot U}
f
\right)
\;,
\qquad
\Rightarrow
\qquad
\overline{\partial_{x}}
=
\partial_{x}
-
\epspot U'
\;.
\end{eqnarray*}
Then, for $\Lsb_{\rm rev}$ we have
\[
\Lsb_{\rm rev}
=
-
\underbrace{\left(b+b^{+}\right)}_{v}
\underbrace{
\left(
\partial_{x}-\epspot U'
\right)
}_{\overline{\partial_{x}}}
\underbrace{-b^{+}}_{\overline{\partial_{v}}}
U'
=
-b
\left(
\partial_{x}-\epspot U'
\right)
-
b^{+}
\left[
\partial_{x}+(1\!-\!\epspot)U'
\right]
\;.
\]
Calling $D_{+}$ to the ``coefficient'' of $b$ and $D_{-}$ to
that of $b^{+}$, we finally have
%
\begin{equation}
\label{Lbarrev}
\left\{
\begin{array}{rcl}
D_{+}
&=&
\partial_{x}-\epspot U'
\\
D_{-}
&=&
\partial_{x}+(1\!-\!\epspot)U'
\end{array}
\right.
\quad
\Rightarrow
\quad
\boxequation{
\Lsb_{\rm rev}
=
-
b
D_{+}
-
b^{+}
D_{-}
\;.
}
\end{equation}


\paragraph*{Calculation of $\hat{Q}_{n,m}$: the Brinkman hierarchy.}

Recall that we need $\LFPb=\Lsb_{\rm rev}+\Lsb_{\rm irr}$, to
obtain its matrix elements $\hat{Q}_{n,m} = \int\!\D v\, \psi_{n}
\LFPb \psi_{m}$, which enter in the equation of motion for the
coefficients $\dot{c}_{n} = \sum_{m} \hat{Q}_{n,m}c_{m}$.
This is an easy task now, since we have written $\Lsb_{\rm irr}$ and
$\Lsb_{\rm rev}$ in terms of $b$ and $b^{+}$.
%
Then, using the ``ladder'' properties of $b$ and $b^{+}$, that is,
$b^{+}\psi_{n}=\sqrt{n+1}\psi_{n+1}$ and
$b\psi_{n}=\sqrt{n}\psi_{n-1}$, plus the orthogonality of the
$\psi_{n}$, we obtain
\[
\hat{Q}_{n,m}
=
\int\!\D v\,
\psi_{n}
\left(
\Lsb_{\rm rev}+\Lsb_{\rm irr}
\right)
\psi_{m}
=
-\sqrt{n}
D_{-}
\delta_{n-1,m}
-
n
\gamma
\delta_{n,m}
-
\sqrt{n+1}
D_{+}
\delta_{n+1,m}
\;.
\]
Therefore, the sum in $\dot{c}_{n} = \sum_{m}\hat{Q}_{n,m}c_{m}$, is
trivially done, and one finally gets the so-called Brinkman hierarchy (1956)
%
\begin{equation}
\label{brinkman}
\boxequation{
\dot{c}_{n}
=
-
\left(
\sqrt{n}
D_{-}\,
c_{n-1}
+
\gamma
n
c_{n}
+
\sqrt{n+1}
D_{+}\,
c_{n+1}
\right)
\;.
}
\end{equation}

We see that, due to (i) the choice of the basis functions to expand
$P$ {\em and\/} (ii) the extraction of the factor $\exp(-v^{2}/4)$,
only the nearest neighbours of $c_{n}$ contribute in
$\dot{c}_{n}=\sum_{m}\hat{Q}_{n,m}c_{m}$.
Writing explicitly the equations, we see that this fact results in a
tri-diagonal structure of the system
\[
\boxequation{
\begin{array}{ccccccccccc}
-\dot{\ec}_{0}
&
=
&
0\gamma \ec_{0}
&
+
&
\sqrt{1}\,D_{+}\ec_{1}
&
+
&
0
&
+
&
\cdots
&
&
\\[1.5ex]
-\dot{\ec}_{1}
&
=
&
\sqrt{1}\,D_{-}\ec_{0}
&
+
&
1\gamma \ec_{1}
&
+
&
\sqrt{2}\,D_{+}\ec_{2}
&
+
&
0
&
+
&
\cdots
\\[1.5ex]
-\dot{\ec}_{2}
&
=
&
0
&
+
&
\sqrt{2}\,D_{-}\ec_{1}
&
+
&
2\gamma \ec_{2}
&
+
&
\sqrt{3}\,D_{+}\ec_{3}
&
+
&
0
\\[1.5ex]
-\dot{\ec}_{3}
&
=
&
\cdots
&
+
&
0
&
+
&
\sqrt{3}\,D_{-}\ec_{2}
&
+
&
3\gamma \ec_{3}
&
+
&
\sqrt{4}\,D_{+}\ec_{4}
\\[1.5ex]
\vdots
&
&
\vdots
&
&
\vdots
&
&
\vdots
&
&
\vdots
&
&
\vdots
\end{array}
}
\]
This equation is completely equivalent to the Klein--Kramers equation
(\ref{KK:Lrev:Lirr:scaled}) and valid for any potential.


\paragraph*{Continued fractions.}

Why are we so happy for having transformed the Klein--Kramers equation
into an infinity hierarchy for the $c_{n}$?
First, by taking the Laplace transform
$\tilde{f}(s)\equiv\int_{0}^{\infty}{\D t}\,e^{-st}f(t)$ a
differential-recurrence relation of the general form
\begin{equation}
\label{dRR}
\frac{\D \ec_{\ir}}{\D t}
+
\eQ_{\ir}^{-}\ec_{\ir-1}
+
\eQ_{\ir}\ec_{\ir}
+
\eQ_{\ir}^{+}\ec_{\ir+1}
=
\eF_{\ir}
\;,
\end{equation}
can be reduced to (we omit the tildes on the Laplace transforms)
\begin{equation}
\label{RR}
\boxequation{
\eQ_{\ir}^{-}\ec_{\ir-1} + \hat{\eQ}_{\ir}\ec_{\ir}
+ \eQ_{\ir}^{+}\ec_{\ir+1}
=
\hat{\eF}_{\ir}
\;,
}
\end{equation}
where $\hat{\eQ}_{\ir}=\eQ_{\ir}+s$ and
$\hat{\eF_{\ir}}=\eF_{\ir}+\ec_{\ir}(0)$.
Then, this relation can be solved by introducing the ansatz
$\ec_{\ir}=\eS_{\ir}\ec_{\ir-1}+\ea_{\ir}$, obtaining
\begin{equation}
\label{C:risken:com}
\boxequation
{
\ec_{\ir}
=
\eS_{\ir}\ec_{\ir-1}+\ea_{\ir}
\quad
\mbox{with}
\quad
\eS_{\ir}
=
-\frac
{\eQ_{\ir}^{-}}
{\hat{\eQ}_{\ir}+\eQ_{\ir}^{+}\eS_{\ir+1}}
\quad
\ea_{\ir}
=
-\frac
{\eQ_{\ir}^{+}\ea_{\ir+1}-\hat{\eF}_{\ir}}
{\hat{\eQ}_{\ir}+\eQ_{\ir}^{+}\eS_{\ir+1}}
}
\end{equation}
It is to be remarked that the quantities involved in the relation
(\ref{RR}) do not need to be scalars, but they can be vectors
($\ec_{\ir}$ and $\eF_{\ir}$) and the coefficients $\eQ_{\ir}$
matrices.
The only change in the solution (\ref{C:risken:com}) is that the
fraction bar then means matrix inversion.

The reason for the name ``continued fraction'', is that, if we consider
for instance $\eS_{\ir}$, it is given in terms of $\eS_{\ir+1}$ in the
denominator.
But this can be expressed in terms of $\eS_{\ir+2}$, and so on,
leading to an expression of the form
%
\begin{equation}
\label{continued-fraction}
K
=
\cfrac{p_{1}}
{q_{1}
+
\cfrac{p_{2}}
{q_{2}
+
\cfrac{p_{3}}
{q_{3}
+
\dotsb
}}}
\;,
\end{equation}
which is called a continued fraction.
Well, if the $\ec_{\ir}$ are judiciously chosen to decrease with
increasing $\ir$, we can truncate at some large $\N$, setting
$\ec_{\N}=0$.
This leads all the quantities to vanish at $\ir=\N$, and we can
iterate downward the continued fractions in Eq.\ (\ref{C:risken:com})
down to $\ir=0$, storing the successive $\eS_{\ir}$ and $\ea_{\ir}$.
Then, starting from $\ec_{0}$ (usually known by some means, e.g.,
normalisation of the distribution), we iterate upwards with
$\ec_{\ir}=\eS_{\ir}\ec_{\ir-1}+\ea_{\ir}$, obtaining the solution to the
recurrence-relation (\ref{RR}).
To ensure convergence, one can repeat the calculation with a
truncation index $2\N$, $4\N$, etc. and check that the results do not
change.
Note that the steps described above are very easy to code in a
computer program (even in a programmable pocket calculator).
This is the answer to the question of why it was important to
transform the Klein--Kramers equation into the Brinkman hierarchy.


\paragraph*{Solving the Brinkman hierarchy.}

As written, the hierarchy (\ref{brinkman}), involves coefficients that
still contain the operators $D_{\pm}$ over the $x$ variables, instead
of true coefficients.
But we only need to find a representation of those operators in a
basis of functions of $x$, say $u_{p}(x)$.
Then we just calculate the matrix elements of any operator $\hat{A}$
in that basis $A_{p\,q}=\int\!\D x\,u_{p}^{\ast}\hat{A}u_{q}$, while
the expansion coefficients of $c_{n}(x,t)$ are expressed as a column
vector
\[
c_{n}(x,t)=\sum_{p}c_{n}^{p}(t)u_{p}(x)
\;,
\qquad
\Rightarrow
\qquad
\mc_{n}
=
\left(
\begin{array}{c}
\ec_{n}^{-P}
\\
\vdots
\\
\ec_{n}^{P}
\end{array}
\right)
\;.
\]
Then, the Brinkman hierarchy (\ref{brinkman}) is reduced to the
following differential recurrence relation
%
\begin{equation}
\label{brinkman:matrix}
\boxequation{
\dot{\mc}_{n}
=
\mQ_{n}^{-}
\mc_{n-1}
+
\mQ_{n}
\mc_{n}
+
\mQ_{n}^{+}
\mc_{n+1}
\;,
}
\end{equation}
whose coefficients are not any more operators but ordinary matrices,
with elements
%
\begin{equation}
\label{brinkman:matrix:elements}
\begin{array}{ccl}
\left(
\mQ_{n}^{-}
\right)_{p\,q}
&=&
-
\sqrt{n}
\left[
\left(\partial_{x}\right)_{p\,q}
+
(1\!-\!\epspot) U_{p\,q}'
\right]
\\
\left(
\mQ_{n}
\right)_{p\,q}
&=&
-
n\,
\gamma\,
\delta_{p\,q}
\\
\left(
\mQ_{n}^{+}
\right)_{p\,q}
&=&
-
\sqrt{n+1}
\left[
\left(\partial_{x}\right)_{p\,q}
-
\epspot U_{p\,q}'
\right]
\end{array}
\end{equation}
The choice of the basis functions $u_{p}(x)$ is naturally dictated by
the symmetry of the potential.

In this form, we can directly use the method of continued fractions to
solve the Brinkman hierarchy.
In principle, with the $c_{n}^{p}$ obtained we can construct the
probability distribution and compute any observable.
Nevertheless, this is not even required, since common observables are
directly obtained from the expansion coefficients.
For instance, if $\epspot=0$ in $W\propto\exp(-v^{2}/4)$, this
quantity is after normalisation $W=\psi_{0}$ (see the definition of
the Hermite functions (\ref{hermite})] so that for the averaged
velocity $\llangle
v
\rrangle
=
\int\!\D x\,
\int\!\D v\,
v
P(x,v,t)$ we get
\[
\llangle
v
\rrangle
=
\sum_{n\,p}
c_{n}^{p}
\int\!\D x\,
u_{p}(x)
\underbrace{
\int\!\D v\,
\overbrace{
\underbrace{v}_{b+b^{+}}
\psi_{0}(v)
}^{\psi_{1}}
\psi_{n}(v)
}_{\delta_{n,1}}
=
\sum_{p}
c_{1}^{p}
\int\!\D x\,
u_{p}(x)
=
c_{1}^{0}
\;,
\]
where he have just assumed that the $0$th element of the basis
$u_{p}(x)$ is constant (say $u_{0}=1$), so that the result follows
from orthogonality ($u_{0}\perp u_{n}$).


\paragraph*{Example: particles in a periodic potential.}

Let us consider the case of a general periodic potential
$U'=\sum_{q}U_{q}'e^{\iu q x}$,
%
where the $U_{q}'$ are the Fourier components of the potential
derivative.
In this case the natural choice of the basis functions $u_{p}(x)$ is
that of plane waves $u_{p}(x)=e^{\iu q x}/\sqrt{2\pi}$.
Then the matrix elements needed to construct the matrices $\mQ_{n}$
in Eq.\ (\ref{brinkman:matrix:elements}) are simply
%
\begin{equation}
\left(\partial_{x}\right)_{p\,q}
=
\iu
p
\delta_{p\,q}
\;,
\qquad
(U')_{p\,q}
=
U_{p-q}'
\;.
\end{equation}
Then, the number of diagonals occupied in the matrices $\mQ_{n}$
below and above the main diagonal is equal to the number of harmonics
in the potential.
In the example of the cosine potential, this number is only one, while
in the ratchet potential (\ref{Vratchet}), this number is two.

\begin{figure}[!t]
\centerline{\epsfig{figure=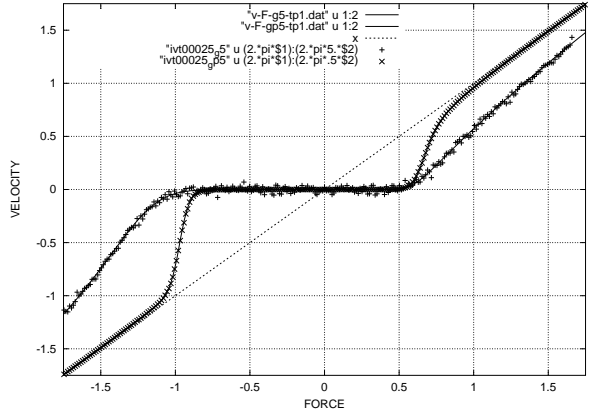,width=8.cm}}
\caption{
Depinning curves in a ratchet potential of Fig.\ \ref{depinning},
obtained with Langevin simulation (symbols), together with the results
obtained with the continued-fraction method (thick lines).
}
\label{depinning-CF}
\end{figure}

Well, in the example of the ratchet potential we computed the average
velocity vs. force by Langevin simulation (Fig.\ \ref{depinning}).
If we compute now the same curves with the continued-fraction method,
we get a complete agreement shown in Fig.\ \ref{depinning-CF} (without
any free parameter).

\begin{figure}[!tbh]
\centerline{\epsfig{figure=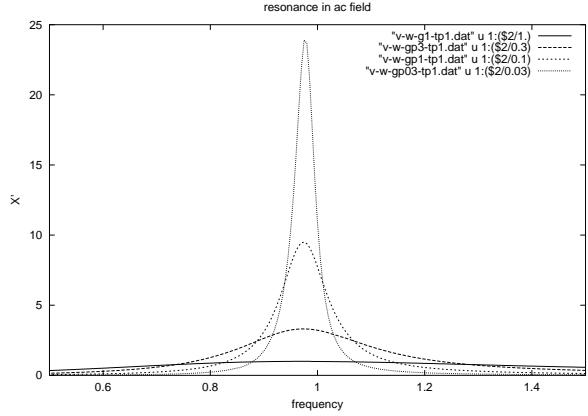,width=8.cm}}
\caption{
Real part of the dynamical susceptibility (ac conductivity)
vs. frequency.
The curves have been obtained with the continued-fraction method for
$T/d=0.1$ and various values of the damping $\gamma=1$,
$0.3$, $0.1$, $0.03$.
The resonance at $\omega\sim1$ is clearly seen.
}
\label{v-resonance}
\end{figure}

As an additional example, in Fig.\ \ref{v-resonance}, we show the real
part of the dynamical susceptibility as a function of the frequency in
an ordinary (non-ratchet) periodic potential (thus, the curves could
correspond to a Josephson junction, or to the ac conductivity in a
transport problem).
We see a peak about the frequency of oscillations in the bottom of the
potential wells (which defines the frequency scale in the graph).
The peak becomes more and more sharp and high the lower the damping
is, showing the typical features of a resonant system.

When the continued-fraction method can be used its advantages are: (i)
it has no statistical error bars, (ii) it is in general extremely
efficient, (iii) it outputs the complete distribution if required, and
(iv) it may be extended to problems of quantum Brownian motion.
The drawbacks are that it is quite specific of the problem to be
solved (we need to compute the matrix elements for each potential, and
the relation between the coefficients and the observables), and in
this respect the Langevin simulations do not have this problem.
Besides, the simulations output trajectories, which are lost in the
continued fraction approach.
Finally, it can only be implemented for systems with a few variables
(e.g., independent particles), while the Langevin simulations do not
suffer this limitation.


\subsubsection{Solution of the Debye equation by
continued--fractions}
\label{CF:debye}

This continued-fraction method can also be applied to rotational
problems.
We shall illustrate this with the \FP\ equation for a dipole
(\ref{debye}).
First, using the definition of the Debye relaxation time
$\tD=\zeta/2\kT$ [Eq.\ (\ref{tau:debye})], the abbreviation
$\alpha=pE/\kT$ and changing to Cartesian coordinates
$z=\cos\vartheta$, the Debye equation can be written as
\begin{equation}
\label{debye:cartesian}
\boxequation{
2\tD
\frac{\partial P}{\partial t}
=
\frac{\partial }{\partial z}
\left[
(1-z^{2})
\left(
\frac{\partial P}{\partial z}
-
\alpha\,
P
\right)
\right]
\;.
}
\end{equation}
\begin{figure}[!thb]
\centerline{\epsfig{figure=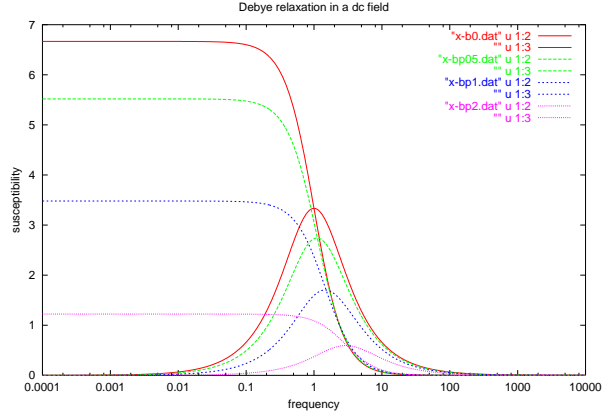,width=8.cm}}
\caption{
Dynamical susceptibility (polarisability) of a dipole in various
static fields.
The curves have been obtained with the continued-fraction method for
$T=0.1$ and the fields used are $pE/\kT=0$, $1$, $2$ and $4$.
}
\label{debye-field}
\end{figure}

In this problem the natural choice for basis function is the Legendre
polynomials $p_{n}(\cos\vartheta)$ [cf.\ Eq.\ (\ref{P:KK:expansion})]
\begin{equation}
\label{P:debye:expansion}
P
=
\sum_{n}
c_{n}(t)
p_{n}(z)
\;.
\end{equation}
Due to the orthogonality of the $p_{n}(z)$,
$\int_{-1}^{1}\!\D z\,
p_{n}(z)p_{m}(z)
=
2\delta_{n,m}/(2n+1)$,
we have $c_{n}=[(2n+1)/2]\int\!\D z\,p_{n}(z)\,P(z)$.
Then the equation of motion for the $c_{n}$ reads [cf.\ Eq.\
(\ref{cn:Qnm:Lbar})]
\[
\frac{4\tD}{2n+1}
\dot{c}_{n}
=
\int\!\D z\,
p_{n}
\overbrace{
\LFP
\underbrace{P}_{\lefteqn {\scriptstyle \sum_{m}c_{m}p_{m}}}
}^{
\partial_{t}P
}
=
\sum_{m}
\hat{Q}_{n,m}
c_{m}
\;,
\quad
\mbox{with}
\quad
Q_{n,m}
=
\int\!\D z\,
p_{n}
\LFP
p_{m}
\;.
\]
To calculate $Q_{n,m}$ (which is not an operator here), we use
relations obeyed by the Legendre polynomials%
\footnote{
Specifically, the Legendre equation, and a sort of first integral:
\[
\frac{\D}{\D z}
\left[
(1-z^{2})
\frac{\D p_{n}}{\D z}
\right]
+
n(n+1)
p_{n}
=
0
\;,
\qquad
(1-z^{2})
\frac{\D p_{n}}{\D z}
=
\frac{n(n+1)}{2n+1}
\left(
p_{n-1}
-
p_{n+1}
\right)
\;.
\]
} 
and an integration by parts:
\begin{eqnarray*}
Q_{n,m}
&=&
\int\!\D z\,
p_{n}
\frac{\D }{\D z}
\left[
(1-z^{2})
\left(
\frac{\D p_{m}}{\D z}
-
\alpha\,
p_{m}
\right)
\right]
\\
&=&
-m(m+1)
\int\!\D z\,
p_{n}
p_{m}
+
\alpha
\int\!\D z\,
\frac{\D p_{n}}{\D z}
(1-z^{2})
p_{m}
\\
&=&
-n(n+1)
\frac{2\delta_{nm}}{2n+1}
+
\alpha\,
\frac{n(n+1)}{2n+1}
\left(
\frac{2\delta_{n-1,m}}{2n-1}
-
\frac{2\delta_{n+1,m}}{2n+3}
\right)
\;.
\end{eqnarray*}
Inserting this result into the equation for $\dot{c}_{n}$, we finally find
%
\begin{equation}
\label{brinkman:dipole}
\boxequation{
\frac{2\tD}{n(n+1)}
\dot{c}_{n}
=
-c_{n}
+
\frac{pE}{\kT}
\left(
\frac{c_{n-1}}{2n-1}
-
\frac{c_{n+1}}{2n+3}
\right)
\;.
}
\end{equation}
which is a tridiagonal differential-recurrence relation for the dipole
analogous to the Brinkman hierarchy (\ref{brinkman}).

Solving Eq.\ (\ref{brinkman:dipole}) by continued fractions, we can
obtain any observable.
For instance, we can compute the linear susceptibility in a static
field (remember that the solved this problem in the absence of bias
field), by setting $E=E_{0}+\Delta E\cos(\omega t)$.
The results are shown in Fig.\ \ref{debye-field}, for various $E_{0}$.
It is seen that the equilibrium susceptibility (real part at
$\omega\to0$) decreases with $E_{0}$, since the susceptibility
measures the slope of the static polarization, which saturates at
large fields.
Besides the relaxation time (roughly, the inverse of the location of
the maximum of the imaginary part) decreases with $E_{0}$, so that the
relaxation is faster in a static field.

\mynewpage
\renewcommand{\ila}{{\rm a}}
\renewcommand{\ilb}{{\rm b}}
\renewcommand{\tp}{t'}
\renewcommand{\Vint}{W}
\renewcommand{\tL}{\Lambda}
\renewcommand{\TL}{\hat{\Lambda}}
\renewcommand{\gmr}{}

\section{Derivation of Langevin equations in\\the bath-of-oscillators
formalism}
\label{calleg}
\markboth{Introduction to the theory of stochastic processes}{Bath of
oscillators formalism}

Now we shall discuss the foundations of the Langevin equations by
showing how they can be ``derived'' from more or less microscopic
description of a system coupled with its environment.


\subsection{Dynamical approaches to the Langevin equations}

The Langevin equations we have discussed for Brownian particles and
spins, are phenomenological inasmuch as they constructed by augmenting
known phenomenological equations by fluctuating terms.
For subsequent reference, let us first rewrite these equations in the
notation we shall employ later:

\noindent
$\bullet$
{\em Brownian particle}
\begin{equation}
\label{langevin:hamilton:II}
\left\{
\begin{array}{rcl}
\displaystyle
\frac{\D \sQ}{\D t}
&=&
\displaystyle
\frac{\sP}{m}
\\[2.ex]
\displaystyle
\frac{1}{m}
\frac{\D \sP}{\D t}
&=&
\displaystyle
-\frac{\partial V}{\partial q}
-\gamma\,\frac{\D\sQ}{\D t}
+
\Lan(t)
\;,
\qquad
\llangle
\Lan(t)\Lan(\tp)
\rrangle
=
2(\gamma\kT/m)
\delta(t-\tp)
\;.
\end{array}
\right.
\end{equation}
Note that the fluctuating terms only enter in the equation for the
momentum.

\noindent
$\bullet$
{\em Brownian spins and dipoles}
\begin{equation}
\label{stolleq:mod:III}
\frac{\D\m}{\D t}
=
\m
\vecpro
\left[\Beff+\bfl(t)\right]
-\lambda\,
\m
\vecpro
\left(\m\vecpro\Beff\right)
\;,
\quad
\llangle
\Lan_{k}(t)\Lan_{\ell}(\tp)
\rrangle
=
2(\lambda\kT)
\delta_{k\ell}
\delta(t-\tp)
\;.
\end{equation}
This {\em stochastic Landau--Lifshitz equation\/} is equivalent to the
Gilbert equation where the damping term is
$\propto-\lambda\m\vecpro(\D\m/\D t)$.
Recall that Eq.\ (\ref{stolleq:mod:III}) contains as a special case
the stochastic dynamical equation for the electrical dipole.

Note that in both equations the fluctuating and dissipation terms are
not independent: $\Dcoeff=\gamma\kT/m$ for the particle, which
corresponds to $\Dcoeff=\lambda\kT$ for the spin.
Besides, the force $F=-\partial V/\partial\sQ$, corresponds to
$\Beff=-\partial\Hs/\partial\m$, while the damping $-\gamma(\D\sQ/\D
t)$, corresponds to $-\lambda\m\vecpro\left(\m\vecpro\Beff\right)$,
which as mentioned above is related with
$\propto-\lambda\m\vecpro(\D\m/\D t)$.

There have been several attempts to justify, starting from dynamical
descriptions of a system coupled to its surroundings, these important
Langevin equations.
The effort was first directed to the Langevin equations for the
Brownian particle (translational problems)
and then to rotational Brownian motion of spins and dipoles.

In most of those studies, the environment is represented as a set of
independent harmonic oscillators.
%
The oscillators are somehow ``projected out'' and an equation for the
system variables is derived.
The final equation has the form of a generalized Langevin equation
(i.e., containing ``memory" terms), whose fluctuating and dissipative
terms naturally obey fluctuation-dissipation relations.
For instance, for the particle problem one gets
\begin{equation}
\label{brownian_particle}
\frac{\D\sP}{\D t}
=
-\frac{\partial\Hs}{\partial\sQ}
+\ffl(t)
-\int_{t_{0}}^{t}\!\!\D{\tp}\,
\Ker(t-\tp)\frac{\D\sQ}{\D t}(\tp)
\;,
\qquad
\llangle \ffl(t)\ffl(\tp)\rrangle
=
\kT\,\Ker(t-\tp)
\;.
\end{equation}
Thus the relaxation (damping) term, involves a memory integral taken
along the past history of the system.
Besides, the memory kernel $\Ker(t-\tp)$ is determined by the
correlation properties of the fluctuating force $\ffl(t)$.
Thus, if the autocorrelation of the fluctuating terms is very short
$\llangle \ffl(t)\ffl(\tp)\rrangle\propto\delta(t-\tp)$, the damping
term reduces to minus the velocity $-(\D\sQ/\D t)$ of the particle.
Similar results are obtained for the spin and dipole problems.

In what follows we shall discuss the bath of oscillators formalism.
First, since we shall use a Hamiltonian formalism throughout, we
shall start with a brief review of the main results from Hamiltonian
mechanics that we shall need.
Subsequently, we shall introduce the model for the system coupled with
its environment, deduce the corresponding dynamical equations, and
finally discuss some examples.


\subsection{Quick review of Hamiltonian dynamics}
\label{subsect:hamilton}

The dynamical equations for a system with Hamiltonian $\Hs(\sP,\sQ)$ are
\begin{equation}
\label{hamilton:eqs}
\frac{\D\sQ}{\D t}
=
\frac{\partial\Hs}{\partial\sP}
\;,
\qquad
\frac{\D\sP}{\D t}
=
-\frac{\partial\Hs}{\partial\sQ}
\;.
\end{equation}
Then, the time evolution of an arbitrary dynamical variable of the
system $A(\sP,\sQ)$ (assumed not explicitly time dependent), is $\D
A/\D t=(\partial A/\partial \sQ)(\D\sQ/\D t)+(\partial A/\partial
\sP)(\D\sP/\D t)$.
Then, using for $\D\sQ/\D t$ and $\D\sP/\D t$ the Hamilton equations
and introducing the {\em Poisson bracket\/} of two arbitrary dynamical
variables
\begin{equation}
\label{poisson}
\pbra A, B\pket
\equiv
\frac{\partial A}{\partial\sQ}
\frac{\partial B}{\partial\sP}
-\frac{\partial A}{\partial\sP}
\frac{\partial B}{\partial\sQ}
\;,
\end{equation}
we have for $\D A/\D t$ the basic Hamiltonian evolution equation
\begin{equation}
\label{hamilton:dA/dt}
\boxequation{
\frac{\D A}{\D t}
=
\pbra A,\Hs\pket
\;.
}
\end{equation}
Finally, by using this equation for $A=\sQ$, with
$\partial\sQ/\partial\sQ=1$, $\partial\sQ/\partial\sP=0$, and for
$A=\sP$ with $\partial\sP/\partial\sQ=0$, and
$\partial\sP/\partial\sP=1$, we see that the Hamilton equations
(\ref{hamilton:eqs}) are a particular case of Eq.\
(\ref{hamilton:dA/dt}).
For a system with variables $(\sP_{a},\sQ_{a})$ $a=1,\dots,N$, the
above results are the same, with the only change of introducing a sum
over $a$ in the definition of the Poisson bracket.

Two more results we shall need are the {\em product rule\/} of the Poisson
bracket
\begin{equation}
\label{productrule}
\pbra A,BC\pket
=
\pbra A,B\pket
C
+
B
\pbra A, C\pket
\;,
\end{equation}
and the {\em chain rule\/}
\begin{equation}
\label{chainrule}
\pbra f,g\pket
=
\sum_{i,k}
\frac{\partial f}{\partial x_{i}}
\frac{\partial g}{\partial x_{k}}
\pbra x_{i},x_{k}\pket
\;,
\qquad
x_{i}=x_{i}(\sP,\sQ)
\;,
\end{equation}
which immediately follow from the ordinary differentiation rules.


\paragraph{Spin dynamics case.}

The equations for an isolated classical spin (not subjected to
fluctuations and dissipation)
\begin{equation}
\label{eqmot_free}
\frac{\D\m}{\D t}
=
\gmr\m\vecpro\Beff
\;,
\qquad
\Beff
=
-\frac{\partial\Hs}{\partial\m}
\;.
\end{equation}
can also be written in Hamiltonian form.
To this end, let us write the formula for the gradient operator in
spherical coordinates
\begin{equation}
\label{gradient}
\frac{\partial u}{\partial\vec{s}}
=
\hat{s}\frac{\partial u}{\partial s}
+
\hat{\vartheta}\frac{1}{s}\frac{\partial u}{\partial\vartheta}
+
\hat{\varphi}\frac{1}{s\sin\vartheta}\frac{\partial u}{\partial\varphi}
\;,
\end{equation}
where $\varphi$ and $\vartheta$ are the azimuthal and polar angles of
$\m$.
Since the length of $\m$ is constant, the set vectorial equations
(\ref{eqmot_free}), can be written as
\begin{equation}
\label{eqmot_free_polar}
\frac{\D\varphi}{\D t}
=
-
\frac{1}{\mm\sin\vartheta}
\frac{\partial\Hs}{\partial\vartheta}
\;,
\qquad
\frac{\D\vartheta}{\D t}
=
\frac{1}{\mm\sin\vartheta}
\frac{\partial\Hs}{\partial\varphi}
\;,
\end{equation}
which correspond to the Hamilton equations (\ref{hamilton:eqs}) with
the conjugate canonical variables
\begin{equation}
\label{canonical_variables}
\boxequation{
\sQ
=
\varphi
\;,
\qquad
\sP
=
\mz
\;.
}
\end{equation}

In terms of the variables (\ref{canonical_variables}) the Cartesian
components of the spin are given by
\begin{equation}
\label{m(p,q)}
\mx
=
\sqrt{\mm^{2}-\sP^{2}}\cos\sQ
\;,
\qquad
\my
=
\sqrt{\mm^{2}-\sP^{2}}\sin\sQ
\;,
\qquad
\mz
=
\gmr\sP
\;.
\nonumber
\end{equation}
From these $\mi(\sP,\sQ)$ and the definition of the Poisson bracket of
two arbitrary dynamical variables [Eq.\ (\ref{poisson})], one can
readily obtain the customary Poisson-bracket (``commutation")
relations among the spin variables
\begin{equation}
\label{commutation}
\pbra\mi,\mj\pket
=
\gmr\sum_{k}\epsilon_{ijk}\mk
\;,
\end{equation}
where $\epsilon_{ijk}$ is the Levi--Civita symbol.%
\footnote{
To illustrate, from
\[
\begin{array}{rclrclr}
\partial\mx/\partial \sQ
&
=
&
-\left[\mm^{2}-\sP^{2}\right]^{1/2}\sin\sQ
\;,
&
\quad
\partial\mx/\partial \sP
&
=
&
-
\sP\left[\mm^{2}-\sP^{2}\right]^{-1/2}\cos\sQ
\;,
\\
\partial\my/\partial \sQ
&
=
&
\hspace{0.75em}
\left[\mm^{2}-\sP^{2}\right]^{1/2}\cos\sQ
\;,
&
\quad
\partial\my/\partial \sP
&
=
&
-
\sP\left[\mm^{2}-\sP^{2}\right]^{-1/2}\sin\sQ
\;,
\end{array}
\]
one gets
$\pbra\mx,\my\pket
=
\sP\sin^{2}\!\sQ
+
\sP\cos^{2}\!\sQ
=
\gmr\mz$.
\qed
} In addition, on using the chain rule of the Poisson bracket [Eq.\
(\ref{chainrule})], one gets the useful relation
\begin{equation}
\label{MpoissonV}
\pbra\mi,\Vint(\m)\pket
=
-\gmr 
\bigg(
\m\vecpro\frac{\partial\Vint}{\partial\m}
\bigg)_{i}
\;,
\end{equation}
which is valid for any function of the spin variables $\Vint(\m)$.%
\footnote{
Note that one can conversely {\em postulate\/} the relations
$\{\mi,\mj\}=\gmr\sum_{k}\epsilon_{ijk}\mk$, and then {\em derive\/}
Eq.\ (\ref{eqmot_free}) starting from the basic Hamiltonian evolution
equation $\D\mi/\D t=\{\mi,\Hs\}$ and using Eq.\
(\ref{MpoissonV}).
This can be considered as a justification of the presence of the
expression $\Beff=-\partial\Hs/\partial\m$ in the dynamical equations
for a classical spin.
} 


\subsection{Dynamical equations in the bath-of-oscillators formalism}

We shall now study a classical system surrounded by an environment
that can be represented by a set of independent classical harmonic
oscillators.
In spite of its academic appearance, those oscillators could
correspond to the {\em normal modes\/} of an electromagnetic field,
the lattice vibrations of a crystal (in the harmonic approximation),
or they can be an effective low-energy description of a more general
surrounding medium (Caldeira and Leggett, \cite{calleg83}).


\subsubsection{The system-environment Hamiltonian}

The total system consisting of the ``system of interest" plus the
oscillators representing the environment forms a {\em closed\/}
dynamical system that we shall describe by augmenting the
isolated-system Hamiltonian as follows
\begin{equation}
\label{hamiltonian:L:1}
\Ham_{\rm T}
=
\Hs(\sP,\sQ)
+
\sum_{\alpha}
\half
\Big\{
\eP_{\alpha}^{2}\overMa
+\Ma\omega_{\alpha}^{2}
\Big[
\eQ_{\alpha}
+
\frac{\coupling}{\Ma\omega_{\alpha}^{2}}
\Fint_{\alpha}(\sP,\sQ)
\Big]^2
\Big\}
\;.
\end{equation}
Here, $\alpha$ is an oscillator index [e.g., the pair $(\vec{k},s)$
formed by the wave-vector and branch index of a normal mode of the
environment], and the coupling terms $\Fint_{\alpha}(\sP,\sQ)$ are
arbitrary functions of the system variables.
These terms may depend on the parameters of the oscillators
$\omega_{\alpha},\Ma$ but not on their dynamical variables
$\eP_{\alpha},\eQ_{\alpha}$.
On the other hand, we have introduced a system-environment coupling
constant $\coupling$ for the sake of convenience in keeping track of
the orders of the various contributions.

The terms proportional to $\Fint_{\alpha}^2$, which emerge when
squaring
$\eQ_{\alpha}+(\coupling/\Ma\omega_{\alpha}^{2})\Fint_{\alpha}$, are
``counter-terms" introduced to balance the coupling-induced
renormalization of the Hamiltonian of the system.
The formalism takes as previously considered whether such a
renormalization actually occurs for a given interaction, so that $\Hs$
would already include it (whenever exists).
An advantage of this convention is that one deals with the
experimentally accessible energy of the system, instead of the ``bare"
one, which might be difficult to determine.


\subsubsection{Dynamical equations}

Let us first cast the Hamiltonian (\ref{hamiltonian:L:1}) into the
form
\begin{equation}
\label{hamiltonian:L:2}
\Ham_{\rm T}
=
\Hsm(\sP,\sQ)
+\sum_{\alpha}
\half
\left(
\eP_{\alpha}^{2}\overMa
+\Ma\omega_{\alpha}^{2}\eQ_{\alpha}^{2}
\right)
+
\coupling\sum_{\alpha}\eQ_{\alpha}\Fint_{\alpha}(\sP,\sQ)
\;,
\end{equation}
where $\sQ$ and $\sP$ are the canonical coordinate and conjugate
momentum of a system with Hamiltonian $\Hs(\sP,\sQ)$ and the
``modified" system Hamiltonian $\Hsm$ augments $\Hs$ by the
aforementioned counter-terms
\begin{equation}
\label{modified_hamiltonian:L}
\Hsm
=
\Hs
+\frac{\coupling^{2}}{2}
\sum_{\alpha}\frac{\Fint_{\alpha}^{2}}{\Ma\omega_{\alpha}^{2}}
\;.
\end{equation}
Besides, in the above expression for $\Ham_{\rm T}$ the Hamiltonian of
the oscillators is clearly recognised
$\Ham_{{\rm E}}
=
\sum_{\alpha}
\half
\left(
\eP_{\alpha}^{2}\overMa
+
\Ma\omega_{\alpha}^{2}\eQ_{\alpha}^{2}
\right)$,
and the same for the coupling term 
$\Ham_{{\rm int}}
=
\coupling
\sum_{\alpha}
\eQ_{\alpha}\Fint_{\alpha}(\sP,\sQ)$.

The equation of motion for any dynamical variable $C$ without explicit
dependence on the time, $\partial C/\partial t\equiv0$, is given by
the basic Hamiltonian evolution equation (\ref{hamilton:dA/dt}) with
$\Hs\to\Ham_{\rm T}$, with the whole Poisson bracket is given by
\[
\pbra A, B\pket
\equiv
\frac{\partial A}{\partial\sQ}
\frac{\partial B}{\partial\sP}
-\frac{\partial A}{\partial\sP}
\frac{\partial B}{\partial\sQ}
+\sum_{\alpha}
\frac{\partial A}{\partial\eQ_{\alpha}}
\frac{\partial B}{\partial\eP_{\alpha}}
-\frac{\partial A}{\partial\eP_{\alpha}}
\frac{\partial B}{\partial\eQ_{\alpha}}
\;.
\]
Therefore, the (coupled) equations of motion for {\em any\/} dynamical
variable of the system $A(\sP,\sQ)$ and the environment variables read
($C=A,\eP_{\alpha}$, and $\eQ_{\alpha}$)
\begin{eqnarray}
\label{eqmot_A:L:1}
\frac{\D A}{\D t}
&
=
&
\pbra A, \Hsm\pket+\coupling
\sum_{\alpha}\eQ_{\alpha}\pbra A, \Fint_{\alpha}\pket
\;,
\\
\label{eqmot_bath:L}
\frac{\D\eQ_{\alpha}}{\D t}
&
=
&
\eP_{\alpha}\overMa
\;,
\qquad
\frac{\D\eP_{\alpha}}{\D t}
=
-\Ma\omega_{\alpha}^{2}\eQ_{\alpha}
-\coupling\Fint_{\alpha}
\;,
\end{eqnarray}
where we have used $\pbra\eQ_{\alpha},\eP_{\alpha}\pket=1$.
The goal is to derive a dynamical equation for $A(\sP,\sQ)$ involving
the system variables only ({\em reduced\/} dynamical equation).

On considering that in Eqs.\ (\ref{eqmot_bath:L}) the term
$-\coupling\Fint_{\alpha}(t)=-\coupling\Fint_{\alpha}[\sP(t),\sQ(t)]$
plays the r\^{o}le of a time-dependent forcing on the oscillators,
those equations can be explicitly integrated, yielding
\begin{equation}
\label{bath_coord:L:1}
\eQ_{\alpha}(t)
=
\eQ_{\alpha}^{\h}(t)
-\frac{\coupling}{\Ma\omega_{\alpha}}
\int_{t_{0}}^{t}\!\!\D{\tp}\,
\sin[\omega_{\alpha}(t-\tp)]
\Fint_{\alpha}(\tp)
\;,
\end{equation}
where
\begin{equation}
\label{qh}
\eQ_{\alpha}^{\h}(t)
=
\eQ_{\alpha}(t_{0})\cos[\omega_{\alpha}(t-t_{0})]
+[\eP_{\alpha}(t_{0})/\Ma\omega_{\alpha}]\sin[\omega_{\alpha}(t-t_{0})]
\;,
\end{equation}
are the solutions of the {\em homogeneous\/} system of equations for
the oscillators in the absence of the system-environment interaction
(proper modes of the environment).
Then, on integrating by parts in Eq.\ (\ref{bath_coord:L:1}) one gets
for the combination $\coupling\eQ_{\alpha}$ that appears in Eq.\
(\ref{eqmot_A:L:1})
\begin{equation}
\label{bath_coord:L:2}
\coupling\eQ_{\alpha}(t)
=
\ffl_{\alpha}(t)
-\left[
\Ker_{\alpha}(t-\tp)\Fint_{\alpha}(\tp)
\right]_{\tp=t_{0}}^{\tp=t}
+\int_{t_{0}}^{t}\!\!\D{\tp}\, 
\Ker_{\alpha}(t-\tp)\frac{\D\Fint_{\alpha}}{\D t}(\tp)
\;,
\end{equation}
where
\begin{equation}
\label{fluct-kernel:L:precursor}
\ffl_{\alpha}(t)
=
\coupling \eQ_{\alpha}^{\h}(t)
\;,
\qquad
\Ker_{\alpha}(\tau)
=
\frac{\coupling^{2}}{\Ma\omega_{\alpha}^{2}}\cos(\omega_{\alpha}\tau)
\;.
\end{equation}

Next, in order to eliminate the environment variables from the
equation for $A(\sP,\sQ)$, one substitutes Eq.\ (\ref{bath_coord:L:2})
back into Eq.\ (\ref{eqmot_A:L:1}), getting
\begin{eqnarray*}
\frac{\D A}{\D t}
=
\pbra A, \Hsm\pket
&-&
\sum_{\alpha}
\pbra A, \Fint_{\alpha}\pket
\Ker_{\alpha}(0)\Fint_{\alpha}(t)
+
\sum_{\alpha}
\pbra A, \Fint_{\alpha}\pket
\Ker_{\alpha}(t-t_{0})\Fint_{\alpha}(t_{0})
\\
&+&
\sum_{\alpha}
\pbra A, \Fint_{\alpha}\pket
\bigg[
\ffl_{\alpha}(t)
+
\int_{t_{0}}^{t}\!\!\D{\tp}\, 
\Ker_{\alpha}(t-\tp)\frac{\D\Fint_{\alpha}}{\D t}(\tp)
\bigg]
\;.
\end{eqnarray*}
The term $\sum_{\alpha}\pbra
A,\Fint_{\alpha}\pket\Ker_{\alpha}(t-t_{0})\Fint_{\alpha}(t_{0})$
depends on the initial state of the system $(\sP(t_{0}),\sQ(t_{0}))$
and produces a transient response that can be ignored in the long-time
dynamics (we shall return to this question below).
The parallel term $-\sum_{\alpha}\pbra
A,\Fint_{\alpha}\pket\Ker_{\alpha}(0)\Fint_{\alpha}(t)$ is derivable
from a Hamiltonian and balances exactly the term due to the
counter-terms in $\pbra A, \Hsm\pket$.
This can be shown by using
\[
-\sum_{\alpha} \pbra A,\Fint_{\alpha}\pket
\Ker_{\alpha}(0)\Fint_{\alpha}
=
\pbra A,
-\half\sum_{\alpha}\Ker_{\alpha}(0) \Fint_{\alpha}^{2} \pket
\;,
\]
which follows from the product rule (\ref{productrule}) of the Poisson
bracket and then using
$\Ker_{\alpha}(0)=\coupling^{2}/\Ma\omega_{\alpha}^{2}$ [see Eq.\
(\ref{fluct-kernel:L:precursor})].
Therefore, one is left with the {\em reduced\/} dynamical equation
\begin{equation}
\label{eqmot_A:L}
\frac{\D A}{\D t}
=
\pbra A,\Hs\pket
+\sum_{\alpha}\pbra A, \Fint_{\alpha}\pket
\bigg[
\ffl_{\alpha}(t)
+\int_{t_{0}}^{t}\!\!\D{\tp}\,
\Ker_{\alpha}(t-\tp)
\frac{\D\Fint_{\alpha}}{\D t}(\tp)
\bigg]
\;,
\end{equation}
where the first term yields the free (conservative) time evolution of
the system, whereas the second term incorporates the effects of the
interaction of the system with its environment.

To conclude, let us decompose the coupling functions as
\begin{equation}
\label{W:L}
\Fint_{\alpha}(\sP,\sQ)
=
\sum_{\ila}
\Cint_{\alpha}^{\ila}
\Vint_{\ila}(\sP,\sQ)
\;.
\end{equation}
Here ``$\ila$'' stands for a general index depending on the type of
interaction.
The idea is to split the part of the coupling $\Vint_{\ila}(\sP,\sQ)$
which is common to all the modes, so that $\Fint_{\alpha}$ is obtained
multiplying that part by certain mode-dependent system-environment
coupling constants $\Cint_{\alpha}^{\ila}$.
For instance, if $\alpha$ is a mode of wave-vector $\vec{k}$, and
$\Fint_{\vec{k}}=\vec{k}\cdot\vec{r}$, then $\ila=i$ (the Cartesian
index) with $\Cint_{\vec{k}}^{i}=k_{i}$ and $\Vint_{i}=r_{i}$.
Introducing the above coupling function into Eq.\ (\ref{eqmot_A:L}),
we have
\[
\frac{\D A}{\D t}
=
\pbra A,\Hs\pket
+\sum_{\alpha}
\pbra
A,
\sum_{\ila}
\Cint_{\alpha}^{\ila}
\Vint_{\ila}
\pket
\bigg[
\ffl_{\alpha}(t)
+\int_{t_{0}}^{t}\!\!\D{\tp}\,
\Ker_{\alpha}(t-\tp)
\sum_{\ilb}
\Cint_{\alpha}^{\ilb}
\frac{\D\Vint_{\ilb}}{\D t}(\tp)
\bigg]
\;.
\]
Therefore, we finally have
\begin{equation}
\label{eqmot_A:L:ab}
\boxequation{
\frac{\D A}{\D t}
=
\pbra A,\Hs\pket
+
\sum_{\ila}
\pbra
A,
\Vint_{\ila}
\pket
\bigg[
\ffl_{\ila}(t)
+
\int_{t_{0}}^{t}\!\!\D{\tp}\,
\sum_{\ilb}
\Ker_{\ila\ilb}(t-\tp)
\frac{\D\Vint_{\ilb}}{\D t}(\tp)
\bigg]
}
\end{equation}
where
\begin{equation}
\label{fluct-kernel:L:ab}
\boxequation{
\ffl_{\ila}(t)
=
\sum_{\alpha}
\Cint_{\alpha}^{\ila}
\ffl_{\alpha}(t)
\;,
\qquad
\Ker_{\ila\ilb}(\tau)
=
\sum_{\alpha}
\Cint_{\alpha}^{\ila}
\Cint_{\alpha}^{\ilb}
\Ker_{\alpha}(\tau)
\;.
}
\end{equation}
The terms $\ffl_{\ila}(t)$ are customarily interpreted as {\em
fluctuating\/} ``forces" (or ``fields").
Indeed $\ffl_{\ila}(t)$ is a sum of a large number of sinusoidal terms
with different frequencies and phases; this can give to
$\ffl_{\ila}(t)$ the form of a highly irregular function of $t$ that
is expected for a fluctuating term (see below).%
\footnote{
Explicit expressions for the $\ffl_{\ila}$ and the kernels in terms of
the proper modes are
\begin{equation}
\label{fluct-kernel:L:ab:explicit}
\ffl_{\ila}(t)
=
\coupling\sum_{\alpha}\Cint_{\alpha}^{\ila}\eQ_{\alpha}^{\h}(t)
\;,
\qquad
\Ker_{\ila\ilb}(\tau)
=
\coupling^{2}
\sum_{\alpha}
\frac{\Cint_{\alpha}^{\ila}\Cint_{\alpha}^{\ilb}}{\Ma\omega_{\alpha}^{2}}
\cos(\omega_{\alpha}\tau)
\end{equation}
} 
The integral term keeps in general memory of the previous history of
the system, and provides the {\em relaxation\/} due to the interaction
with the surrounding medium.

The origin of both types of terms can be traced back as
follows. Recall that in Eq.\ (\ref{bath_coord:L:1}) the time evolution
of the oscillators has formally been written as if they were driven by
(time-dependent) forces $-\coupling\Fint_{\alpha}[\sP(\tp),\sQ(\tp)]$
depending on the state of the system.  Therefore, $\eQ_{\alpha}(t)$
consists of the sum of the proper (free) mode $\eQ_{\alpha}^{\h}(t)$
and the driven-type term, which naturally depends on the ``forcing"
(state of the system) at previous times. Then, the replacement of
$\eQ_{\alpha}$ in the equation for the system variables by the
driven-oscillator solution incorporates:
\begin{enumerate}
\item
The time-dependent modulation due to the proper modes of the environment.
\item
The ``back-reaction" on the system of its preceding action on the
surrounding medium.
\end{enumerate}
Thus, the formalism leads to a description in terms of a reduced
number of dynamical variables at the expense of both explicitly
time-dependent (fluctuating) terms and history-dependent (relaxation)
terms.


\subsubsection{Statistical properties of the fluctuating terms}

In order to determine the statistical properties of the fluctuating
sources $\ffl_{\ila}(t)$, one usually assumes that the environment was
in thermodynamical equilibrium at the {\em initial\/} time (recall
that no statistical assumption has been explicitly introduced until
this point):
\[
\Weq(\ePm(t_{0}),\eQm(t_{0}))
\propto
\exp
\big[
-\beta\Ham_{{\rm E}}(t_{0})
\big]
\;,
\quad
\Ham_{{\rm E}}(t_{0})
=
\sum_{\alpha}
\half
\left[
\eP_{\alpha}(t_{0})^{2}\overMa
+\Ma\omega_{\alpha}^{2}\eQ_{\alpha}(t_{0})^{2}
\right]
\;.
\]
The initial distribution is therefore Gaussian and one
has for the first two moments
\begin{eqnarray*}
\llangle\eQ_{\alpha}(t_{0})\rrangle
=
0
\;,
&
&
\llangle\eP_{\alpha}(t_{0})\rrangle
=
0
\;,
\\
\llangle\eQ_{\alpha}(t_{0})\eQ_{\beta}(t_{0})\rrangle
=
\delta_{\alpha\beta}\frac{\kT}{\Ma\omega_{\alpha}^{2}}
\;,
&
\llangle\eQ_{\alpha}(t_{0})\eP_{\beta}(t_{0})\rrangle
=
0
\;,
&
\llangle\eP_{\alpha}(t_{0})\eP_{\beta}(t_{0})\rrangle
=
\delta_{\alpha\beta}\Ma\kT
\;.
\end{eqnarray*}
From these results one readily gets the averages of the proper modes
over initial states of the environment (ensemble averages):
\begin{eqnarray*}
\llangle\eQ_{\alpha}^{\h}(t)\rrangle
&
=
&
\underbrace{
\llangle\eQ_{\alpha}(t_{0})\rrangle
}_{0}
\cos[\omega_{\alpha}(t-t_{0})]
+\underbrace{
\llangle\eP_{\alpha}(t_{0})\rrangle
}_{0}
\frac{1}{\Ma\omega_{\alpha}}\sin[\omega_{\alpha}(t-t_{0})]
\;,
\\
\llangle\eQ_{\alpha}^{\h}(t)\eQ_{\beta}^{\h}(\tp)\rrangle
&
=
&
\underbrace{
\llangle\eQ_{\alpha}(t_{0})\eQ_{\beta}(t_{0})\rrangle
}_{\delta_{\alpha\beta}\kT/\Ma\omega_{\alpha}^{2}}
\cos[\omega_{\alpha}(t-t_{0})]\cos[\omega_{\beta}(\tp-t_{0})]
\\
&
& {}+
\underbrace{
\llangle\eQ_{\alpha}(t_{0})\eP_{\beta}(t_{0})\rrangle
}_{0}
\frac{1}{\Mb\omega_{\beta}}
\cos[\omega_{\alpha}(t-t_{0})]\sin[\omega_{\beta}(\tp-t_{0})]
\\
&
& {}+
\underbrace{
\llangle\eP_{\alpha}(t_{0})\eQ_{\beta}(t_{0})\rrangle
}_{0}
\frac{1}{\Ma\omega_{\alpha}}
\sin[\omega_{\alpha}(t-t_{0})]\cos[\omega_{\beta}(\tp-t_{0})]
\\
&
& {}+
\underbrace{
\llangle\eP_{\alpha}(t_{0})\eP_{\beta}(t_{0})\rrangle
}_{\delta_{\alpha\beta}\Ma\kT}
\frac{1}{\Ma\omega_{\alpha}\Mb\omega_{\beta}}
\sin[\omega_{\alpha}(t-t_{0})]\sin[\omega_{\beta}(\tp-t_{0})]
\\
&
=
&
\kT
\frac{\delta_{\alpha\beta}}{\Ma\omega_{\alpha}^{2}}
\{
\cos[\omega_{\alpha}(t-t_{0})]\cos[\omega_{\alpha}(\tp-t_{0})]
\\
&
&
\hspace{4em}
+\sin[\omega_{\alpha}(t-t_{0})]\sin[\omega_{\alpha}(\tp-t_{0})]
\}
\;,
\end{eqnarray*}
so that
\begin{equation}
\label{stats:qh}
\llangle\eQ_{\alpha}^{\h}(t)\rrangle
=
0
\;,
\qquad
\llangle\eQ_{\alpha}^{\h}(t)\eQ_{\beta}^{\h}(\tp)\rrangle
=
\kT\frac{\delta_{\alpha\beta}}{\Ma\omega_{\alpha}^{2}}
\cos[\omega_{\alpha}(t-\tp)]
\;.
\end{equation}

\begin{figure}[!tbh]
\centerline{\epsfig{figure=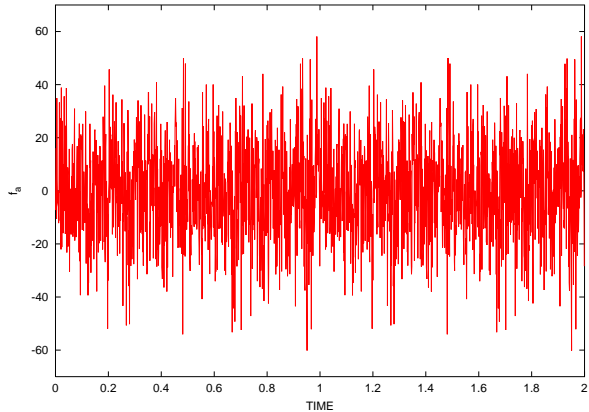,width=8.cm}}
\caption{
The quantity $\ffl(t)$ obtained by summing over $1000$
oscillators with initial conditions drawn from a Gaussian
distribution.
}
\label{noise}
\end{figure}

Then, since Eq.\ (\ref{fluct-kernel:L:ab:explicit}) says that
$\ffl_{\ila}(t)
=
\coupling\sum_{\alpha}\Cint_{\alpha}^{\ila}\eQ_{\alpha}^{\h}(t)$
and
$\Ker_{\ila\ilb}(\tau)
=\coupling^{2}
\sum_{\alpha}
(\Cint_{\alpha}^{\ila}\Cint_{\alpha}^{\ilb}/\Ma\omega_{\alpha}^{2})
\cos(\omega_{\alpha}\tau)$, the equations
(\ref{stats:qh}) give for the averages of the fluctuating terms
$\ffl_{\ila}(t)$:
\begin{equation}
\label{stats:L}
\boxeqnarray{0.5\textwidth}{
\llangle\ffl_{\ila}(t)\rrangle
&
=
&
0
\;,
\\
\llangle\ffl_{\ila}(t)\ffl_{\ilb}(\tp)\rrangle
&
=
&
\kT\,\Ker_{\ila\ilb}(t-\tp)
\;.
}
\end{equation}
The second equation relates the statistical time correlation of the
fluctuating terms $\ffl_{\ila}(t)$ with the relaxation memory kernels
$\Ker_{\ila\ilb}(\tau)$ occurring in the dynamical equations ({\em
fluctuation-dissipation\/} relations).
Short (long) correlation times of the fluctuating terms entail
short-range (long-range) memory effects in the relaxation term, and
vice versa.
The emergence of this type of relations is not surprising in this
context, since fluctuations and relaxation arise as different
manifestations of the {\em same\/} interaction of the system with the
surrounding medium.%
\footnote{
If one assumes that the environment is at $t=t_{0}$ in thermal
equilibrium {\em in the presence of the system}, which is however
taken as {\em fastened\/} in its initial state,
the corresponding initial distribution of the
environment variables is $\Weq\propto\exp[-\Ham_{{\rm
SE}}(t_{0})/\kT]$, with
\[
\textstyle
\Ham_{{\rm SE}}(t_{0})
=
\sum_{\alpha}
\half
\big\{
\eP_{\alpha}(t_{0})^{2}\overMa
+\Ma\omega_{\alpha}^{2}
\big[
\eQ_{\alpha}(t_{0})
+(\coupling/\Ma\omega_{\alpha}^{2})\Fint_{\alpha}(t_{0})
\big]^2
\big\}
\;.
\]
In this case, the dropped terms
$\Ker_{\alpha}(t-t_{0})\Fint_{\alpha}(t_{0})$, which for
$\Fint_{\alpha}=\sum_{\ila}\Cint_{\alpha}^{\ila}\Vint_{\ila}$ lead to
$\sum_{\ilb}\Ker_{\ila\ilb}(t-t_{0})\Vint_{\ilb}(t_{0})$, are included
into an alternative definition of the fluctuating sources, namely
$\tilde{f}_{\ila}(t)
=
\ffl_{\ila}(t)+\sum_{\ilb}\Ker_{\ila\ilb}(t-t_{0})\Vint_{\ilb}(t_{0})$.
The statistical properties of these terms, as determined by the above
distribution, are given by expressions {\em identical\/} with Eqs.\
(\ref{stats:L}).
Then, with both types of initial conditions one obtains the {\em
same\/} Langevin equation after a time of the order of the width of
the memory kernels $\Ker_{\ila\ilb}(\tau)$, which is the
characteristic time for the ``transient" terms
$\sum_{\ilb}\Ker_{\ila\ilb}(t-t_{0})\Vint_{\ilb}(t_{0})$ to die out.
} 
To conclude, we show in Fig.\ \ref{noise}, the quantity $\ffl(t)
=
\coupling
\sum_{k}
\Cint_{k}
\left[\eQ_{k}(t_{0})\cos(\omega_{k}t)
+[\eP_{k}(t_{0})/\Ma\omega_{k}]\sin(\omega_{k}t)
\right]$,
with $\Cint_{k}\propto k$, $\omega_{k}=ck$, and the
$(\eP(t_{0}),\eQ(t_{0}))$ drawn from a Gaussian distribution.
The graph shows that a quantity obtained by adding many sinusoidal
terms with different frequencies and phases can actually be a highly
irregular function of $t$.
%


\subsubsection{Markovian regime}

We shall now study the form that the dynamical equations derived
exhibit in the absence of memory effects.
This occurs when the memory kernels are sharply peaked at $\tau=0$,
the remainder terms in the memory integrals change slowly enough in
the relevant range, and the kernels enclose a finite non-zero
algebraic area.
Under these conditions, one can replace the kernels by Dirac deltas
and no memory effects occur.

Doing this with the memory kernel (\ref{fluct-kernel:L:ab}), we write
\begin{equation}
\label{kernel:L:markov}
\Ker_{\ila\ilb}(\tau)
=
2
\gamma_{\ila\ilb}
\delta(\tau)
\;,
\end{equation}
where the $\gamma_{\ila\ilb}$ are {\em damping coefficients\/}
related with the strength and characteristics of the coupling (see
below).
Then, on using
$\int_{0}^{\infty}\!\D\tau\,\delta(\tau)h(\tau)=h(0)/2$, equation
(\ref{eqmot_A:L:ab}) reduces to
\begin{equation}
\label{eqmot_A:L:ab:markov}
\boxequation{
\frac{\D A}{\D t}
=
\pbra A,\Hs\pket
+
\sum_{\ila}
\pbra
A,
\Vint_{\ila}
\pket
\bigg[
\ffl_{\ila}(t)
+
\sum_{\ilb}
\gamma_{\ila\ilb}
\frac{\D\Vint_{\ilb}}{\D t}
\bigg]
\;,
}
\end{equation}
with
\begin{equation}
\label{stats:L:markov}
\llangle\ffl_{\ila}(t)\rrangle
=
0
\;,
\qquad
\llangle\ffl_{\ila}(t)\ffl_{\ilb}(\tp)\rrangle
=
2\gamma_{\ila\ilb}\kT\,\delta(t-\tp)
\;.
\end{equation}

Inspecting Eq.\ (\ref{kernel:L:markov}), one sees that the damping
coefficients can be obtained from the area enclosed by the memory
kernels or, alternatively, by inserting the definition of the kernel
(\ref{fluct-kernel:L:ab:explicit}) into the corresponding integral and
then using
$\int_{0}^{\infty}\!\D\tau\,\cos(\omega\tau)=\pi\delta(\omega)$:
\begin{equation}
\label{lambdas:kernel}
\gamma_{\ila\ilb}
=
\int_{0}^{\infty}\!\!\D\tau\,\Ker_{\ila\ilb}(\tau)
\;,
\qquad
\gamma_{\ila\ilb}
=
\pi\coupling^{2}
\sum_{\alpha}
\frac
{\Cint_{\alpha}^{\ila}\Cint_{\alpha}^{\ilb}}
{\Ma\omega_{\alpha}^{2}}\delta(\omega_{\alpha})
\;.
\end{equation}
The area $\int_{0}^{\infty}\!\!\D\tau\,\Ker_{\ila\ilb}(\tau)$ must
be: (i) {\em finite\/} and (ii) {\em different from zero}, for the
Markovian approximation to work.
The second expression gives the damping coefficients in terms of the
distribution of normal modes and system-environment coupling
constants, and could be useful in cases where it could be difficult to
find the kernels exactly.
%


\subsection{Examples: Brownian particles and spins}

In order to particularize the general expressions to definite
situations, we only need to specify the structure of the coupling
terms $\Fint_{\ila}$.


\paragraph*{Brownian particle.}

For instance, let us set $\Fint_{\alpha}(\sP,\sQ)=-\Cint_{\alpha}\sQ$
(bilinear coupling), and write down Eq.\ (\ref{eqmot_A:L}) for
$A=\sQ$ and $A=\sP$ with help from $\pbra\sP,B\pket=-\partial
B/\partial\sQ$ and $\pbra\sQ,B\pket=\partial B/\partial\sP$.
Then one gets $\D\sQ/\D t=\partial\Hs/\partial\sP$ plus
Eq.\ (\ref{brownian_particle}), which is the celebrated generalized
Langevin equation for a ``Brownian" particle.
The fluctuating force is explicitly given by
$f(t)=\sum_{\alpha}\Cint_{\alpha}\ffl_{\alpha}(t)$ and the memory
kernel by
$\Ker(\tau)=\sum_{\alpha}\Cint_{\alpha}^{2}\Ker_{\alpha}(\tau)$.
Naturally, in the Markovian limit $\Ker(\tau)=2m\gamma\delta(\tau)$ we
have
\begin{equation}
\label{brownian_particle:markov}
\frac{\D\sQ}{\D t}
=
\frac{\partial\Hs}{\partial\sP}
\;,
\qquad
\frac{\D\sP}{\D t}
=
-\frac{\partial\Hs}{\partial\sQ}
+\ffl(t)
-\gamma
\frac{\D\sQ}{\D t}
\;,
\end{equation}
whose relaxation term comprises minus the velocity $-(\D\sQ/\D t)$ of
the particle.

In general, when $\pbra A,\Fint_{\ila}\pket$ in Eq.\
(\ref{eqmot_A:L:ab:markov}) is not constant, the fluctuating terms
$\ffl_{\ila}(t)$ enter multiplying the system variables ({\em
multiplicative\/} fluctuations).
In this example, owing to the fact that
$\pbra\sQ,-\Cint_{\alpha}\sQ\pket=0$ and
$\pbra\sP,-\Cint_{\alpha}\sQ\pket=\Cint_{\alpha}$, the fluctuations
are {\em additive}.


\paragraph*{Spin-dynamics.}

Let us now particularize the above results to the dynamics of a
classical spin.
To do so, we merely put $A=\mi$, $i=x,y,z$, in Eq.\ (\ref{eqmot_A:L}),
and then use Eq.\ (\ref{MpoissonV}) to calculate the Poisson brackets
required.
Using also $\D\Vint_{\ilb}/\D
t=(\partial\Vint_{\ilb}/\partial\m)\cdot(\D\m/\D t)$, we have
\[
\frac{\D\mi}{\D t}
=
-
\bigg(
\m\vecpro\frac{\partial\Hs}{\partial\m}
\bigg)_{i}
-
\sum_{\ila}
\bigg(
\m\vecpro\frac{\partial\Vint_{\ila}}{\partial\m}
\bigg)_{i}
\bigg[
\ffl_{\ila}(t)
+
\sum_{\ilb}
\gamma_{\ila\ilb}
\frac{\partial\Vint_{\ilb}}{\partial\m}
\cdot
\frac{\D\m}{\D t}
\bigg]
\;,
\]
On gathering these results for $i=x,y,z$ in vectorial form and
recalling the definition of the effective field
$\Beff=-\partial\Hs/\partial\m$, we arrive at
\begin{eqnarray*}
\frac{\D\m}{\D t}
=
\m\vecpro\Beff
-
\m\vecpro
\Bigg(
\sum_{\ila}
\ffl_{\ila}(t)
\frac{\partial\Vint_{\ila}}{\partial\m}
+
\bigg[
\sum_{\ila\ilb}
\gamma_{\ila\ilb}
\frac{\partial\Vint_{\ila}}{\partial\m}
\frac{\partial\Vint_{\ilb}}{\partial\m}
\bigg]
\cdot
\frac{\D\m}{\D t}
\Bigg)
\;.
\end{eqnarray*}
Then, defining the {\em fluctuating magnetic field\/}
\begin{equation}
\label{bfl:L}
\bfl(t)
=
-\sum_{\ila}\ffl_{\ila}(t)
\frac{\partial\Vint_{\ila}}{\partial\m}
\;,
\end{equation}
and the second-rank tensor $\TL$ with elements
\begin{equation}
\label{tensor:L}
\tL_{ij}
=
\sum_{\ila,\ilb}\gamma_{\ila\ilb}
\frac{\partial\Vint_{\ila}}{\partial\mi}
\frac{\partial\Vint_{\ilb}}{\partial\mj}
\;,
\end{equation}
we finally obtain the Langevin equation for the spin%
\footnote{
Although we omit the symbol of scalar product, the action of a dyadic
$\vec{A}\,\vec{B}$ on a vector $\vec{C}$ is the standard one:
$(\vec{A}\,\vec{B})\vec{C}\equiv\vec{A}(\vec{B}\cdot\vec{C})$.
}
\begin{equation}
\label{eqmot_M_Giltyp_markov}
\boxequation{
\frac{\D\m}{\D t}
=
\gmr
\m\vecpro 
\left[
\Beff+\bfl(t)
\right]
-
\m\vecpro 
\TL\frac{\D\m}{\D t}
\;.
}
\end{equation}
Equation (\ref{eqmot_M_Giltyp_markov}) contains $\D\m/\D t$ on its
right-hand side, so it will be referred to as a {\em Gilbert-type\/}
equation.
For $\coupling\ll1$, on replacing perturbatively that derivative by
its conservative part, $\D\m/\D t\simeq\gmr\m\vecpro\Beff$, one gets
the weak-coupling {\em Landau--Lifshitz-type\/} equation
\begin{equation}
\label{eqmot_M_LLtyp_markov:L}
\boxequation{
\frac{\D\m}{\D t}
=
\gmr\m\vecpro 
\left[
\Beff+\bfl(t)
\right]
-
\m\vecpro\TL
\left(\m\vecpro\Beff\right)
\;.
}
\end{equation}
which describes weakly damped precession.
From the statistical properties (\ref{stats:L:markov}) of the
fluctuating sources $\ffl_{\ila}(t)$, one gets
\begin{equation}
\label{stats:bfl:L}
\llangle \Lan_{i}(t)\rrangle
=
0
\;,
\quad
\llangle \Lan_{i}(t)\Lan_{j}(\tp)\rrangle
=
2\tL_{ij}
\kT\delta(t-\tp)
\;,
\end{equation}
which relates the structure of the correlations of the fluctuating
field and the relaxation tensor.

For a general form of the spin-environment interaction, due to the
occurrence of the tensor $\TL$ the structure of the relaxation terms
in the above equations deviates from the forms proposed by Gilbert and
Lan\-dau and Lif\-shitz.
However, if the spin-environment interaction yields uncorrelated {\em
and\/} isotropic fluctuations ($\tL_{ij}=\lambda\delta_{ij}$), one
finds that: (i) the statistical properties (\ref{stats:bfl:L}) reduce
to those in (\ref{stolleq:mod:III}) and (ii) the Langevin equation
(\ref{eqmot_M_LLtyp_markov:L}) reduces to the stochastic
Landau--Lifshitz equation (\ref{stolleq:mod:III}).

We remark in closing that the occurrence of the vector {\em product\/}
$\m\vecpro\bfl$ in the dynamical equations entails that the
fluctuating terms enter in a {\em multiplicative\/} way.
%
%
In the spin-dynamics case, in analogy with the results obtained for
mechanical rigid rotators, 
the multiplicative character of the fluctuations is an inevitable
consequence of the Poisson bracket relations for angular-momentum-type
dynamical variables $\pbra\mi,\mj\pket=\gmr\sum_{k}\epsilon_{ijk}\mk$,
which, even for $\Fint_{\ila}$ linear in $\m$, lead to non-constant
$\pbra A,\Fint_{\ila}\pket$ in Eq.\ (\ref{eqmot_A:L}).
In our derivation this can straightly be traced back by virtue of the
Hamiltonian formalism employed.


\subsection{Discussion}

We have seen that starting from a Hamiltonian description of a
classical system interacting with the surrounding medium, one can
derive generalized Langevin equations, which, in the Markovian
approach, reduce to known phenomenological Langevin equations.

Note however that the presented derivation of the equations is formal
in the sense that one must still investigate specific realizations of
the system-plus-environment whole system, and then prove that the
assumptions employed (mainly that of Markovian behavior) are at least
approximately met.
Let us give an example for a particle coupled to the elastic waves
(phonons) of the substrate where it moves.
The interaction would be proportional to the deformation tensor
$\Ham_{\rm SE}\propto\partial u/\partial x$ in one dimension.
Expanding the displacement field in normal modes
$u(x)=\sum_{k}u_{k}\exp(\iu kx)$, where the $u_{k}$ are the
coordinates of the environment variables (our $\eQ_{\alpha}$), we have
$\Ham_{\rm SE}\propto\sum_{k}\iu k\exp(\iu kx)u_{k}$, so that
$\Cint_{k}\propto\iu k\exp(\iu kx)$.
If we had allowed complex $\Cint_{\alpha}$, the products
$\Cint_{\alpha}^{2}$ would had been replaced by
$|\Cint_{\alpha}|^{2}$.
Well, the corresponding memory kernel [Eq.\
(\ref{fluct-kernel:L:ab:explicit})], then gives
\[
\Ker(\tau)
=
\coupling^{2}
\sum_{\alpha}
\frac{|\Cint_{\alpha}|^{2}}{\Ma\omega_{\alpha}^{2}}
\cos(\omega_{\alpha}\tau)
\stackrel{\omega_{k}=ck}{\to}
\int_{0}^{k_{\rm D}}
\D k
\frac{\cross k^{2}}{c^{2}\cross k^{2}}
\cos(c k\tau)
\propto
\frac{\sin(\omega_{\rm D}\tau)}{\tau}
\;.
\]
But, $\sin(\Omega\tau)/\tau$ plays the role of a Dirac delta for any
process with time-scales much larger than $1/\Omega$.
Thus, taking the Markovian limit is well justified in this case.

On the other hand, we have considered the classical regime of the
environment and the system.
A classical description of the environment is adequate, for example,
for the coupling to low-frequency ($\hbar\omega_{\alpha}/\kT\ll1$)
normal modes.
Nevertheless, the fully Hamiltonian formalism used, allows to guess
the structure of the equations in the quantum case (just replacing
Poisson brackets by commutators).

\mynewpage

\providecommand{\bysame}{\leavevmode\hbox to3em{\hrulefill}\thinspace}
\providecommand{\MR}{\relax\ifhmode\unskip\space\fi MR }
\providecommand{\MRhref}[2]{%
  \href{http://www.ams.org/mathscinet-getitem?mr=#1}{#2}
}
\providecommand{\href}[2]{#2}

\addcontentsline{toc}{section}{Bibliography}


\mynewpage

\medskip
\medskip
\medskip
\centerline{\bf{\Large APPENDICES}}

\appendix


\section*{Dynamical equations for the averages: macroscopic equation}

From the master equation one can derive the dynamical equations for the
averages of a Markov stochastic process.
We shall write down the corresponding derivations directly in the
multivariate case.

Let us first write the equation for the time evolution of an arbitrary
function $\llangle f(\multi{y})\rrangle$.%
\footnote{Here we use the same notation for the stochastic process and its
realisations.}
First, one has
\[
\Der{}{t}\av{f(\multi{y})}
=
\Der{}{t}\Intdef{}{}{\multi{y}}
f(\multi{y})P(\multi{y},t)
=
\Intdef{}{}{\multi{y}}f(\multi{y})\Dpar{P(\multi{y},t)}{t}
\;.
\]
Then, by using the master equation to express $\dpar{P}{t}$, one has
\begin{eqnarray}
\label{avdyneq:f:n-dim}
\Der{}{t}\av{f(\multi{y})}
&=&
\underbrace{
\Intdef{}{}{\multi{y}}
f(\multi{y})
\Intdef{}{}{\multi{y}'}
W(\multi{y}|\multi{y}')
P(\multi{y}',t)}
_{\multi{y}'\leftrightarrow\multi{y}}
-\Intdef{}{}{\multi{y}}
f(\multi{y})
\Intdef{}{}{\multi{y}'}
W(\multi{y}'|\multi{y})P(\multi{y},t)
\nonumber\\[-1.ex]
&=&
\Intdef{}{}{\multi{y}'}f(\multi{y}')
\Intdef{}{}{\multi{y}}
W(\multi{y}'|\multi{y})P(\multi{y},t)
-\Intdef{}{}{\multi{y}}
f(\multi{y})
\Intdef{}{}{\multi{y}'}
W(\multi{y}'|\multi{y})P(\multi{y},t)
\nonumber\\
&=&
\Intdef{}{}{\multi{y}}P(\multi{y},t)
\Intdef{}{}{\multi{y}'}
[f(\multi{y'})-f(\multi{y})]
W(\multi{y}'|\multi{y})
\;.
\end{eqnarray}
On applying now this equation to $f(\multi{y})=y_{i}$, and defining
[cf.\ Eq.\ (\ref{a_0&W})]
\begin{equation}
\label{a_1&W:n-dim}
a_{i}^{(1)}(\multi{y},t)=\Intdef{}{}{\multi{y}'}
(y_{i}'-y_{i})W(\multi{y}'|\multi{y})
\;,
\end{equation}
the last line in Eq.\ (\ref{avdyneq:f:n-dim}) is the average of
$a_{i}^{(1)}(\multi{y},t)$, so we can finally write
\begin{equation}
\label{avdyneq:n-dim}
\boxequation{
\Der{}{t}\av{y_{i}}=\bbav{a_{i}^{(1)}(\multi{y},t)}
\qquad
(i=1,2,\ldots)
\;.
}
\end{equation}
This is an exact consequence of the master equation and therefore holds
for any Markov process.

Note that when $a_{i}^{(1)}$ is {\em linear\/} function of $\multi{y}$
one has
$\bbav{a_{i}^{(1)}(\multi{y},t)}=a_{i}^{(1)}(\av{\multi{y}},t)$, whence
\[
\Der{}{t}\av{y_{i}}=a_{i}^{(1)}(\av{\multi{y}},t)
\;,
\]
which is a system of ordinary differential equations for $\av{\multi{y}}$
and can be identified with the {\em macroscopic\/} equation of the
system.
For instance in the decay problem
(\ref{masterequation:discrete:decay}), since $W_{n',n}$ is non-zero
for $n'=n-1$, we have
\[
a^{(1)}(n,t)
=
\sum_{n'}
(n'-n)
\underbrace{
W_{n',n}
}_{\gamma n\delta_{n',n-1}}
=
[(\cross n-1)-\cross n]
\,
\gamma n
=
-\gamma n
\;.
\]
Therefore, from the general result (\ref{avdyneq:n-dim}) we have
$\D\llangle n\rrangle/\D t
=
\langle
a^{(1)}(n,t)
\rangle
=
-\llangle\gamma n\rrangle$,
in agreement with Eq. (\ref{average:decay}).

On the other hand, when $a_{i}^{(1)}$ is a {\em non-linear\/} function
of $\multi{y}$, Eq.\ (\ref{avdyneq:n-dim}) is not a differential
equation for $\av{y_{i}}$.
Then Eq.\ (\ref{avdyneq:n-dim}) is not a closed equation for
$\av{y_{i}}$ but higher-order moments enter as well.
Thus, the evolution of $\av{y_{i}}$ in the course of time is not determined
by $\av{y_{i}}$ itself, but is also influenced by the fluctuations around
this average.
To get equations for higher-order moments we proceed analogously.
For instance, for $\av{y_{i}(t)y_{j}(t)}$, we can use Eq.\
(\ref{avdyneq:f:n-dim}) with $f(\multi{y})=y_{i}y_{j}$.
Writting 
$(y_{i}'y_{j}'-y_{i}y_{j})
=
(y_{i}'-y_{i})(y_{j}'-y_{j})
     +y_{i}(y_{j}'-y_{j})
     +y_{j}(y_{i}'-y_{i})$,
and defining, analogously to Eqs.\ (\ref{a_0&W}) and
(\ref{a_1&W:n-dim}),
\begin{equation}
\label{a_2&W:n-dim}
a_{ij}^{(2)}(\multi{y},t)
=
\Intdef{}{}{\multi{y}'}
(y_{i}'-y_{i})(y_{j}'-y_{j})W(\multi{y}'|\multi{y})
\;.
\end{equation}
we finally have
\begin{equation}
\label{avdyneq:2nd:n-dim}
\boxequation{
\Der{}{t}\av{y_{i}y_{j}}
=\bbav{a_{ij}^{(2)}(\multi{y},t)}
+\bbav{y_{i}a_{j}^{(1)}(\multi{y},t)}
+\bbav{y_{j}a_{i}^{(1)}(\multi{y},t)}
\;.											}
\end{equation}
which is also an exact consequence of the master equation.%
\footnote{
Note that for one variable (or for $i=j$) Eq.\
(\ref{avdyneq:2nd:n-dim}) reduces to
\[
\Der{}{t}\av{y^{2}}=\bbav{a^{(2)}(y,t)}+2\bbav{ya^{(1)}(y,t)}
\;,
\]
where [cf.\ Eq.\ (\ref{a_0&W})]
\begin{equation}
\label{a1:a2}
a^{(1)}(y,t)=\Intdef{}{}{y'}(y'-y)W(y'|y)
\;,
\qquad
a^{(2)}(y,t)=\Intdef{}{}{y'}(y'-y)^{2}W(y'|y)
\;,
\end{equation}
are the one-variable counterparts of Eqs.\ (\ref{a_1&W:n-dim}) and
(\ref{a_2&W:n-dim}), respectively.
} 
However, if $a_{ij}^{(2)}$ is a non-linear function of $\multi{y}$,
the equation involves even higher order moments
$\llangle y_{i}y_{j}y_{k}\rrangle$, so what we have is an infinite
hierarchy of coupled equations for the moments.


\section*{The Langevin process $\Lan(t)$ as the derivative of the
\WL\ process}

Let us formally write $\D W/\D t=\Lan(t)$, and see which are the
properties of the $W(t)$ so defined.
On integrating over an interval $\tau$, we have
\begin{equation}
\label{wienner:increment}
w(\tau)\equiv\Delta W(\tau)\equiv W(t+\tau)-W(t)
=\int_{t}^{t+\tau}\!\! \Lan(s)ds
\;.
\end{equation}
Let us show that this $w(\tau)$ is indeed a \WL\ process.
Firstly, $w(\tau)$ is Gaussian because $\Lan(t)$ is so.
Furthermore, on using the statistical properties (\ref{langevin:moments})
one gets ($\tau,\tau_{1},\tau_{2}\geq0$)
\begin{equation}
\label{wienner:increment:moments}
w(0)=0
\;,
\qquad
\av{w(\tau)}=0
\;,
\qquad
\av{w(\tau_{1})w(\tau_{2})}
=2D\min(\tau_{1},\tau_{2})
\;.
\end{equation}
Proof: $w(0)=0$ follows immediately from the definition
(\ref{wienner:increment}), while for the average $\av{w(\tau)}$, one
gets
$\av{w(\tau)}=\int_{t}^{t+\tau}\!\!\,\underbrace{\av{\Lan(s)}}_{0}ds=0$.
On the other hand, for $\av{w(\tau_{1})w(\tau_{2})}$, one finds
\begin{eqnarray*}
\av{w(\tau_{1})w(\tau_{2})}
&=&
\int_{t}^{t+\tau_{1}}
\int_{t}^{t+\tau_{2}}\,
\overbrace{\av{\Lan(s)\Lan(s')}}^{2D\delta(s-s')}
ds'ds
\\
&=&
2D\int_{t}^{t+\min(\tau_{1},\tau_{2})}\,
\overbrace{\int_{t}^{t+\max(\tau_{1},\tau_{2})}\,\,
\delta(s-s')ds'}^{1}ds
\\
&=&
2D\min(\tau_{1},\tau_{2})
\;,
\end{eqnarray*}
where we have sorted the integrals to ensure that, when using the Dirac
delta to take one of them, the location of the ``maximum" of the delta is
inside the corresponding integration interval, and the result is therefore
unity.

Now on comparing these results with those for the {\em increment\/} of
the \WL\ process, whose average is zero since that of the \WL\ process
is zero and the second moment is given by Eqs.\
(\ref{wienerlevy:moment:a}), one realises that the process defined by
Eq.\ (\ref{wienner:increment}) coincides with the increment of a \WL\
process.%
\footnote{They exactly coincide if $w(\tau)$ is multiplied by
$1/\sqrt{2D}$.}


\section*{Proof of the convergence of the Heun scheme}

We shall check that the Heun scheme correctly generates the \KM\
coefficients, by carrying out the Taylor expansion of Eq.\
(\ref{heun:scheme}), accounting for Eq.\ (\ref{euler:support}).
Concerning the terms involving $\drift_{i}$, one has
\begin{eqnarray*}
\lefteqn{\half 
         \lrc{\drift_{i}(\tilde{\multi{y}},t+\Delt)
         +\drift_{i}[\multi{y}(t),t]}\Delt}
\hspace{3em}
\\[-3ex]
& &
=\half 
\bgbgc{
					\drift_{i}[\multi{y}(t),t]
     +\Dpar{\drift_{i}}{t}\Delt				
+\sum_{j}\Dpar{\drift_{i}}{y_{j}}
\overbrace{[\tilde{y}_{j}-y_{j}(t)]}%
^{\lefteqn{\qquad\sy{\drift_{j}\Delt+\sum_{k}\diff_{jk}\DelW_{k}}}}
+\cdots
					+\drift_{i}[\multi{y}(t),t]
					}\Delt
\\
& &
=\drift_{i}[\multi{y}(t),t]\Delt +{\cal O}[(\Delt)^{3/2}]
\;,
\end{eqnarray*}
whereas, the terms involving $\diff_{ik}$ can be expanded as
\begin{eqnarray*}
\lefteqn{\half 
         \lrc{\diff_{ik}(\tilde{\multi{y}},t+\Delt)
         +\diff_{ik}[\multi{y}(t),t]}\DelW_{k}}
\hspace{0.5em}
\\[-3ex]
&=&
\half 
\bgbgc{
	\diff_{ik}[\multi{y}(t),t]
     +\Dpar{\diff_{ik}}{t}\Delt
					+\sum_{k}\Dpar{\diff_{ik}}{y_{j}}
     \overbrace{[\tilde{y}_{j}-y_{j}(t)]}%
     ^{\lefteqn{\qquad
                \sy{\drift_{j}\Delt+\sum_{\ell}\diff_{j\ell}\DelW_{\ell}}}}
     +\cdots
		+\diff_{ik}[\multi{y}(t),t]
		}\DelW_{k}
\\
&=&
\diff_{ik}[\multi{y}(t),t]\DelW_{k}
+\sum_{k}\Dpar{\diff_{ik}}{y_{j}}[\multi{y}(t),t]
\sum_{\ell}\diff_{j\ell}[\multi{y}(t),t]\DelW_{\ell}\DelW_{k}
+{\cal O}[(\Delt)^{3/2}]
\;.
\end{eqnarray*}
In this case we have retained in $\tilde{y}_{j}-y_{j}(t)$ terms up to
order $(\Delt)^{1/2}$, which in the corresponding expansion of
$\drift_{i}$ are omitted since they yield terms of order
$(\Delt)^{3/2}$.
Finally, on inserting these expansions in Eq.\ (\ref{heun:scheme}), on
gets
\begin{eqnarray}
\label{ramsanher:scheme}
	y_{i}(t+\Delt)
\simeq
y_{i}(t) +\drift_{i}[\multi{y}(t),t]\Delt
+\sum_{k}\diff_{ik}[\multi{y}(t),t]\DelW_{k}
\nonumber\\
{}+\frac{1}{2}\sum_{k\ell}
\bgbgc{
	\sum_{j}\diff_{j\ell}[\multi{y}(t),t]
	\Dpar{\diff_{ik}}{y_{j}}[\multi{y}(t),t]
	}
\DelW_{k}\DelW_{\ell}
\;,
\end{eqnarray}
which corresponds to Eq.\ (2.8) of Ram{\'{\i}}rez-Piscina, Sancho and
Hern{\'{a}}ndez-Machado. 
Finally, to obtain the \KM\ coefficients, we have to average
Eq. (\ref{ramsanher:scheme}) for fixed initial values $\multi{y}(t)$
(conditional average).
To do so, one can use $\av{\DelW_{k}}=0$ and
$\av{\DelW_{k}\DelW_{\ell}}=(2D\Delt)\delta_{k\ell}$, to get
\begin{eqnarray*}
\bgbgav{
	\sum_{k}\diff_{ik}\DelW_{k}
				}
&=&
0
\;,
\\
\bgbgav{
				\frac{1}{2}\sum_{jk\ell}
				\diff_{j\ell}\Dpar{\diff_{ik}}{y_{j}}
				\DelW_{k}\DelW_{\ell}
				}
&=&
D\bgbg{
						\sum_{jk}
						\diff_{jk}
						\Dpar{\diff_{ik}}{y_{j}}
						}\Delt
\;,
\\
\bgbgav{
				\sum_{k}\diff_{ik}\DelW_{k}
    \sum_{\ell}\diff_{j\ell}\DelW_{\ell}
				}
&=&
2D
\bgbg{
  				\sum_{k}\diff_{ik}\diff_{jk}
						}\Delt
\;.
\end{eqnarray*}
Therefore, from Eq.\ (\ref{ramsanher:scheme}) one obtains
\begin{eqnarray*}
& &
\av{y_{i}(t+\Delt)-y_{i}(t)}
=
\bgbg{
 	\drift_{i}
  	+D\sum_{jk}
 	\diff_{jk}\Dpar{\diff_{ik}}{y_{j}}
	}\Delt
+{\cal O}[(\Delt)^{3/2}]
\\
& &
\av{\lrs{y_{i}(t+\Delt)-y_{i}(t)}\lrs{y_{j}(t+\Delt)-y_{j}(t)}}
=
2D
\bgbg{
  				\sum_{k}\diff_{ik}\diff_{jk}
						}\Delt
+{\cal O}[(\Delt)^{3/2}]
\;,
\end{eqnarray*}
which lead to the \KM\ coefficients (\ref{a_nu:langevin:n-dim}) via Eq.\
(\ref{jumpmoments:derivative:n-dim}).\qed

\newpage
\section*{Proof of the Box--Muller algorithm.}

We can verify that the transformation (\ref{boxmuller}) leads to a
pair of independent Gaussian random numbers as an exercise of {\em
transformation of variables\/} as introduced in Sec.\
\ref{transformation_variables}:
\begin{eqnarray*}
P_{W_{1},W_{2}}(w_{1},w_{2})
&=&
\Intdef{0}{1}{r_{1}}\!
\Intdef{0}{1}{r_{2}}
\delta[w_{1}-\sqrt{-2\ln(r_{1})}\cos(2\pi r_{2})]
\\[-4ex]
& &
\hspace{5.5em}{}\times
\delta[w_{2}-\sqrt{-2\ln(r_{1})}\sin(2\pi r_{2})]
\overbrace{P_{R_{1},R_{2}}(r_{1},r_{2})}^{1~\text{by~hypothesis}}
\;.
\end{eqnarray*}
Let us now introduce the substitution
\[
u_{1}(r_{1},r_{2})
=\sqrt{-2\ln(r_{1})}\cos(2\pi r_{2})
\;,
\quad
u_{2}(r_{1},r_{2})
=\sqrt{-2\ln(r_{1})}\sin(2\pi r_{2})
\;,
\]
the Jacobi matrix of which reads
\[
\left(
\begin{array}{cc}
\Dpar{u_{1}}{r_{1}} & \Dpar{u_{1}}{r_{2}}
\\[2ex]
\Dpar{u_{2}}{r_{1}} & \Dpar{u_{2}}{r_{2}}
\end{array}
\right)
=\left(
\begin{array}{cc}
-\frac{1}{r_{1}}\frac{1}{\sqrt{-2\ln(r_{1})}}\cos(2\pi r_{2})
&
-2\pi\sqrt{-2\ln(r_{1})}\sin(2\pi r_{2})
\\
-\frac{1}{r_{1}}\frac{1}{\sqrt{-2\ln(r_{1})}}\sin(2\pi r_{2})
&
\quad
2\pi\sqrt{-2\ln(r_{1})}\cos(2\pi r_{2})
\end{array}
\right)
\]
and the corresponding Jacobian (the determinant of this matrix) is given by
$\dpar{(u_{1},u_{2})}{(r_{1},r_{2})}=-\tyfrac{2\pi}{r_{1}}$.
Nevertheless, when changing the variables in the above integrals one needs
the {\em absolute value\/} of the Jacobian of the {\em inverse\/}
transformation, which is given by
$|\dpar{(r_{1},r_{2})}{(u_{1},u_{2})}|=\tyfrac{r_{1}}{2\pi}$.
Besides, $r_{1}(u_{1},u_{2})$ can be obtained from the above transformation:
$-2\ln(r_{1})=u_{1}^{2}+u_{2}^{2}$
$\Rightarrow$
$r_{1}=\exp[-\frac{1}{2}\left(u_{1}^{2}+u_{2}^{2}\right)]$.
On using all these results the probability distribution of $(w_{1},w_{2})$
is finally given by
\begin{eqnarray*}
P_{W_{1},W_{2}}(w_{1},w_{2})
&=&
\Intdef{-\infty}{\infty}{u_{1}}
\Intdef{-\infty}{\infty}{u_{2}}
\delta(w_{1}-u_{1})
\delta(w_{2}-u_{2})
\frac{1}{2\pi}\exp\lrs{-\half \left(u_{1}^{2}+u_{2}^{2}\right)}
\\
&=&
\frac{1}{\sqrt{2\pi}}\exp\lr{-\half w_{1}^{2}}
\frac{1}{\sqrt{2\pi}}\exp\lr{-\half w_{2}^{2}}
\;.
\end{eqnarray*}
This expression demonstrates that when $r_{1}$ and $r_{2}$ are
independent random numbers uniformly distributed in the interval
$(0,1)$, the random variables $w_{1}$ and $w_{2}$ given by the
transformation (\ref{boxmuller}) are indeed independent and
Gaussian-distributed with zero mean and variance unity.


\end{document}

%% file: commands.tex
\newcommand{\B}{\vec{B}}
\newcommand{\Beff}{\vec{B}_{\eff}}	
\newcommand{\bfl}{\vec{b}_{\fl}}	

\newcommand{\blangle}{\big\langle}
\newcommand{\brangle}{\big\rangle}

\newcommand{\Cint}{c}
\newcommand{\coupling}{\varepsilon}

\newcommand{\cross}{{{}\not}}
\newcommand{\D}{{\rm d}}

\newcommand{\Dcoeff}{D}

\newcommand{\Delt}{\Delta t}

\newcommand{\DelW}{\Delta W}
\newcommand{\diff}{B}
\newcommand{\drift}{A}

\newcommand{\dU}{\Delta U}

\newcommand{\eff}{{\rm e}}

\newcommand{\eq}{{\rm eq}}
\newcommand{\ffl}{f}
\newcommand{\Fint}{F}

\newcommand{\fp}{{\rm FP}}
\newcommand{\FP}{Fok\-ker--Planck}
\newcommand{\gmr}{\gamma}
\newcommand{\h}{{\rm h}}
\newcommand{\half}{\tfrac{1}{2}}

\newcommand{\Hs}{{\cal H}}
\newcommand{\Ham}{{\cal H}}
\newcommand{\Hsm}{\Hs^{({\rm m})}}	

\newcommand{\ila}{{\bf l}}
\newcommand{\ilb}{{\bf l'}}

\newcommand{\ir}{i}

\newcommand{\iu}{\mathrm{i}}
\newcommand{\Ker}{{\cal K}}

\newcommand{\kT}{k_{{\rm B}}T}

\newcommand{\Ls}{{\cal L}}
\newcommand{\Lsb}{\overline{\cal L}}

\newcommand{\LFP}{{\cal L}_{\fp}}
\newcommand{\LFPo}{{\cal L}_{\fp\,,0}}
\newcommand{\LFPext}{{\cal L}_{\fp\,{\rm ext}}}
\newcommand{\LFPb}{\overline{{\cal L}}_{\fp}}
\newcommand{\Lan}{\xi}

\newcommand{\llangle}{\left\langle}

\newcommand{\Ma}{}

\newcommand{\overMa}{}
\newcommand{\mm}{M}
\newcommand{\m}{\vec{\mm}}

\newcommand{\mi}{\mm_{i}}
\newcommand{\mj}{\mm_{j}}
\newcommand{\mk}{\mm_{k}}
\newcommand{\ml}{\mm_{\ell}}

\newcommand{\mr}{\mm_{r}}
\newcommand{\ms}{\mm_{s}}
\newcommand{\mx}{\mm_{x}}
\newcommand{\my}{\mm_{y}}
\newcommand{\mz}{\mm_{z}}

\newcommand{\Mb}{m_{b}}
\newcommand{\ea}{a}
\newcommand{\ec}{c}
\newcommand{\eF}{f}
\newcommand{\eP}{P}
\newcommand{\ePm}{{\bf P}}
\newcommand{\eQ}{Q}
\newcommand{\eQm}{{\bf Q}}
\newcommand{\eS}{S}

\newcommand{\mc}{\boldsymbol{c}}

\newcommand{\mQ}{{\bf Q}}

\newcommand{\N}{N}
\newcommand{\Order}{{\cal O}}
\newcommand{\pbra}{\big\{}
\newcommand{\pket}{\big\}}
\newcommand{\qed}{\quad{\rm Q.E.D.}\quad{}}

\newcommand{\rat}{\rho}

\newcommand{\rrangle}{\right\rangle}
\newcommand{\sP}{p}

\newcommand{\sQ}{q}
\newcommand{\smag}{s}
\newcommand{\s}{\vec{\smag}}

\newcommand{\si}{\smag_{i}}

\newcommand{\sy}{\smag_{y}}

\newcommand{\tD}{\tau_{{\rm D}}}

\newcommand{\tL}{\Lambda^{\lin}}
\newcommand{\TL}{\hat{\Lambda}^{\lin}}

\newcommand{\tp}{s}

\newcommand{\vecpro}{\wedge}

\newcommand{\Vint}{V}

\newcommand{\W}{P}
\newcommand{\wa}{\omega_{a}}
\newcommand{\wb}{\omega_{b}}

\newcommand{\Weq}{\W_{\eq}}

\def\case#1#2{{\textstyle \frac{#1}{#2}}}

\def\multi#1{{\bf{#1}}}

\def\boxequation#1{{\fboxsep 0.6\topsep\fbox{$\displaystyle{#1}$}}}

\def\boxeqnarray#1#2%
{{\fboxsep
-1.ex\fbox{\parbox{#1}{\begin{eqnarray*}#2\end{eqnarray*}}}}}

\newlength{\minuslength}



%% file: newcommands_tesis
\def\multi#1{{\bf{#1}}}

\def\FP{Fok\-ker--Planck}
\def\WL{Wie\-ner--L\'{e}\-vy}
\def\ChK{Chapman--Kolmogorov}
\def\OU{Ornstein--Uhlenbeck}
\def\KM{Kram\-ers--Moyal}

\def\tyfrac#1#2{{#1}/{#2}}

\def\Dpar#1#2{\frac{\partial{#1}}{\partial{#2}}}
\def\dpar#1#2{\partial{#1}/\partial{#2}}
\def\Der#1#2{\frac{d{#1}}{d{#2}}}

\def\Intdef#1#2#3{{\displaystyle \int_{#1}^{#2}}\! \D{#3}\,}

\def\eval#1#2{\left.{#1}\right|_{{#2}}}

\def\bgbg#1{\bigg({#1}\bigg)}

\def\BBs#1{\Big[{#1}\Big]}
\def\bgbgs#1{\bigg[{#1}\bigg]}

\def\bgbgc#1{\bigg\{{#1}\bigg\}}

\def\bbav#1{\big\langle{#1}\big\rangle}
\def\BBav#1{\Big\langle{#1}\Big\rangle}
\def\bgbgav#1{\bigg\langle{#1}\bigg\rangle}

\def\lr#1{\left({#1}\right)}
\def\lrs#1{\left[{#1}\right]}
\def\lrc#1{\left\{{#1}\right\}}

\def\av#1{\left\langle{#1}\right\rangle}

\def\cum#1{\left\langle{\left\langle{#1}\right\rangle}\right\rangle}

\def\sy#1{{\scriptstyle {#1}}}